\documentclass[12pt]{article}
\usepackage[margin=1in]{geometry}
\usepackage{amssymb}
\usepackage{amsmath}
\usepackage{amsthm}
\usepackage{gensymb}
\usepackage{dsfont}
\usepackage{xcolor}
\usepackage{todonotes}
\usepackage{url}
\usepackage{float}
\usepackage{graphicx}

\usepackage[font=scriptsize]{caption}
\usepackage{wrapfig}
\usepackage{tocloft}
\theoremstyle{definition}

\usepackage{color}
\usepackage{xkeyval,xcolor}

\DeclareMathOperator{\var}{Var}
\begin{document}

\noindent \begin{minipage}[h]{\textwidth}
\centering

{\large{\textbf{Scale-Rich Network-Based Metamaterials}}}

{\small
{Csaba Both\textsuperscript{1,\textdagger}, Andrew Yen-Jong Chen\textsuperscript{2,\textdagger}, Ting-Ting Gao\textsuperscript{1}, Niek Mooij\textsuperscript{3,4}, Mohammad Charara\textsuperscript{2},  Carlos M. Portela\textsuperscript{2,$\ast$}, Albert-László Barabási\textsuperscript{1,5,6,$\ast$}\\

{\footnotesize $^1$ Network Science Institute, Northeastern University, Boston, MA, USA},\\
{\footnotesize $^2$ Department of Mechanical Engineering, Massachusetts Institute of Technology, Cambridge, MA, USA},\\
{\footnotesize $^3$ Mathematical Institute, Utrecht University, Utrecht, Netherlands},\\
{\footnotesize $^4$ Centre for Complex Systems Studies, Utrecht University, Utrecht, Netherlands},\\
{\footnotesize $^5$  Department of Medicine, Brigham and Women's Hospital, Harvard Medical School, Boston, MA, USA},\\
{\footnotesize $^6$  Department of Data and Network Science, Central European University, Hungary}

{\scriptsize(\textsuperscript{$\ast$}Corresponding authors: cportela@mit.edu, barabasi@gmail.com)}

{\scriptsize(\textsuperscript{\textdagger}These authors contributed equally to this work.)}
}
}

\begin{abstract}
\fontsize{12}{16}\selectfont

Materials, at their essence, are networks defined by homogeneity: uniform bonds, fixed thicknesses, and discrete length scales.
Mechanical metamaterials, while representing structurally more diverse microstructures, remain defined by the homogeneity of their unit cells, pore sizes, or repeating features.
In contrast, as network science has revealed, real-world and biological systems---from the Internet to the brain---derive their function from broad, multiscale variability in connectivity and link length.
Here we introduce Scale-Rich (SR) metamaterials, a design framework that embeds network heterogeneity into mechanical metamaterials, achieving order-of-magnitude heterogeneity in ligament lengths, thicknesses, and connectivity.
Governed by only two parameters, SR networks span orders of magnitude in structural features, overcoming prior constraints in metamaterial design.
Translating these network models into physically realizable materials, we use simulations and experiments to show that SR metamaterials exhibit properties inaccessible to traditional single-scale systems, including highly tunable elastic anisotropy, delocalized nonlinear deformation with high energy absorption, and programmable acoustic wave control.
This network-science-based paradigm establishes a minimal yet universal framework for engineering multifunctional materials whose mechanical and acoustic behavior emerge directly from scale diversity itself.

\end{abstract}

\end{minipage}

\clearpage

\section*{Introduction}
At their core, materials are networks: in crystalline solids, atoms form regular lattices defined by uniform bonds, while in mechanical metamaterials, a distinct graph encodes the interactions among structural components.
These graphs share a critical feature---an exceptional homogeneity of their local properties.
Crystalline lattices are defined by unique bond lengths and interaction strengths, and even disordered systems such as amorphous microstructures display narrow distributions of connectivity and bond strength.
Mechanical metamaterials, or architected materials, inherit this homogeneity: despite their structural diversity, they are typically governed by one or a few discrete length scales
\cite{surjadiPortela2025,Bauer2017}, even in the presence of hierarchy \cite{meza2015resilient,Vigliotti2013} or modest heterogeneity \cite{schumacher2015microstructures,medina2023nonlinear,lew2023designing}.
Unit-cell design further enforces uniform connectivity \cite{mirzaali2019auxeticity,deshpande2001foam,Mousanezhad2015} and geometric compatibility constraints \cite{liu2022growth,zheng2023unifying}, while plate- and shell-based architectures too rely on fixed thicknesses and uniform pore sizes \cite{BergerEtAl2017,portela2020extreme}.

However, as network science has revealed in the past two decades, real-world networks rarely exhibit the homogeneity of crystalline or amorphous solids.
Instead, systems as diverse as protein–protein interaction networks, social networks, and global infrastructures such as the Internet display remarkable structural variability: node degrees span orders of magnitude and interaction strengths cover broad ranges
\cite{cui2010complex, barabasi2016networkscience,caldarelli2007scale,bullmore2009complex,bardoscia2017pathways, menczer2020first}.
This heterogeneity underlies their robustness \cite{albert2000error,cohen2000resilience,cohen2001breakdown,motter2004cascade} and resilience \cite{gao2016universal}, facilitates dynamical processes \cite{pastor2001epidemic}, and enables control through a small subset of nodes \cite{liu2011controllability,cornelius2013realistic,d2023controlling}.
Even \textit{physical networks} \cite{dehmamy2018structural,posfai2024impact,pete2024physical}, like neuronal and vascular networks, whose nodes and links have material presence, are highly heterogeneous, with degrees, link lengths, and strengths spanning several orders of magnitude \cite{piazza2025physical,paetzold2021whole}.

Elements of multiscale organization, while largely absent from most natural and engineered materials, are a hallmark of many biological systems.
Dragonfly wings, for example, display vein lengths spanning more than three orders of magnitude, from $\ell_{\min} = 0.1\,\mathrm{mm}$ to $\ell_{\max} \approx 50\,\mathrm{mm}$ \cite{eshghi2024allometric}.
Leaves exhibit equally complex venation patterns, with length scales ranging from $\ell \approx 10^{-1}$ to $10^{3}\,\mathrm{mm}$ and connectivity spanning roughly three decades \cite{price2011leaf,blonder2020linking}.
Spider webs are made of string whose lengths extend from $\ell \approx 1$ to $10^{3}\,\mathrm{mm}$ \cite{xie2023investigation,soler2016secondary}, and bones represent hierarchical architectures built from rod- and plate-like elements, whose lengths range from $\ell \approx 100$ to $1500$ \textmu{}m and thicknesses from $\lambda \approx 40$ to $400$ \textmu{}m \cite{lee2017three}.

While in the past two decades numerous models have been proposed to capture and analyze the broad range of length scales that emerge in complex networks \cite{barabasi1999emergence, dorogovtsev2000structure, bianconi2007entropy,voitalov2019scale}, no comparable framework exists to understand the role of such heterogeneity in material systems.
To address this gap, here we introduce a new design paradigm for mechanical metamaterials that leverages a network-science-based approach to generate architectures with a controlled, yet broadly varying, span of connectivities, length scales, and interaction strengths.
We refer to these materials as \textit{Scale-Rich (SR) networks}, as they allow the simultaneous coexistence of lengths, thicknesses, and connectivities that span orders of magnitude within the same system.
SR networks offer a simple but systematic framework for designing architected materials beyond conventional local geometric rules, enabling a controlled exploration of the role of heterogeneity and disorder.
Using both computational modeling and experiments, we show that SR metamaterials, governed by only two parameters, enable properties unattainable in single-scale systems: a 24-fold tunable range of elastic anisotropy across densities, stable and delocalized nonlinear deformation with high energy absorption, and programmable acoustic responses that allow the design of mechanical-wave lenses with tailored refractive indices.
Taken together, these results establish Scale-Rich metamaterials as a versatile platform for engineering multifunctional systems with programmable performance across scales and offer a minimal yet powerful framework for inverse design.

\section*{The Scale-Rich model}
\begin{figure}[htpb]
    \centering
    \includegraphics[width=1\linewidth]{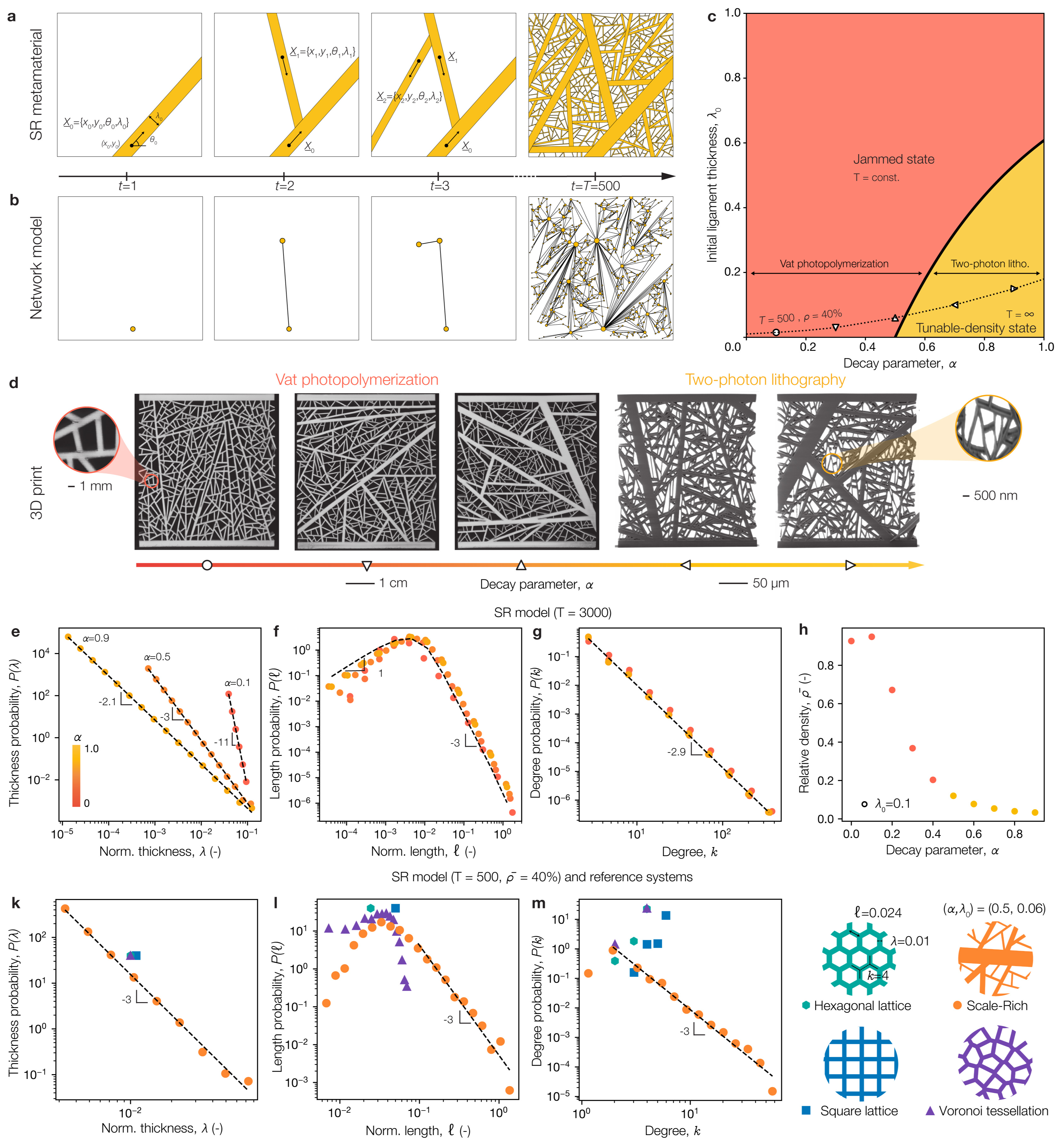}
    \caption{\textbf{Scale-Rich networks.} \textbf{(a)} Generation process of an SR metamaterial, starting at time step $t=1$ where an initial ligament of thickness $\lambda_0$ is generated from a randomized coordinate $(x_0,y_0)$ in a randomized orientation $\theta_0$.
    Additional ligaments are then generated at each time step $t$ (nucleated with randomized coordinates and orientations) with thickness $\lambda_t = \lambda_0 t^{-\alpha}$, terminating at the intersection with another ligament or the domain boundary.
    \textbf{(b)} The network behind the SR metamaterial, inspired by road networks \cite{rosvall2005networks}, represents ligaments as nodes and intersections as links.
    The panels show the growth of the network corresponding to the configurations shown in (a).
    \textbf{(c)} The analytically derived ($\alpha, \lambda_0$) phase diagram (two-parameter design space) of SR metamaterials.
    For small $\alpha$ the system gets jammed, i.e. it does not allow the introduction of new ligaments.
    For large $\alpha$, the system can indefinitely accommodate additional ligaments, reaching a phase where the density of the system is tunable (see SI ~\ref{subsec: phase_diagram}).
    Experimentally realized SR metamaterials with $T=500$ and relative density $\bar{\rho}=40\%$, via two types of 3D-printing techniques, are depicted with white symbols.
    \textbf{(d)}
    Images of 3D-printed SR metamaterial specimens, at centimetre (vat photopolymerization) and micrometre (two-photon lithography) scales, corresponding to the white symbols in (b).
    Distributions of \textbf{(e)}, ligament thickness $P(\lambda)$; \textbf{(f)}, ligament length $P(\ell)$; and \textbf{(g)}, ligament degrees (number of connections per ligament) $P(k)$, for $\lambda_0 = 0.1$, all exhibit a power-law tail.
    \textbf{(h)} Relative density modulation as a function of $\alpha$ for $\lambda_0 = 1$, averaged over 20 runs for $T = 30000$.
    \textbf{(k)--(m)} Thickness, length, and degree distribution of SR metamaterials ($T=500$, $\alpha=0.5$, $\lambda_0=0.06$) compared to reference systems: hexagonal lattice, square lattice, and Voronoi tessellation (all at $\bar{\rho}=40\%$).}

    \label{fig:model}
\end{figure}

To construct an SR metamaterial, we start with a unit square domain ($L = 1$) and iteratively add ligaments whose thickness $\lambda$ decreases with time $t$ as $\lambda_t = \lambda_0 t^{-\alpha}$, where $\lambda_0$ is the initial thickness and
$\alpha$ is the decay exponent.
The generation process consists of two steps (Fig.~\ref{fig:model}a): (i) choose a random nucleation point $(x_t, y_t)$ within the $L^2$ domain and a random angle $\theta_t$; (ii) grow a ligament of thickness $\lambda_t$ from the nucleation point in both directions, defined by $\theta_t$, until it intersects other ligaments or the boundary.
We repeat (i) and (ii) until a predefined time $t=T$ or until we are unable to add new ligaments, leading to a state of jamming \cite{posfai2024impact,baule2018edwards} (see SI \ref{subsec: model_limit} for more examples).

In the model, the length of newly added ligaments decreases with time as $\ell_t \approx ((1-\sum\limits_{j=0}^{t-1} \ell_j\lambda_j)/t)^{1/2}$, as the remaining undivided area follows the recursive relation $A_t = A_{t-1} - \lambda_0 A_{t-1}^{1/2} (t-1)^{-(\alpha + 1/2)}$ (SI \ref{subsec: phase_diagram}).
This reduction is governed by the convergence of the sum
$A_{t}^\mathrm{total} = 1- \sum\limits^{t-1}_{j=0}\ell_j \lambda_j \approx 1 - \lambda_0 \sum\limits^{t-1}_{j=0}A_{j-1}^{1/2}(j-1)^{-(\alpha +1/2)},$ whose asymptotic behavior defines two distinct phases (Fig.~\ref{fig:model}c): jammed and tunable-density states.
If $\alpha <  1/2$, or if $\alpha > 1/2$ and $\lambda_0 > (\ell_0 \zeta(2\alpha))^{-1}$, where $l_0 \approx 0.946$ \cite{keller1971}, the sum diverges, hence \( A^{\text{total}}_{\infty} \to 0 \) resulting in complete coverage and jamming (Fig.~\ref{fig:model}c, orange area).
For other parameters, the sum converges, and the system reaches a finite tunable relative density (i.e., fill fraction) $\bar{\rho} \approx l_0 \lambda_0 \zeta(\alpha + \beta)$, where $\beta$ is the ligament length decay rate (Fig.~\ref{fig:model}c, yellow area), a phase of particular interest for lightweight materials \cite{meza2014strong}.
As we show in Fig.~\ref{fig:model}d, systems with small $\alpha$ are physically realizable across length scales via additive 
manufacturing processes like vat photopolymerization, while higher $\alpha$ designs require sub-micron resolution offered by two-photon lithography.

Both regular and irregular architectures have unique characteristic length scales, characterized by either a unique or narrowly varying coordination number (degree, $k$), length ($\ell$), and ligament thickness ($\lambda$) \cite{kiang1966random, portela2020extreme}.
The distinguishing feature of an SR network is its lack of characteristic scale for any of these three defining dimensions.
Indeed, as the ligament thickness follows $\lambda_t = \lambda_0 t^{-\alpha}$, the overall thickness distribution $P(\lambda)$ follows the power-law distribution $P(\lambda) \approx \lambda^{-(1+1/\alpha)}$ (Fig.~\ref{fig:model}e and SI~\ref{sec:line_thickness_distribution}).

To determine the ligament length distribution, we must consider that the ligament lengths decay as $\ell_t = \ell_0 t^{-\beta}$, where $\beta = 0.5$, except at the boundary, where $\beta = \alpha$ (SI~ \ref{sec:expected_length}).
As we derive in SI~\ref{sec:length_distribution}, the resulting ligament lengths follow an Erlang distribution:
for high $\ell$  ($\ell \approx \sqrt{2}$) the distribution decays as a power-law tail $P(\ell) \approx\ell^{-\mu}$ with $\mu = 1 + 1/\beta$, while for low  $\ell$ it increases with exponent $\mu = -1$ (Fig.~\ref{fig:model}f).

A fundamental network characteristic is the degree (or coordination number) distribution \cite{barabasi1999emergence,barabasi2016networkscience,caldarelli2007scale}.
It is tempting to consider each ligament intersection a node and each connecting segment a link, an approach often adopted in physical networks \cite{paetzold2021whole, dorkenwald2024neuronal}.
However, in the SR model, ligament thickness spans multiple orders of magnitude, prompting us to adopt a representation commonly used in transportation networks \cite{rosvall2005networks,pete2024physical},
where each ligament (or road) is conceptualized as a single node, and intersections are represented as edges (Fig.~\ref{fig:model}b).
In this representation, the degree distribution $P(k)$ follows the power law  $P(k) \approx k^{-\gamma}$ with $\gamma = 3$ (Fig.~\ref{fig:model}g, SI~\ref{fig:sup_thickness} and \ref{degree_distribution}).
The power law emerges thanks to a spatial preferential attachment \cite{barabasi1999emergence}: longer ligaments have linearly more opportunities to intersect with newly added ligaments (SI~\ref{fig:sup_thickness}).

Taken together, we find that in SR networks all three salient features of the system's architecture---coordination number (degree, $k$), length  ($\ell$), and ligament thickness ($\lambda$)---follow power law distributions.
In other words, while in conventional material systems $k$, $\ell$, and $\lambda$ are constant or vary in a very narrow range, in SR materials they span several orders of magnitude within the same sample, encoding the coexistence of many different length scales.
The unique heterogeneity of the SR network raises an important question: do SR metamaterials posses physical properties distinct from those of single-scale natural and architected materials?
\section*{Scale-Rich material microstructures}

We begin by determining the mechanical characteristics of SR microstructures, comparing them against three single-scale microstructures commonly used in engineering applications: the square lattice,  the hexagonal lattice, and the irregular Voronoi tessellation.
By generating finite domains of each microstructure with a sufficiently large number of characteristic features, we confirm that the $P(\lambda)$, $P(\ell)$, and $P(k)$ for square and hexagonal lattices result in delta functions, while the Voronoi tessellation is characterized by narrow distributions (Fig.~\ref{fig:model}k-m).

To extend the comparison beyond theoretical differences between architectures, we experimentally and numerically determined the mechanical properties of these metamaterial microstructures, using vat photopolymerization 3D-printing (VPP) to fabricate centimeter-scale specimens out of elastic-plastic polymer resin (Fig.~\ref{fig:model}d, left; SI~\ref{sec:SI-material_properties}), with a feature resolution of ${\sim}100$ \textmu{}m.
We focused on SR systems with $T = 500$---a point where the homogenized mechanical properties show convergence and boundary effects are minimized (SI Fig.~\ref{fig:SI-SR-convergence}).
Given this feature resolution and the 500-ligament constraint, the relative densities achievable with VPP ranged between $20\% \leq \bar{\rho} \leq 75\%$; we can demonstrate realization of lower fill fractions within the tunable density phase ($\bar{\rho} < 20\%$, Fig.~\ref{fig:model}c, SI~\ref{sec:SI-fabrication_limit}) through the use of two-photon lithography (Fig.~\ref{fig:model}d, right) with sub-micron minimum feature sizes.
Next, we characterized the linear and nonlinear mechanical responses of these SR microstructures using the macroscopic samples, in combination with computational mechanics models.

\begin{figure}[H]
    \centering
     \includegraphics[width=1\linewidth]{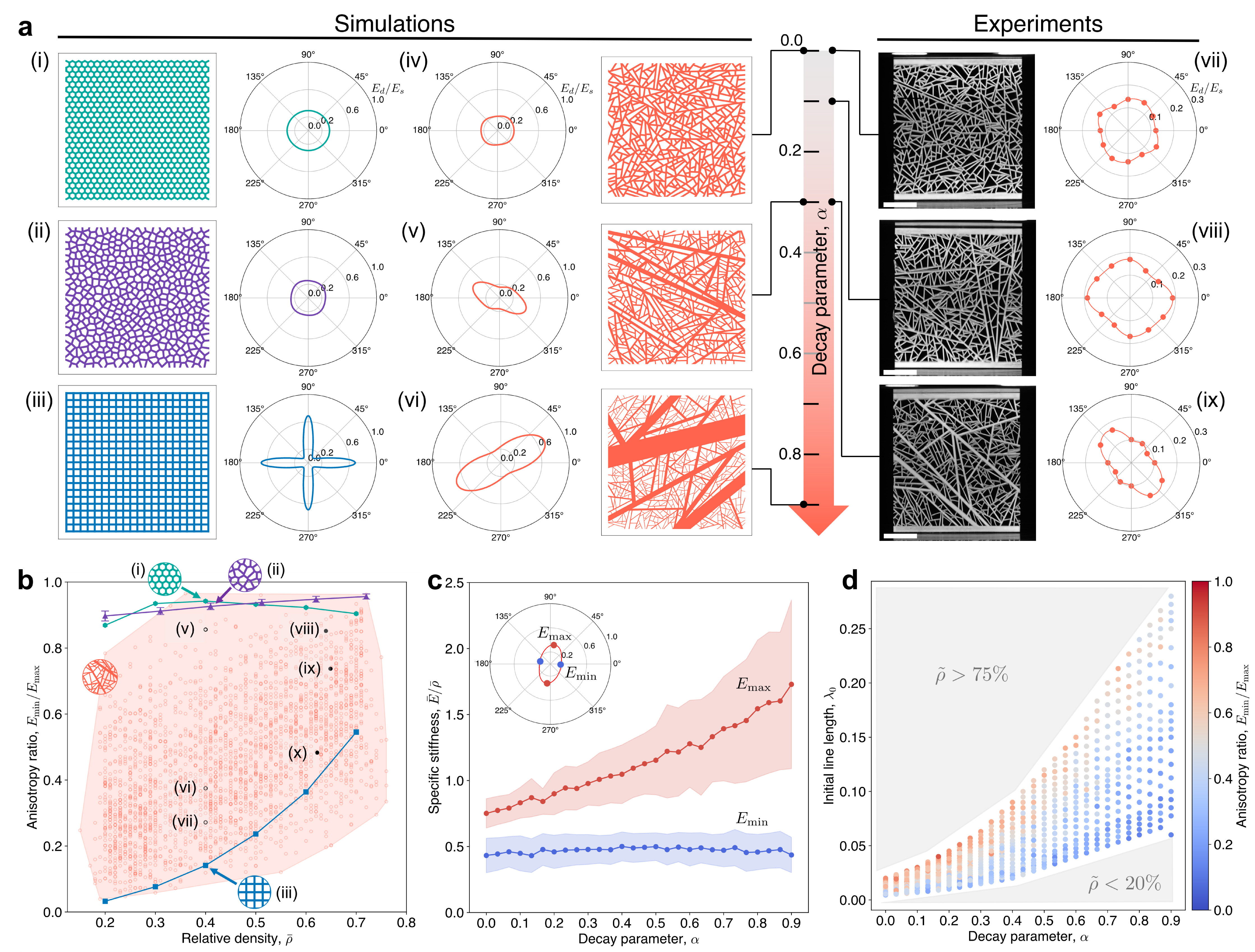}
    \caption{\textbf{Linear-elastic behavior of the Scale-Rich microstructure.}
    \textbf{(a)} Directional stiffness plots of the reference hexagonal (i), Voronoi (ii), and square (iii) lattices from finite element simulation, compared with SR samples with $\alpha = 0, 0.3, 0.9$ at the same relative density (v)-(vii). Panels (vii)-(ix) show the experimentally measured directional stiffness for SR materials with $\alpha = 0, 0.1, 0.3$, validating the theoretically predicted stiffness anisotropy. Scale bars in (vii)-(ix) measure 10 mm. \textbf{(b)} Anisotropy-density plot, illustrating the broad range of anisotropy achievable with the SR framework compared to the narrow range of values attained by the reference samples.
    The black filled circles (vii)-(ix) show the experimentally measured anisotropy.
    \textbf{(c)} As the decay parameter $\alpha$ increases, anisotropy arises because the maximum specific directional stiffness increases, whereas the minimum specific directional stiffness is constant. The solid lines represent the mean across all samples, and the shaded regions correspond to a $\pm 1$ standard deviation from the mean value.
    \textbf{(d)} Phase diagram in $(\alpha, \lambda_0)$ space of the anisotropy, which functions as a design map for a tunable anisotropic response.}
    \label{fig:elastic_alt}
\end{figure}

\paragraph*{Highly tunable linear-elastic responses.}
To determine the linear-elastic response of SR metamaterials, we performed computational homogenization using the finite-element method (SI~\ref{sec:homgenization}), extracting a directional modulus $E_d$ in each direction $\mathbf{e}_d$ and calculating a non-dimensional modulus $\bar{E} \equiv E_d / E_s$, where $E_s$ is the Young's modulus of the constituent material.
At a relative density of $\bar{\rho} = 40\%$, we performed this analysis on the three conventional geometries (square, honeycomb, and Voronoi) and 47 SR geometries, demonstrated by the representative realizations in Fig.~\ref{fig:elastic_alt}a, via polar representations of $\bar{E}$ as a function of the probing direction.
While the honeycomb and Voronoi structures display a near-constant $\bar{E}$ in all orientations (i.e., an isotropic stiffness response, as expected \cite{Symons2008}, Fig.~\ref{fig:elastic_alt}a, (i, ii)), the square lattice is highly anisotropic, displaying high stiffness parallel to its struts and minimal stiffness in any other direction (Fig.~\ref{fig:elastic_alt}a, (iii)).
In contrast to these near extremes, we find the behavior of the SR structures to be modulated by $\alpha$: for $\alpha \approx 0$ the stiffness is nearly isotropic (Fig.~\ref{fig:elastic_alt}a, (iv)), and as $\alpha \to 1$ a principal orientation of maximum stiffness emerges (Fig.~\ref{fig:elastic_alt}a, (v, vi)). Altogether, the full set of SR architectures evidenced highly tunable (an)isotropy exceeding that of the three conventional geometries combined.

To quantify the mechanical implications of this observed range in anisotropy across the full space of SR architectures, we plot the elastic anisotropy ratio $\eta_E \equiv E_\mathrm{min} / E_\mathrm{max}$, where $E_\mathrm{min} \equiv \mathrm{min}(\bar{E})$ and $E_\mathrm{max} \equiv \mathrm{max}(\bar{E})$, as a function of relative density for the reference geometries and 2600 distinct SR microstructures with different $\alpha$ and $\lambda_0$ (Fig.~\ref{fig:elastic_alt}b).
We confirm that the hexagonal lattice ($0.87 \leq \eta_E \leq 0.94$) and the Voronoi tessellation ($0.88 \leq \eta_E \leq 0.97$) remain nearly isotropic across a wide range of relative densities.
For the square lattice, the anisotropy decreases from $\eta_E = 0.03$ at $\bar{\rho} = 20\%$ to $\eta_E = 0.55$ at $\bar{\rho} = 70\%$.
By contrast, we find that the SR materials span a 24-fold range of elastic anisotropies, covering 77\% of the available area on the density-anisotropy plot (Fig.~\ref{fig:elastic_alt}b) and reaching values not accessible to the reference systems.
Indeed, by changing $\alpha$ and $\lambda_0$, we can attain highly anisotropic systems ($\eta_E \leq 0.2$) across a wide range of densities ($17\% \leq \bar{\rho} \leq 64\%$) as well as nearly isotropic systems ($\eta_E \geq 0.85$) in the range $33\% \leq \bar{\rho} \leq 72\%$.
Simultaneously, we can generate systems of widely varying anisotropies at any fixed relative density: at $\bar{\rho} = 20\%$, we span $0.04 \leq \eta_E \leq 0.78$; at $\bar{\rho} = 40\%$, we span $0.12 \leq \eta_E \leq 0.88$; and at $\bar{\rho} = 70\%$, we span $0.26 \leq \eta_E \leq 0.93$.

To elucidate the origin of this tunable anisotropy we plot the mass-normalized (i.e., specific) minimum and maximum stiffness values ($E_\mathrm{min} / \bar{\rho}, E_\mathrm{max} / \bar{\rho}$) as a function of the decay parameter $\alpha$ across the entire sample set (Fig.~\ref{fig:elastic_alt}c).
While the minimum specific stiffness is largely independent of $\alpha$, increasing $\alpha$ while keeping density constant enables an increase in the maximum achievable specific stiffness, leading to robust tunability in the architecture space.
This tunable anisotropy is best captured by the $(\alpha, \lambda_0)$ configuration space (Fig.~\ref{fig:elastic_alt}d), which shows that anisotropy increases with $\alpha$ and decreases with $\lambda_0$.

To validate our finite-element model predictions and confirm the isotropic and anisotropic responses, we performed uniaxial compression experiments across various SR geometries (Fig.~\ref{fig:elastic_alt}a, (vii)-(ix) and Fig.~\ref{fig:elastic_alt}b, filled black circles).
For each choice of $\alpha$ and $\lambda_0$, we generated a sufficiently large domain of size $(\sqrt{2}L)^2$, from which oriented subdomains of size $L^2$, rotated in increments of 30$\degree$ from $0\degree$ to $150\degree$, were extracted and fabricated via VPP (SI~\ref{sec:SI-experiment-stiffness}).
We normalized the experimentally measured stiffness of each oriented subdomain by the modulus of the constituent material, as plotted in Fig.~\ref{fig:elastic_alt}a, (vii)-(ix), confirming that at constant density increasing $\alpha$ directly leads to greater anisotropy (SI Fig.~\ref{fig:experiment-stiffness}).

Using a scalar metric that accounts for alignment of material along a certain loading direction (SI~\ref{sec:SI-geometric anisotropy}) explains this correlation: the origin of the anisotropy in the SR systems can be explained by the inherent heterogeneity in ligament orientations and sizes dictated by $\alpha$. This scalar \textit{geometric anisotropy} metric agrees with the elastic anisotropy $\eta_E$, providing an intuitive, computationally inexpensive predictive metric that can be embedded in the design framework.
Altogether, this highly-tunable anisotropy of SR architectures lends itself toward structural applications where highly tunable and spatially programmable stiffness and mass are required, from ultra-sensitive wearable pressure sensors \cite{Ha2021} to robotic limbs or grippers with programmable compliance \cite{Shintake2018}, or medical implants and cell scaffolds requiring a tunable anisotropic response \cite{Baruffaldi2021}.

\begin{figure}[htpb]
    \centering
    \includegraphics[width=0.85\linewidth]{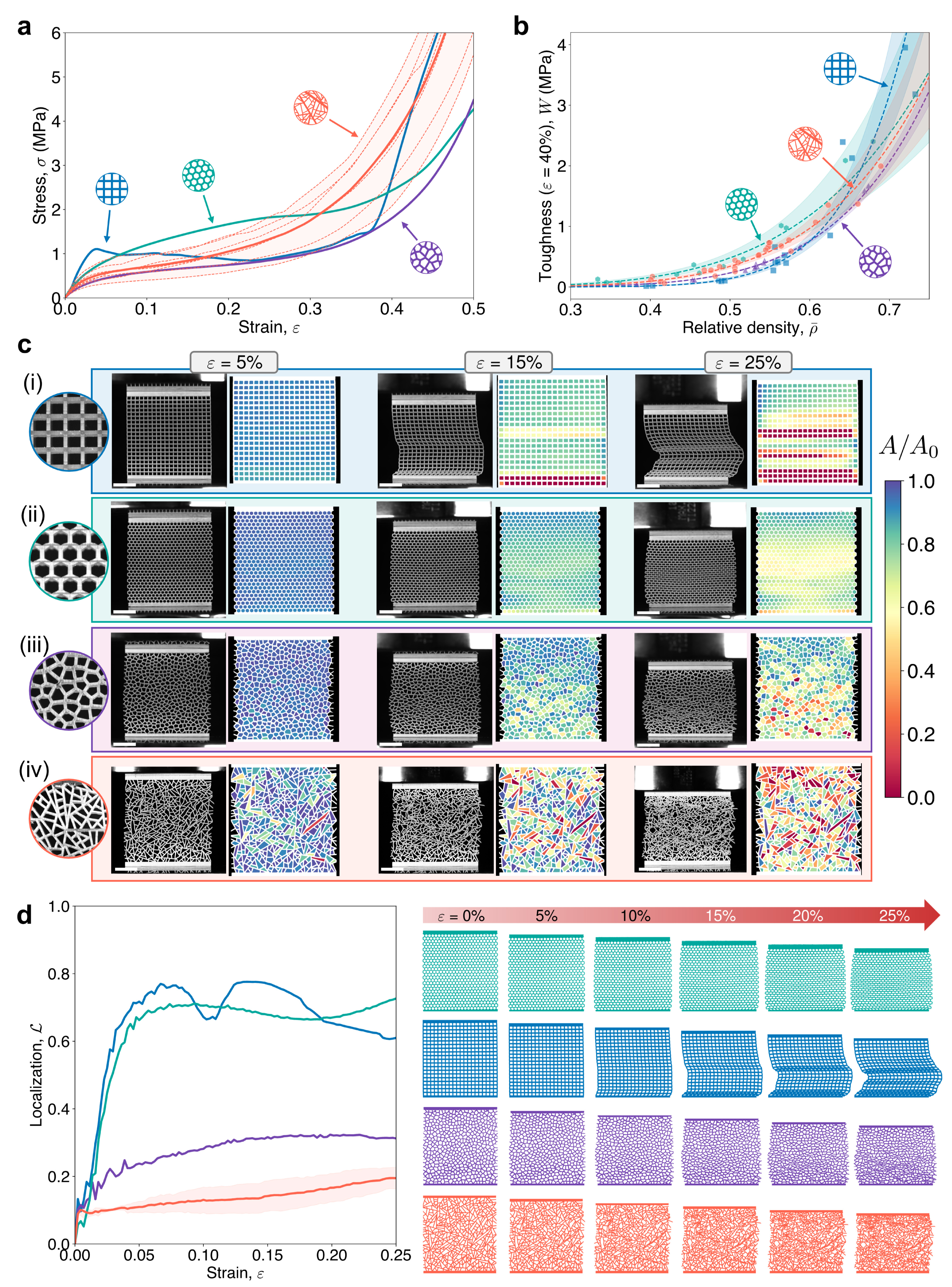}
    \caption{ \textbf{Nonlinear behavior of the Scale-Rich microstructure.}
    \textbf{(a)} Experimental stress–strain curves of Scale-Rich and reference samples, for $\bar{\rho}_{\mathrm{nom}} = 42\%$, under uniaxial compression. For the reference samples, the mean behavior of three replicate specimens is plotted. For the SR geometries, the solid line represents the mean of all samples with $\alpha = 0$, whereas the dashed curves are individual samples. The response of SR samples with $\alpha > 0$ is given in SI Fig.~\ref{fig:SI-high-alpha-stress-strain}.
    \textbf{(b)} Experimental toughness-density Ashby plot, showing the energy absorption capacity across all relative densities.
    \textbf{(c)} Evolution of localized deformation with increasing macroscopic strain: still images together with computed areal Jacobians, $J = A/A_0$, for each cell. The geometries dominated by a characteristic length scale all develop regions of compaction where localization is observed; only the Scale-Rich sample avoids the formation of compaction bands. We tracked images up to a macroscopic strain of 25\%, after which compaction precludes the accurate computation of Jacobian values. All scale bars measure 10 mm.
    \textbf{(d)} Evolution of the localization parameter $\mathcal{L}$ in each geometry together with snapshots of the deformation evolution as recovered from still images.
    Here, the mean and standard deviation of three unique SR specimens is plotted.
    }
    \label{fig:nonlinear}
\end{figure}

\paragraph*{Delocalized nonlinear deformations.}
Next, we characterized the nonlinear response of SR metamaterials by conducting large-deformation uniaxial compression experiments (SI~\ref{sec:fabrication_compression}) on the three reference geometries and SR systems of $\alpha = 0$, chosen to minimize the impact of elastic anisotropy and emphasize the effects of the heterogeneity of ligament lengths (Fig.~\ref{fig:nonlinear}a, 3 specimens each, all at nominal $\bar{\rho} = 42\%$).
Experiments on the square, hexagonal, and Voronoi tessellations showed qualitative differences between each architecture, including differences in densification strain, while the square tessellation exhibited undesirable softening at the onset of nonlinearity (Fig.~\ref{fig:nonlinear}a, blue curve).
Repeated samples of each architecture showed minimal variation (SI Fig.~\ref{fig:errorbars}).
On the other hand, all realizations of SR architectures (Fig.~\ref{fig:nonlinear}a, orange dotted lines) demonstrated a monotonically increasing stress-strain response, coexisting with a wide envelope of responses, elucidating the tunability enabled by the SR framework and breaking the structure-response constraints observed in the single-scale reference geometries.

From each stress-strain response, we computed the toughness, measured as the integral under the stress-strain curve to an engineering strain of 40\%---prior to the onset of complete densification.
These toughness metrics, when plotted against relative density (Fig.~\ref{fig:nonlinear}b), showcase the ability of SR architectures to match or surpass the mechanical resilience of the three reference architectures.
Namely, SR architectures could serve as a replacement material system for protective applications, avoiding undesirable softening responses and minimizing the peak stresses during deformation, without sacrificing ultimate energy absorption.

To understand the deformation mechanisms that lead to these benefits, we extracted still frames from the experiments and computed the areal Jacobian $J_i \equiv A_i/A_{i,0}$ of each $i$-th cell (the relative change in void area, Fig.~\ref{fig:nonlinear}c) where $A_{i,0}$ and $A_i$ denote the undeformed and deformed cell areas, respectively \cite{Bauer2021}.
The square lattice was prone to buckling of vertically aligned ligaments, resulting in the observed local maxima in its stress-strain response (Fig.~\ref{fig:nonlinear}a, blue curve).
The buckling events were observed to begin at a single ligament, followed by a horizontal cascade of buckling events that spanned the whole length of the specimen (Fig.~\ref{fig:nonlinear}c, (i)), leading to a sudden and unpredictable drop in its load capacity.
This process repeated in a layer-by-layer fashion until the entire specimen buckled.
Similarly, the hexagonal lattice (teal; Fig.~\ref{fig:nonlinear}c, (ii)) also developed wide compaction bands that spanned the width of the lattice.
Finally, the Voronoi tessellation did not succumb to well-defined shear or compaction bands due to its lacking periodicity, but exhibited localized compaction near the bottom, ultimately forming a domain of compacted cells that spanned the width of the specimen (purple; Fig.~\ref{fig:nonlinear}c, (iii)), while unit cells near the top remained largely undeformed.

In contrast to these well-localized deformation patterns in the three reference systems, we find that the SR architectures display a spatially homogeneous and delocalized response (Fig.~\ref{fig:nonlinear}c, (iv)).
This lack of compaction or shear bands can be attributed to the local coexistence of ligaments of different lengths and thicknesses, hence the failure of the weakest ligament in one domain will not trigger the propagation of contiguous failure pathways.
This behavior is consistent across all characterized samples, indicating that the SR architecture effectively delocalizes deformation and preserves its global structural integrity under different loading directions.

To quantify this delocalization, we introduce a localization parameter
\[\mathcal{L} \equiv \frac{\sum\limits_{i=1}^{N} (\mu_i - \overline{\mu})^2}{\sum\limits_{i=1}^{N} (J_i - \overline{J})^2},\]

\noindent where $\mu_i \equiv \frac{1}{M} \sum_{j \in \mathcal{B}_r(i)} J_j$ is the local average Jacobian of cell $i$, defined over a neighborhood $\mathcal{B}_r(i)$ of $i$ containing $M$ cells (SI~\ref{sec:curly_L}), and $N$ is the total number of cells.
As a result, $\mathcal{L} \in [0,1]$, where $\mathcal{L} = 0$ implies that the deformation in every neighborhood is identical---the ideal case---and $\mathcal{L} = 1$ indicates that the deformation-variation between neighborhoods is as large as the deformation-variation among all individual voids, corresponding to extreme localization.

For all samples $\mathcal{L}$ initially increases with increasing strain (Fig.~\ref{fig:nonlinear}d); the square and hexagonal lattices exhibit rapidly increasing and high asymptotic $\mathcal{L}$ due to the presence of compaction bands, while the disordered Voronoi lattice has lower $\mathcal{L}$.
Consistently, the SR samples display the lowest value of $\mathcal{L}$---with the mean measured localization within $\mathcal{L} \leq 0.2$ through 25\% macroscopic strain---which confirms the material's ability to uniformly distribute damage and offer a macroscopically stable load-bearing capacity.

Taken together, these results suggest that macroscopically, SR samples can provide an improvement over single-scale lattices in applications where energy absorption is a requirement but the localization of deformation is undesirable, offering advantages in a wide range of engineering applications, from protective headgear to sports equipment and footwear; in energy-absorbing automotive components; as well as in aerospace applications where sensitive instruments need to be mechanically protected \cite{lu2003energy,duncan2018review, jiao2023mechanical}.

\paragraph*{Steering mechanical waves using SR materials.}

\begin{figure}[p]
    \centering
    \includegraphics[width=1\linewidth]{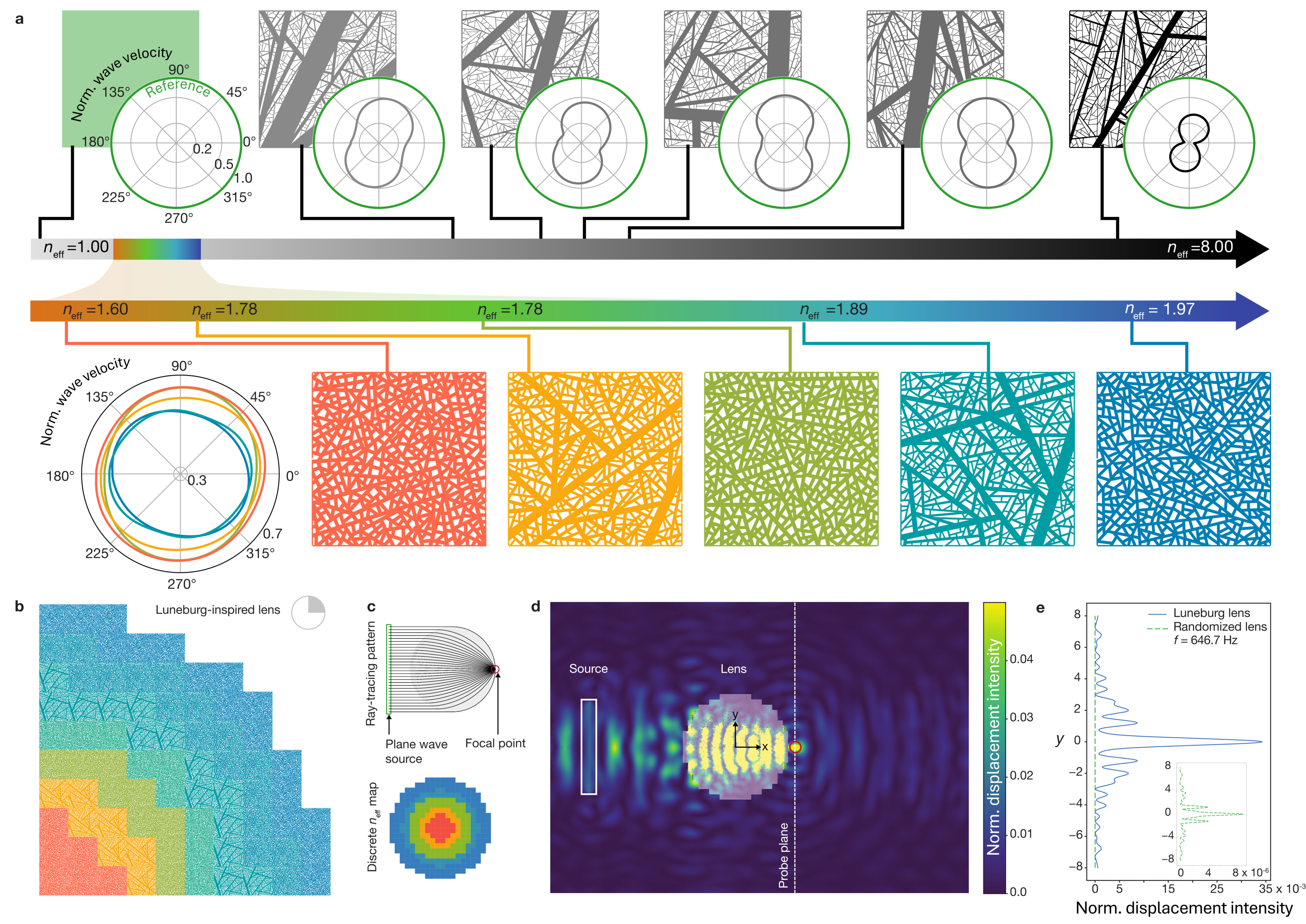}
    \caption{\textbf{Elastic lenses from SR metamaterials.} \textbf{(a)} SR samples with their corresponding directional velocity polar plots, demonstrating a wide range of achievable space of wave velocities and anisotropy. Curves, matching the color of their respective geometries, denote the quasi-longitudinal wave velocities, while the green curve represents the wave velocity of the monolithic material. Samples in the gray-black color scale show the breadth of wave anisotropy achieved, while the samples in the rainbow scale demonstrate isotropic SR unit cells used to construct the elastic lens in (b).
    We use each sample as a unit cell, adding a frame with small gaps to provide boundary conditions for building the superstructure. The effective refractive index, $n_\mathrm{eff}$, is defined as the ratio of the green to blue curves along the $x$-direction.
    \textbf{(b)} One quarter of a Luneburg-inspired lens constructed from SR samples with varying refractive indices, highlighted by colour (blue, high refractive index; orange, low refractive index).
    \textbf{(c)} Schematic illustration of the steady-state wavefield of the SR elastic Luneburg-like lens showing the elastic perturbation plane source (white) and focal point (red). The elastic wave is focused on the side opposite the source, as evidenced by the radial wave pattern emanating from the focal point. On the left, the pixelated refractive index map and analytical solution from optical equations are shown below and above, respectively. \textbf{(d)}
    The normalized displacement intensity along the y-axis cross-section through the lens focal point (white dashed line in panel d).
    The Luneburg-like lens (solid blue curve) exhibits a sharp focal peak, whereas the randomized lens (inset, green dashed curve) shows no noticeable focusing. The residual intensity peak in the randomized lens—three orders of magnitude smaller—arises from weak energy trapping at the elastic circular boundary.
    A suitable excitation frequency was found in the range $f \sim 646.7$Hz, well within the long-wavelength regime of each pixel in the lens, where the refractive indices are still valid.}
    \label{fig:waveProp}
\end{figure}

Beyond static responses, SR metamaterials also support programmable mechanical-wave propagation, due to their widely tunable elastic anisotropy, fully parametrized with only two parameters $(\alpha, \lambda_0)$.
To quantify direction-dependent wave propagation in SR metamaterials, we performed Bloch wave analysis on SR microstructures across the $(\alpha,\lambda_0)$ design space---here reinterpreted as representative volume elements (RVEs) in an infinite tiling (SI~\ref{sec:bloch_waveSpeed}).
Some architectures exhibit perfect isotropy while others feature high direction-dependent wave character (Fig.~\ref{fig:waveProp}a and SI Fig.~\ref{fig:SI-BAWV}). Leveraging this tunability, we designed acoustic metamaterials out of quasi-isotropic SR architectures, which are advantageous for multi-directional wave-control scenarios.
To contextualize the tunable wave responses, we borrow the effective refractive index concept from optics $n_\mathrm{eff}=c_0/v_s$, which compares the wave velocity in the constituent material $c_0$  to the velocity $v_s$ of a wave traveling through an SR metamaterial made of the same constituent material. We find that for SR metamaterials, $n_\mathrm{eff}$ spans $1.5 \leq n_\mathrm{eff} \leq 8.3$ in the $(\alpha,\lambda_0)$ configuration space, demonstrating a substantial range in achievable refractive indices compared to the hexagonal and square lattices, which span $1.0 \leq n_\mathrm{eff} \leq 1.9$ within the same relative density limits ($20\%\leq\bar{\rho}\leq75\%$) (SI~\ref{sec:sqHex_wv}).

The refractive-index tunability of SR microstructures enables multiple elastodynamic functionalities such as wave guiding or lensing, with applications in subwavelength imaging and energy harvesting \cite{tol2017phononic, tol2017structurally, kim2021poroelastic}.
To demonstrate this, we use the discretized gradient-index (GRIN) concept \cite{xie2018acoustic, kim2021poroelastic}, which guides elastic waves via a gradual change in refractive index, to design a Luneburg-like cylindrical lens of diameter $d$, focusing elastic perturbations from an incident plane wave (Fig.~\ref{fig:waveProp}b).
We construct the pixelated elastic lens out of quasi-isotropic SR RVEs tessellated in space to achieve a circular footprint with $n_\mathrm{eff}$ changing radially, selected from an SR library (SI Fig.~\ref{fig:SI-refractiveInd}) to match the required refractive indices for each lens region (SI~\ref{sec:elastic_lens_construction}).
We validate the SR-derived elastic-lens via steady-state dynamic finite-element simulations.
We build the lens into an elastic background, and excite the medium with a plane source at a distance $d/2$ from the edge of the lens (Fig.~\ref{fig:waveProp}c).
Displacement fields reveal that waves traversing the lens focus on the opposite side, producing semi-circular wave fronts at the focal point, and indicating SR lenses qualitatively follow Luneburg patterns~\cite{tol2017phononic}.
To quantify the lens performance, we extracted the displacement intensity (squared displacement magnitude) contour along the vertical line through the focal point, which demonstrates a peak that is an order-of-magnitude larger than the side peaks, indicating energy concentration into a region approximately 0.1$d$ wide  (Fig.~\ref{fig:waveProp}d). For comparison, a randomized reference lens constructed by randomly selecting SR cells at each radius produced a focal intensity roughly three orders of magnitude weaker, with a broader peak and side lobes of comparable magnitude, confirming that focusing arises from the GRIN arrangement of the SR pixels.
Spanning such a breadth of $n_\mathrm{eff}$
enables the generation of graded materials with smoothly changing acoustic properties over a much larger range than any individual lattice family, achieved by tuning ($\alpha, \lambda_0$).
This positions the SR framework beyond the current metamaterials paradigm, enabling impedance matching for ultrasound devices \cite{imani2024advanced} and phase delays for metasurface flat, low-aberration lenses \cite{zhu2016anomalous}.

\section*{Discussion and outlook}
While mechanical metamaterials have significantly expanded the accessible space of material properties \cite{Bauer2017, surjadiPortela2025}, their reliance on a unit cell forces on them characteristic scales that inherently restrict their performance.
Deviating from this single-scale restriction, here we drew inspiration from advances in network science to introduce and experimentally characterize Scale-Rich (SR) network-based metamaterials.
These architectures accommodate orders-of-magnitude variations in ligament thickness, length, and connectivity, while retaining control over global geometric properties such as density and anisotropy.
We find that SR materials have a series of unique mechanical features, like highly tunable directional stiffness across densities, enabling elastic anisotropic responses far beyond those of single-scale systems (Fig.~\ref{fig:elastic_alt})---relevant for adaptive load-bearing structures, wearable sensors, and biomedical implants requiring programmable compliance.
Moreover, they delocalize strain across the entire domain (Fig.~\ref{fig:nonlinear}), mitigating catastrophic failure while preserving high energy absorption, a combination attractive for protective gear, crashworthiness components, and aerospace structures.
Beyond the quasi-static regime, SR metamaterials offer a broad spectrum of refractive indices, enabling the rapid design of acoustic lenses (Fig.~\ref{fig:waveProp}), with potential applications in imaging, energy harvesting, and vibration control.
\begin{figure}[p]
    \centering
    \includegraphics[width=.95\linewidth]{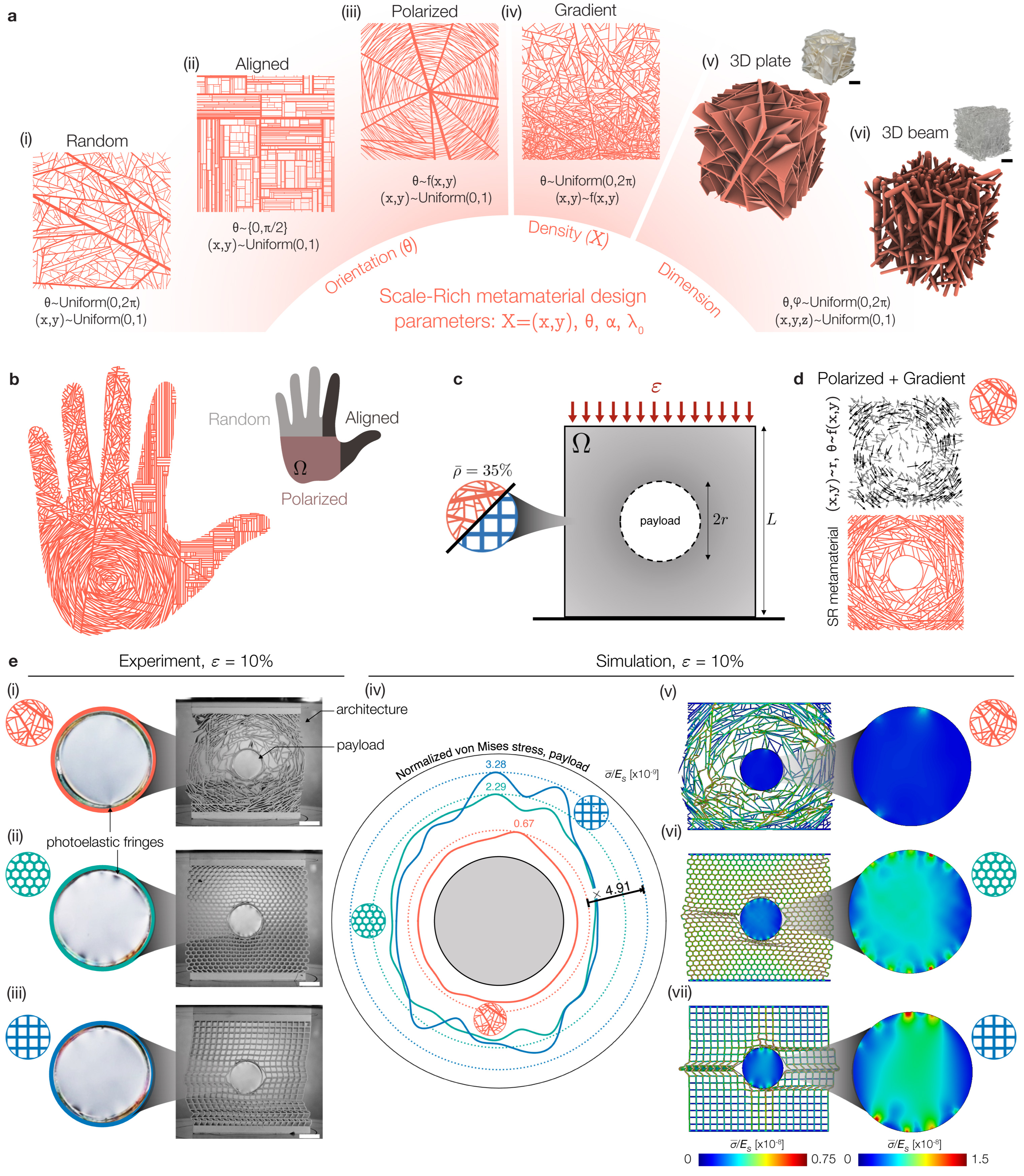}
    \caption{ \textbf{Design versatility of the Scale-Rich framework.} \textbf{(a)} The SR framework enables precise control over ligament orientation, allowing the generation of (i) random configurations where angles are selected from a uniform distribution, (ii) aligned networks using random selections from a prescribed set of angles, and (iii) polarized patterns by sampling angles based on a predefined vector field. (iv) Nucleation point positions can also be controlled to vary density across the domain, enabling gradient patterns. The model further generalizes to three dimensions through the growth of (v) plates and (vi) beams, where two angles specify the orientation of each plate or beam (for more examples, see SI~\ref{subsec: model_limit}). Scale bars, 10 mm.
    \textbf{(b)} SR microstructure generated within a non-convex domain by combining three distinct orientation patterns---random, aligned, and polarized---each applied on a different region of the domain.
    \textbf{(c)} Functional design using the SR framework: a microstructure is chosen to fill a given domain under a relative-density constraint in an effort to protect a ``payload''.
    \textbf{(d)} The SR framework allows simultaneous variation of both the ligament density and orientation. A circular design inspired by the natural microstructure of \emph{Citrus maxima} \cite{le2023influence} is shown, with a linear density gradient in the radial direction combined with a polarized orientation applied to 20\% of the initial lines.
    \textbf{(e)} Stress reduction in SR-protected payloads. (i-iii) Disc-shaped payloads made of poly(methyl methacrylate) (PMMA) were characterized using photoelasticity to visualize stresses under nonlinear deformation.
    Square and hexagonal lattices generate multiple fringes, indicating high internal stress, whereas the SR architecture shows markedly fewer fringes. Scale bars, 10 mm.
    (v - vii) Nonlinear simulations confirm these findings: the SR design reduces the average stress on the embedded payload by a factor of up to $4.91$ compared to the reference geometries (iv).}
    \label{fig:summary}
\end{figure}
Scale-Rich metamaterials also offer remarkable design flexibility (Fig.~\ref{fig:summary}a), obtained
by independently controlling orientation, seed density distribution, and dimensionality.
This allows us to generate microstructures with tailored heterogeneity that expand beyond traditional unit-cell constraints, e.g., filling an arbitrary, non-convex domain with spatially-programmable Scale-Rich networks (Fig.~\ref{fig:summary}b).
Moreover, to demonstrate the facility with which the framework can generate tailorable microstructures,
we designed a SR system with programmed spatial distribution and orientation of ligaments to protect a fragile payload embedded within the microstructure during uniaxial compression (Fig.~\ref{fig:summary}c).
We took advantage of two design elements: a polarized distribution of nucleation angles, $\theta_i = f(x_i, y_i)$, and inspired by the microstructure of \emph{Citrus maxima} \cite{le2023influence}, a radial gradient in the nucleation seed density profile, $(x_i,y_i) \sim r_i$, aiming to divert load-transfer pathways around the payload (Fig.~\ref{fig:summary}d).
We fabricated disc-shaped payloads out of poly(methyl methacrylate) (PMMA) and used photoelastic birefringence to identify stresses induced in the payload upon macroscopic deformation.
We find multiple photoelastic fringes nucleated within the PMMA in the case of the square and hexagonal lattice structures of the same relative density, indicating the development of local stresses within the payload.
In contrast, the payload stress is suppressed in the SR system due to the favorable orientation of its ligaments (Fig.~\ref{fig:summary}e, (i)-(iii)).
We confirm and quantify these results with nonlinear finite-element simulations (Fig.~\ref{fig:summary}e, (v)-(vii); SI~\ref{sec:nonlinear_fem}), which reveal that at identical macroscopic strains, the stresses in the SR-protected payload are reduced by up to a factor of 4.9 compared to the single length-scale systems (Fig.~\ref{fig:summary}e, (iv)).
This example elucidates the SR framework's ability to generate highly programmable architectures for targeted mechanical performance and protection.

Moreover, the SR framework can be naturally extended to 3D by iteratively adding (non-crossing) plates with areas $A_t = A_0 t^{-\alpha_{\mathrm{3D}}}$ (Fig.~\ref{fig:summary}a, (v)), or beams with thicknesses $\lambda_t = \lambda_0 t^{-\alpha_{\mathrm{3D}}}$ (Fig.~\ref{fig:summary}a, (vi)).
The 3D plate system displays a similar and analytically defined phase diagram to the 2D system,
showing that the plate area distribution and degree  distribution follow $P(A) \sim A^{-2.5}$ and $P(k) \sim k^{-2.5}$ respectively (SI ~\ref{sec:3D_plates_model}).
We also find that the linear-elastic behavior of the 3D plate system inherits the anisotropic trends observed in 2D: as $\alpha$ increases, the degree of anisotropy in the stiffness tensor also increases (SI Fig.~\ref{fig:3D_plates}).
Moreover, by generalizing the model to include closed-cell structures, the framework captures a broader class of architected materials, providing a systematic route towards the design of lightweight, mechanically robust, and directionally tunable 3D metamaterials (SI \ref{sec:3D_plates}).
The facile generation method of the SR model, requiring only two input parameters, is particularly suited to leverage current advances in artificial intelligence and inverse-design optimization of metamaterials.
To that end, simple parametrizations of metamaterials will allow optimization of mechanical microstructures for target directional stiffness, large-deformation compaction, or elastodynamic responses.

\clearpage

\section*{Acknowledgments}
A.Y.C. acknowledges financial support from the National Science Foundation through the Graduate Research Fellowship Program. This material is based upon work supported by the National Science Foundation Graduate Research Fellowship Program under Grant No. 2141064. Any opinions, findings, and conclusions or recommendations expressed in this material are those of the authors and do not necessarily reflect the views of the National Science Foundation.
N.M. acknowledges support for a research visit by the European Union’s Horizon 2020 research and innovation programme under the Marie Skłodowska-Curie grant agreement No. 823952 (TREND), and by the Complex Systems Fund, Utrecht University, through a donation from Peter Koeze.
M.C. acknowledges support from MIT's Postdoctoral Fellowship Program for Engineering Excellence.

\section*{Funding}
A.-L.B. was partially supported by the National Science Foundation (NSF) award No. 2243104 – COMPASS and by the European Union’s Horizon 2020 research and innovation programme No. 810115 – DYNASNET. C.M.P. acknowledges financial support from NSF, via CAREER Award CMMI-2142460.

\section*{Author contributions}
A.-L.B. and C.B. conceived the SR model.
C.B., N.M., and T.-T. G. performed the theoretical analysis, developed the algorithm, and ran the SR model simulations.
C.M.P. and A.Y.C. conceived the experimental design.
A.Y.C. and C.B. carried out the finite element simulations and analyzed the data.
A.Y.C. fabricated the samples and performed the mechanical experiments.
M.C. designed and simulated the lenses.
C.B., A.Y.C., M.C., C.M.P., and A.-L.B. wrote the manuscript, with A.-L.B. as the lead writer.
All authors discussed the results and contributed to the final manuscript.

\section*{Competing interests}
A.-L.B. is the founder of Scipher Medicine, a company that explores the role of networks in health.
C.B., A.Y.C., T.-T. G., N.M., M.C., and C.M.P. declare no competing interests.

\section*{Data and materials availability}
The code used to generate the data is available at: \url{https://github.com/Barabasi-Lab/Scale-Rich-Metamaterials.git}

\newpage
\bibliographystyle{unsrt}
\bibliography{bib}
\clearpage

\pagenumbering{arabic}
\setcounter{page}{1}

\begin{center}
    {\LARGE \textbf{Supplementary Information}}
\end{center}
\vspace{1em}

\begin{minipage}[h]{\textwidth}
\centering

{\large{\textbf{Scale-Rich Network-Based Metamaterials}}}

{\small
Csaba Both, Andrew Yen-Jong Chen, Ting-Ting Gao, Niek Mooij, Mohammad Charara, \newline  Carlos M. Portela, Albert-László Barabási}

\end{minipage}

\renewcommand{\contentsname}{}
\tableofcontents

\setcounter{section}{0}
\setcounter{figure}{0}
\setcounter{table}{0}
\setcounter{equation}{0}

\renewcommand{\thesection}{S\arabic{section}}
\renewcommand{\thesubsection}{S\arabic{section}.\arabic{subsection}}
\renewcommand{\thefigure}{S\arabic{figure}}
\renewcommand{\thetable}{S\arabic{table}}
\renewcommand{\theequation}{S\arabic{equation}}

\clearpage
\section{Methods}
\subsection{Elastic homogenization}
\label{sec:homgenization}
We evalulated the directional stiffness of the network and reference geometries in the linear elastic regime by applying computational homogenization using the finite element method (FEM) implemented with the commercial software package ABAQUS/Standard. Samples were discretized using six-node, quadratic plane-strain elements (CPE6) with a minimum of two elements across the width of any strut to properly resolve stresses and strains. To this end, meshing was performed using an adaptive strategy where the seed size for each strut was computed based on its width. Thus, the total number of degrees of freedom and the minimum element size varied across geometries.
We assume the base material is isotropic and linear elastic (with material parameters $E_s$ and $\nu_s$) and compute the effective stiffness tensor of the architecture by applying an orthogonal set of linear perturbations.
Because the network geometries are aperiodic, in order to enforce an average state of strain we apply affine boundary conditions on the edge nodes of each sample by prescribing relevant nodal displacements along the boundaries.

\subsection{Sample fabrication and uniaxial compression experiments}\label{sec:fabrication_compression}
Samples were fabricated using vat photopolymerization using a desktop 3D-printer (Phrozen Sonic Mini 8K). Geometries generated using the model were converted into three-dimensional solid bodies by adding a prescribed, uniform thickness that yielded a plane-strain response (i.e., avoiding out-of-plane deformation). To aid in experimental repeatability, the specimens were printed with monolithic top and bottom plates with a thickness equal to 5\% of the height of the architecture. For the nonlinear experiments, we used a commercially available resin (Anycubic Tough Green) with a high elongation to break. For the experiments within the linear regime, we selected a stiffer but more brittle resin (Formlabs Clear v4) so that the linear-elastic properties could be more readily extracted. In each case, we printed using the manufacturer's recommended settings for exposure parameters at a layer height of 0.025 mm. After printing, samples were washed in isopropanol and dried in air. The dried samples were then kept in storage away from ambient UV exposure and tested 72 h after printing.

Compression experiments were performed on an Instron 5500 series load frame under displacement control. The lower compression platen was held fixed and the upper platen was given a prescribed downward velocity corresponding to a (quasi-static) strain rate of $5 \times 10^{-3}$ s$^{-1}$.

To perform the areal Jacobian calculations, grayscale photographs were taken at regular intervals during loading. These photographs were thresholded to separate the printed beams from void spaces. A custom image tracking script using the Python library OpenCV was used to track the areas of each void as deformation progressed.

\subsection{Elastic lens construction}
\label{sec:elastic_lens_construction}
We assembled a pixelated GRIN lens from SR unit cells to create a Luneburg-like cylindrical lens which focuses a plane wave to a point image on the far side. The lens was simulated in the ABAQUS/Standard FEM software using a linear-elastic constituent material model. The Luneburg lens refractive index profile follows $n=\sqrt{2-(r/R)^2}$, where $r$ is a radial position in the lens and $R$ is the radius of the lens; hence, the refractive index spans $1.0 \leq n \leq \sqrt{2}$. The circular lens footprint with $R=d/2=2.5$ is discretized into smaller pixels of side length $l=0.25$, so that $l/d=0.05$ --- this choice was made to strike a balance between lens profile accuracy and simulation complexity. We discretized the theoretical profile curve to seven distinct concentric regions equally radially spaced, whose refractive index value was calculated at the central radial location of its pixel. We constructed the lens out of quasi-isotropic SR unit cells---with minimum to maximum directional velocity ratio of $v_\mathrm{min}/v_\mathrm{max}>0.85$---to suppress lens aberration as much as possible. Because the SR unit cell refractive index library does not include some of the lower values of $n$ required for the Luneburg lens, we offset the lens $n$ profile and prioritize maintaining its absolute shape when choosing SR unit cells to fill the seven regions; this offset introduces a slight aberration but the focal point is maintained. Rigorously, we use an optimization algorithm to find the profile offset which minimizes the sum of the squared differences between the $n$ values of the chosen unit cells and the ideal Luneburg lens with the added offset--i.e., $\mathrm{min} \sum_{i=1}^7 \left(n_{\mathrm{pixel},i} - n_\mathrm{ideal}(r)\right)^2$ where $n_{\mathrm{pixel},i}$ and $n_\mathrm{ideal}(r)$ are the refractive indices of the pixelated lens at region $i$ and the theoretical refractive index of a Luneburg lens evaluated at that region, respectively. When assembling the lens, we add a thin disconnected border to each unit cell to ensure assembly compatibility and energy propagation between adjacent pixels. The final assembled lens is inserted into a background medium composed of the same constituent isotropic linear elastic material, and the background is surrounded by a moderately dissipative medium which reduces reflections from its edges. The lens is excited by a plane wave of width $d$ placed at a distance $d/2$ away.

\subsection{Nonlinear finite element simulations}\label{sec:nonlinear_fem}
To numerically evaluate the nonlinear behavior of the architecture-payload system in Fig.~\ref{fig:summary}, we used the finite element method (FEM) implemented with the commercial software package ABAQUS/Explicit. As in the elastic case, both the architecture and payload domains were discretized using modified six-node, quadratic plane strain elements (CPE6M) with the same mesh density in the architecture. The architecture was modeled as elastic-plastic with hardening and the material properties were taken from experimental data (SI~\ref{sec:SI-material_properties}). The rigid payload was modeled as isotropic and linear elastic with an elastic modulus $10^{6}$ times that of the architecture. The compression platens were modeled as analytical rigid surfaces, with the bottom surface encastred and the top surface given a downward velocity consistent with the experimental strain rate $\dot{\varepsilon} = 5 \times 10^{-3} \; \mathrm{s}^{-1}$. We used the built-in general contact algorithm between all surfaces (including self-contact within the architecture), with a friction coefficient of 0.2 defining the tangential behavior and a ``hard'' pressure-overclosure condition defining the normal behavior.

\clearpage
\section{The Scale-Rich model}
In this section, we introduce a generative metamaterial model to generate Scale-Rich (SR) network-based materials.
We start with a unit square domain ($L = 1$) and iteratively add ligaments whose thickness $\lambda$ decreases with time $t$ as $\lambda_t = \lambda_0 t^{-\alpha}$, where $\lambda_0$ is the initial thickness and
$\alpha$ is the decay exponent.
The generation process consists of two steps:

(i) choose a random nucleation point $(x_t, y_t)$ within the $L^2$ domain and a random angle $\theta_t$;

(ii) grow a ligament of thickness $\lambda_t$ from the nucleation point in both directions, defined by $\theta_t$, until it intersects other ligaments or the boundary.

\begin{figure}[htpb]
    \centering
    \includegraphics[width=\linewidth]{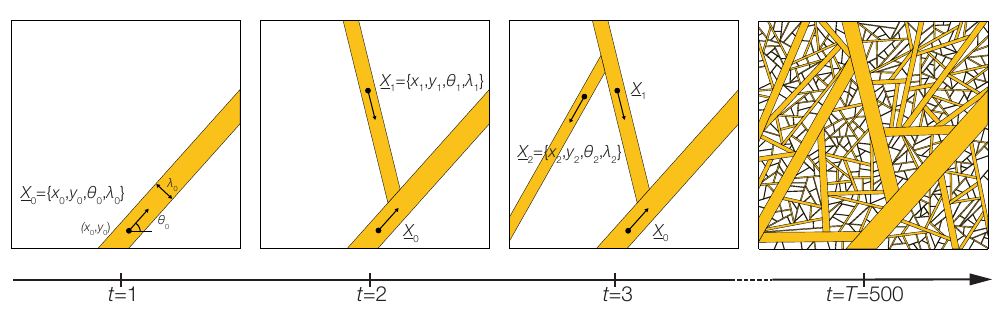}
    \caption{\textbf{Generation process of SR metamaterial.}
    At time step $t=1$ where an initial ligament of thickness $\lambda_0$ is generated from a randomized coordinate $(x_0,y_0)$ in a randomized orientation $\theta_0$.
    Additional ligaments at each time step $t$ (randomized coordinates and orientations) have thickness determined by $\lambda_t = \lambda_0 t^{-\alpha}$, terminating at the intersection with another ligament or the domain boundary.}
    \label{fig:SI_generation}
\end{figure}

We repeat (i) and (ii) until a predefined time $t=T$ or until we are unable to add new ligaments (Fig.~\ref{fig:SI_generation}), leading to a jammed state \cite{posfai2024impact}.

To understand the characteristics of the SR model, we begin by characterizing how ligament length and network density evolve with the decay exponent $\alpha$ and the initial thickness $\lambda_0$.
We first derive the $\lambda_0 = 0$ case, which isolates the geometric fragmentation from the physical effects of the ligaments.
We then extend the analysis to variable thickness, revealing how ligament thickness, length, and degree distribution evolve together to define the structure of SR metamarerials.

\subsection{Phase diagram of Scale-Rich model}\label{subsec: phase_diagram}

\subsubsection{Expected ligament length of the Scale-Rich model}\label{sec:expected_length}
The SR metamaterial formation begins with a unit square domain, where the total area is $M = 1$.
To isolate the purely geometric effects of the ligaments, we first consider the case $\lambda_0 = 0$, where each newly added line has zero thickness and does not change the overall density, allowing the derivation to follow directly from geometric considerations.
Each newly added line divides one existing polygon into two sub-polygons, leading to a linear increase in the number of polygons over time, following $N(t) = t + 1$.
Consequently, the average polygon area scales as $\langle M_t \rangle \sim t^{-1}$.

As lines are added in random positions and orientations, the mean shape of the polygons approaches a roughly square shape.
This leads to an expected line length at time $t$ of

\begin{align}
    \langle \ell_t \rangle &= \langle M_t \rangle^{1/2} = t^{-1/2}. \label{eq:mean_length_thickless}
\end{align}

For $\lambda_0 = 0$, where ligaments carry no thickness, this decay in line length is fully determined by geometric fragmentation and is independent of $\alpha$.
In this case, the system covers zero area, as the lines themselves occupy no space.

We now extend this reasoning to the finite-thickness case. Since the subdivision dynamics remain unchanged, we assume that the characteristic ligament length continues to follow a power-law decay, $\langle \ell_t \rangle = \ell_0 t^{-\beta}$,
with a decay exponent $\beta$ that may vary with the thickness-decay parameter $\alpha$.

We refer to lines with thickness as ligaments.
For $\lambda_0 \neq 0$, adding a new ligament increases the covered area, reducing the remaining available space and, consequently, the expected ligament length at each iteration.
We define the ligament thickness at time $t$ as $\lambda_t=\lambda_0 t^{-\alpha}$.
Let $\langle \rho_t \rangle = \lambda_0 \ell_0 \sum_{j=1}^{t} j^{-\alpha - \beta}$ be the expected covered space at time $t$, and let $\rho_\infty = \lambda_0 \ell_0 \sum_{j=1}^{\infty} j^{-\alpha - \beta}$ be the area of the asymptotically covered space.

\noindent At time $t$, the mean polygon area is given by the total domain area $M$ minus the areas of all previously added ligaments, divided by the number of polygons $t$
\begin{align}
    \langle M_t \rangle = \frac{M - \langle \rho_t \rangle }{t+1}.
\end{align}
The expected length of the $(t+1)$‑th ligament is
\begin{equation}
    \langle \ell_{t+1} \rangle = \langle M_t \rangle^{1/2} = \left( \frac{M - \langle \rho_t \rangle }{t+1} \right)^{1/2}.
    \label{eq:length_expectation}
\end{equation}

\noindent To calculate how $\langle \ell_{t+1} \rangle$ depends on ($\alpha, \lambda_0$), we study how the covered area changes in the limit for large $t$
\begin{equation}
    \rho_\infty = \lambda_0 \ell_0 \sum\limits_{j=1}^{\infty} j^{-\alpha - \beta}
    = \begin{cases}
        \text{finite} & \text{if } \alpha + \beta > 1 \quad \text{ (series converges)},\\[1mm]
        \text{infinite} &\text{if } \alpha + \beta \le 1 \quad \text{ (series diverges)}.
    \end{cases}
\end{equation}

\noindent The empty area after adding the $t$th ligament equals
\begin{equation}
    M - \langle \rho_t \rangle
    = \sum_{j=1}^{\infty} \lambda_0 j^{-\alpha} \ell_0 j^{-\beta} - \sum_{j=1}^t \lambda_0 j^{-\alpha} \ell_0 j^{-\beta}
    = \lambda_0 \ell_0 \sum_{j=t+1}^{\infty} j^{-\alpha-\beta}.\label{eq:2D_left_volume}
\end{equation}

\noindent For $\beta+\alpha > 1$, we obtain
\begin{align}
    \sum_{j=t+1}^{\infty} j^{-\beta-\alpha} \approx \int_{t+1}^{\infty} x^{-\beta-\alpha} \, \mathrm{d}x
    = \frac{(t+1)^{1-\beta-\alpha}}{\beta+\alpha-1}.\label{eq:2D_left_volume_approx}
\end{align}

\noindent Leading to
\begin{align}
    \langle \ell_t \rangle
    = \left( \frac{M - \rho_\infty}{t+1} + \frac{\rho_\infty - \langle \rho_t \rangle}{t+1} \right)^{1/2}
    = \left( \frac{M - \rho_\infty}{t+1} + \frac{\ell_0 \lambda_0}{\alpha + \beta - 1} t^{-\alpha - \beta} \right)^{1/2}.
\end{align}
If the domain is  fully covered ($\rho_{\infty} = M$), the decay rate of $\langle \ell_t \rangle$ is governed by a powerlaw with exponent $-(\alpha + \beta)/2$. Thus, consistency requires $\alpha=\beta$.

If the domain is not fully covered ($\rho_\infty < M$), the term inside the square root is dominated by the $t^{-1}$ contribution (since $M - \rho_\infty > 0$). Thus, by consistency with the form $\langle \ell_t \rangle \sim t^{-\beta}$, we have
\begin{equation}
    t^{-1/2} = t^{-\beta}
    \quad \Rightarrow \quad
    \beta = \frac{1}{2}
    \qquad \text{when } \rho_\infty < M.
    \label{eq:erlang_mean}
\end{equation}

\noindent At the boundary ($\alpha=\beta$) we obtain,
\begin{equation}
    \rho_\infty = \lambda_0 \ell_0 \sum_{j=1}^{\infty} j^{-2\alpha}
    = \lambda_0 \ell_0\,\zeta(2\alpha),
\end{equation}
where $\zeta$ is the Riemann zeta function. Requiring full coverage in the infinite limit ($\rho_\infty = M$) yields the condition
\begin{equation}
    \lambda_0 = \frac{M}{\ell_0 \,\zeta(2\alpha)}.
    \label{eq:tunable_boundary}
\end{equation}

\begin{figure}[htpb]
    \centering
    \includegraphics[width=.8\linewidth]{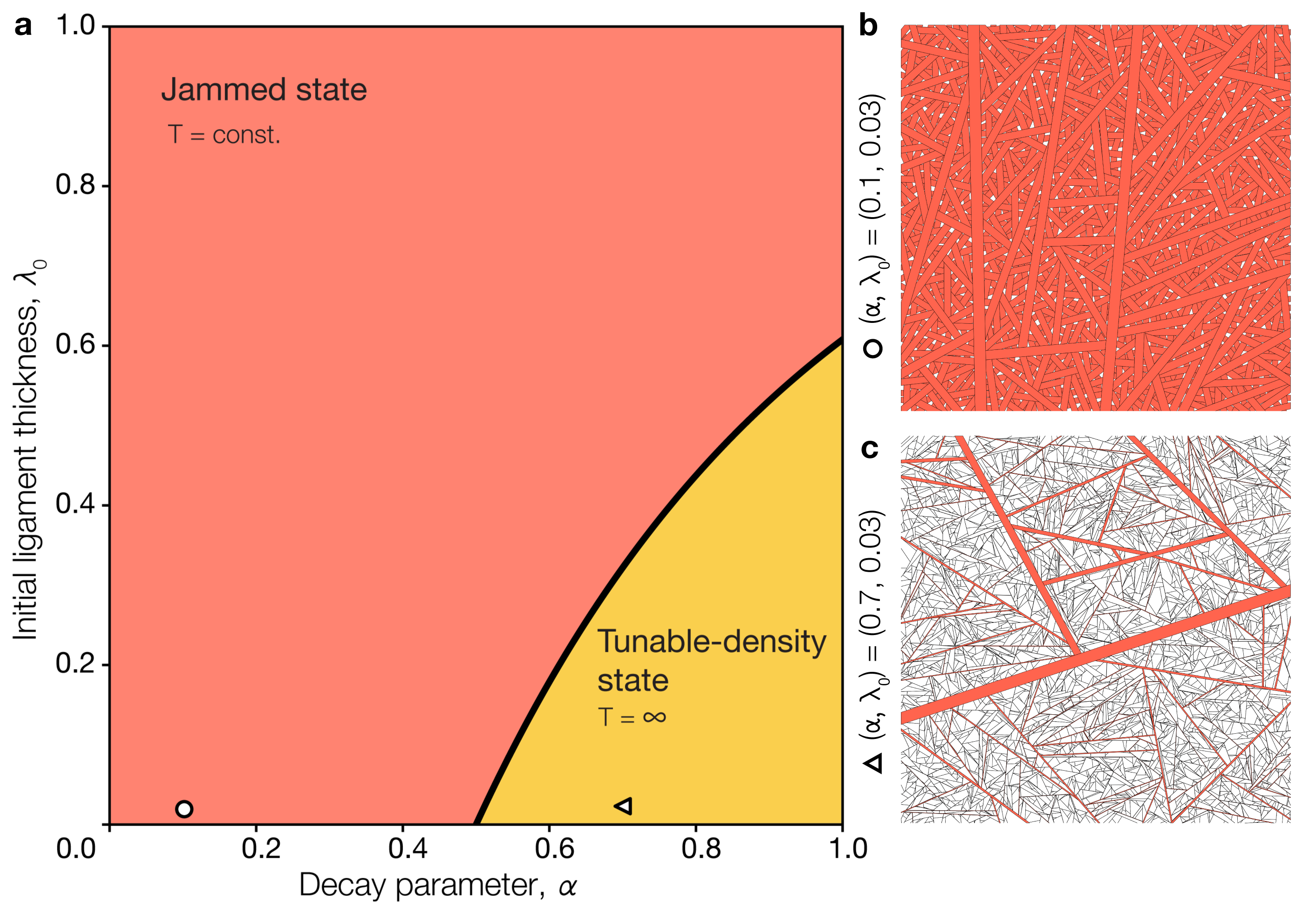}
    \caption{\textbf{($\alpha, \lambda_0$) phase space.} \textbf{(a)} Phase-diagram (two-parameter design space) of SR metamaterials.
    For small $\alpha$ the system gets jammed i.e. it does not allow the introduction of new ligaments. \textbf{(b)} Sample from the jammed regime.
    For large $\alpha$, the system can indefinitely accommodate additional ligaments, reaching a phase where the density of the system is tunable.
    \textbf{(c)} Sample from the tunable-density state after $T=500$.
    }
    \label{fig:SI_phase_space}
\end{figure}

\noindent This defines a phase space (Fig.~\ref{fig:SI_phase_space}a)

\begin{equation}
    \rho_\infty =
    \begin{cases}
        M & \text{if } \lambda_0 > (\ell_0 \zeta(2\alpha)))^{-1} \quad(\text{jammed}),\\[1mm]
        \lambda_0 \ell_0 \,\zeta(\alpha + \beta) < M &\text{if } \alpha > 1/2, \, \lambda_0 \leq (\ell_0 \zeta(2\alpha)))^{-1} \quad(\text{tunable density}).
    \end{cases}
\end{equation}
Fig.~\ref{fig:SI_phase_space}b illustrates a jammed configuration, in which no additional ligaments can be accommodated, whereas Fig.~\ref{fig:SI_phase_space}c shows a tunable-density state, where we can indefinitely add ligaments.

\noindent  Taken together, with finite ligament thickness, the system now accumulates area as it grows.
This leads to two possible behaviors: a jammed state, where the available space eventually vanishes and no new ligaments can be added, and a tunable-density state, where growth continues indefinitely.
Consequently, the average ligament length $\langle \ell_t \rangle$ exhibits two distinct scaling behaviors depending on the $(\alpha, \lambda_0)$ parameters

\begin{equation}
    \langle \ell_t \rangle = \left( \frac{M -  \lambda_0 \ell_0 \sum\limits_{j=1}^{t} j^{-\alpha - \beta} }{t+1} \right)^{1/2} =
    \begin{cases}
        \left( \frac{M -  \lambda_0 \ell_0 \sum\limits_{j=1}^{t} j^{-2\alpha} }{t+1} \right)^{1/2} &\text{}  \beta = \alpha \text{ boundary},\\[1mm]
        \left( \frac{M -  \lambda_0 \ell_0 \sum\limits_{j=1}^{t} j^{-\alpha-1/2} }{t+1} \right)^{1/2} &\text{} \beta = 1/2 \text{ tunable-density} .
    \end{cases}
\end{equation}

\subsubsection{Density evolution of the Scale-Rich model}\label{sec:density-evolution}
\begin{figure}[htbp]
    \centering
    \includegraphics[width=\linewidth]{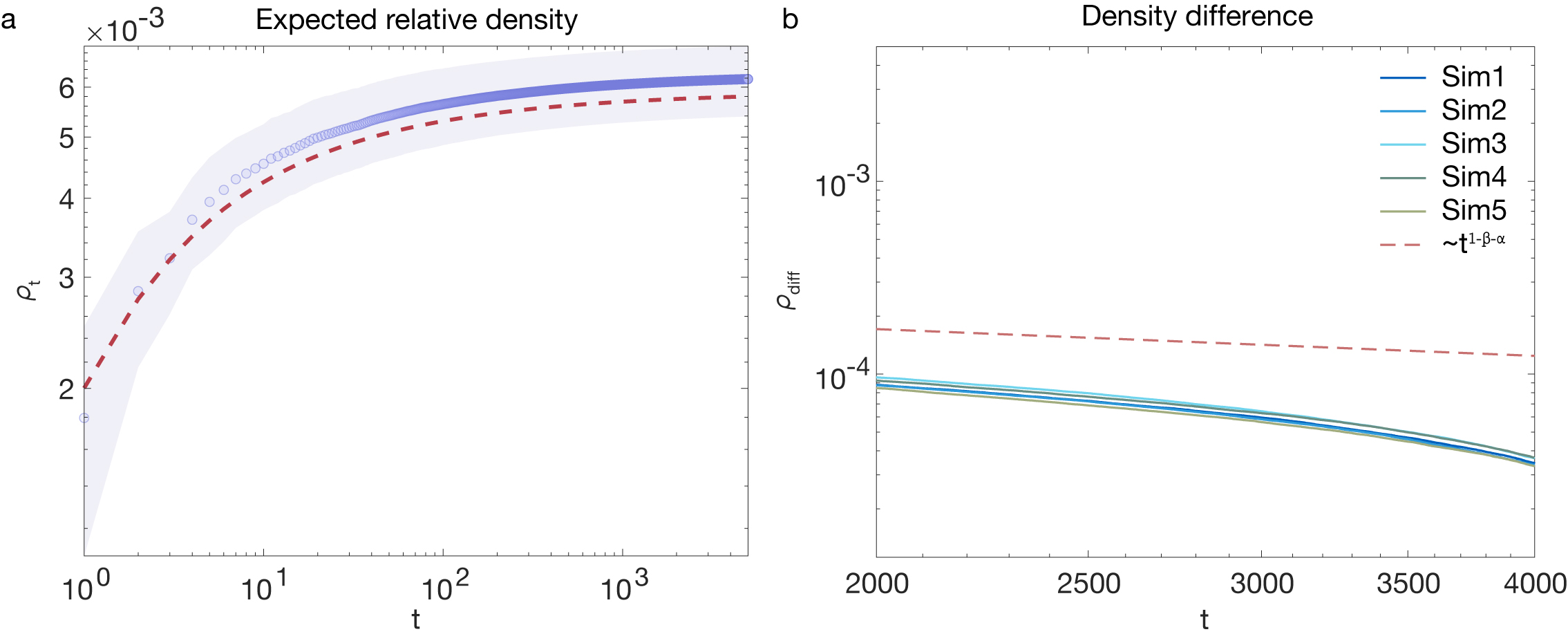}
    \caption{\textbf{Density evolution.} \textbf{(a)}The expected relative density $\rho_t$ as a function of time $t$. Purple circles denote the averaged simulation results over 5 realizations, with the shaded region representing one standard deviation. The red dashed line shows the analytical prediction. \textbf{(b)} Difference between the instantaneous and asymptotic densities, plotted for 5 independent simulations (colored lines). The red dashed line shows the theoretical power-law decay of $\rho_{\rm{diff}}$. All networks are generated with random seeds, with the parameters $\alpha = 0.95$, and $\lambda_0 = 0.02$.}
    \label{fig:density convergence}
\end{figure}

We now turn to the question of how the density evolves as ligaments are progressively added to the system.
Although the asymptotic density $\rho_{\infty}$ is controlled by the exponents $\alpha$ and $\beta$, the approach to this limit can be markedly slow.
To capture this dynamic filling process, we consider the density after $t$ ligament insertions.
Let $\rho_{\infty}$ be the asymptotic density of the system ($\alpha+\beta > 1$). According to Eq. \eqref{eq:length_expectation}, we obtain the expectation of the density
\begin{equation}
   \langle \rho_{t} \rangle = \sum_{j=1}^{t}\langle \ell_j \rangle \lambda_j = \rho_{\infty} - \sum_{j=t+1}^{\infty}\langle \ell_j \rangle \lambda_j \approx \rho_{\infty} - \frac{\ell_0 \lambda_0}{\beta+\alpha-1} (t+1)^{1-\beta-\alpha}.
\end{equation}
This expression shows that the density increases monotonically with $t$ and gradually approaches its asymptotic value $\rho_\infty$.
The deviation from saturation, $\rho_\infty - \rho_t$, decays as a power law with exponent $1 - \beta - \alpha$, indicating a progressively slower filling of the remaining space as the system approaches its converged density.
As shown in Fig.~\ref{fig:density convergence}, the simulated trajectories of $\rho_t$ closely follow the analytical prediction across all realizations, and the residual difference $\rho_{\mathrm{diff}} = \rho_\infty - \rho_t$ exhibits the expected power-law decay, confirming the theoretical scaling behavior of the density convergence.

\subsubsection{The theoretical time of jamming}
\label{sec:jamming-time}
The Scale-Rich model reaches a jammed state for parameters $\alpha < 1/2$, or $\alpha \ge 1/2$ with $\lambda_0 \ge (\ell_0 \zeta(2\alpha))^{-1}$ (Fig.~\ref{fig:SI_phase_space}a, orange).

Jammed state refers to the point at which no additional ligament can be inserted as there is no sufficient free space to accommodate a new ligament (e.g. Fig.~\ref{fig:SI_phase_space}b).
Physically, this reflects space saturation: early on, new ligaments easily fill empty regions, but as the structure grows denser, the remaining gaps shrink until they reach a size comparable to the ligament width. The jamming time is the point where no more ligaments can be added due to lack of available space.

\begin{figure}[h!]
    \centering
    \includegraphics[width=0.9\linewidth]{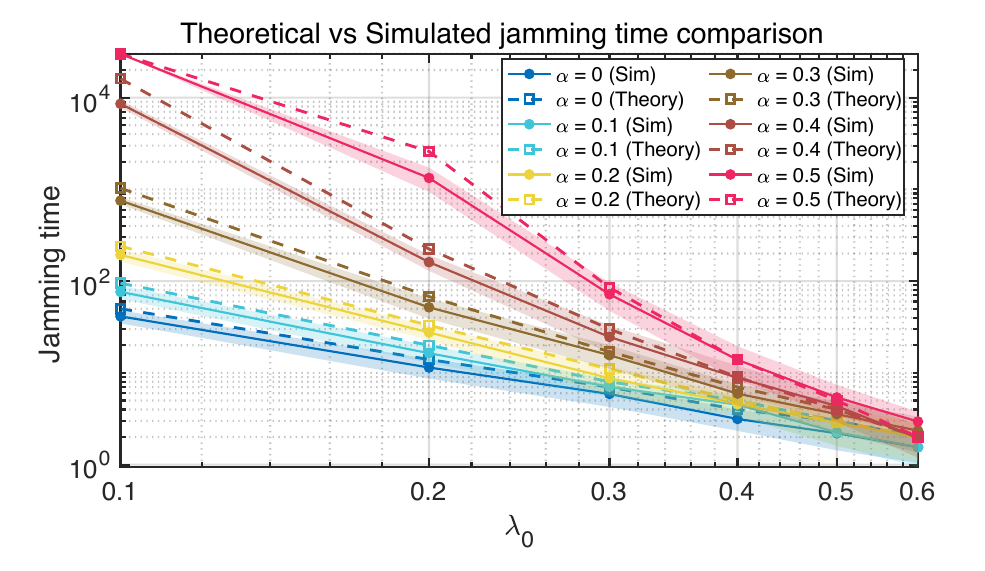}
    \caption{
    \textbf{The comparison between the theoretical jamming time and simulations.} Each network simulation was run for a maximum of 30,000 steps. For each combination of $\lambda_0$ and $\alpha$, 20 independent networks were generated. The solid lines indicate the average jamming time across simulations, with the shaded regions representing the corresponding standard deviation. Dashed lines denote the theoretical jamming time.}
    \label{fig:jammed}
\end{figure}

To estimate the jamming time, we consider the evolution of the empty region’s perimeter as the structure develops.
We have shown that the expected ligament length follows Eq. \eqref{eq:length_expectation}.
Therefore, the expected perimeter of the remaining void when adding the $(t+1)$-th line is
\begin{equation}
    L_{t+1} = 4\ell_{t+1} = 4\left( \frac{M - \sum_{j=1}^{t} \ell_j \lambda_j}{t} \right)^{1/2}.
    \label{perimeter}
\end{equation}
For a ligament with thickness $\lambda_{t}$, assuming its length can be arbitrarily small, the limiting condition for successfully placing it inside an available void of perimeter $L_{t+1}$ is $2\lambda_{t+1} < L_{t+1}$.
When this inequality no longer holds, that is, when $2\lambda_{t+1} \ge L_{t+1}$, the system reaches its theoretical jamming time. This criterion defines the transition between the growth and jammed regimes and allows the computation of the expected jamming time shown in Fig.~\ref{fig:jammed}.
The theoretical value slightly exceeds the simulation results because the generation algorithm samples only a finite number of random insertion attempts before the configuration is considered jammed.

\subsection{Thickness distribution of the Scale-Rich model}
\label{sec:line_thickness_distribution}
During SR metamaterial generation, the ligament thickness follows $\lambda_t = \lambda_0 t^{-\alpha}$, where $\alpha$ is the decay exponent and $\lambda_0$ is the initial ligament thickness.
This decay results in the coexistence of multiple thicknesses in the system.
The parameter $\alpha$ can take any value; in this work we focus on the $\alpha \in [0,1]$ range, but it could be larger or even negative.
Next we derive the resulting thickness distribution in the final structure.

\noindent The thickness of the ligaments follows a power-law decay
\begin{equation}
    \lambda_t = \lambda_0 t^{-\alpha}.
\end{equation}
\noindent Rearranging, we express $t$ in terms of $\lambda$ as $t = (\lambda/\lambda_0)^{-\frac{1}{\alpha}}$.

Ignoring constant factors for simplicity, differentiating $\lambda_t$ gives
\begin{equation}
    \frac{\mathrm{d}\lambda}{\mathrm{d}t} = -\lambda_0 \alpha t^{-\alpha-1}.
\end{equation}
\begin{figure}[htbp]
    \centering
    \includegraphics[width=.5\linewidth]{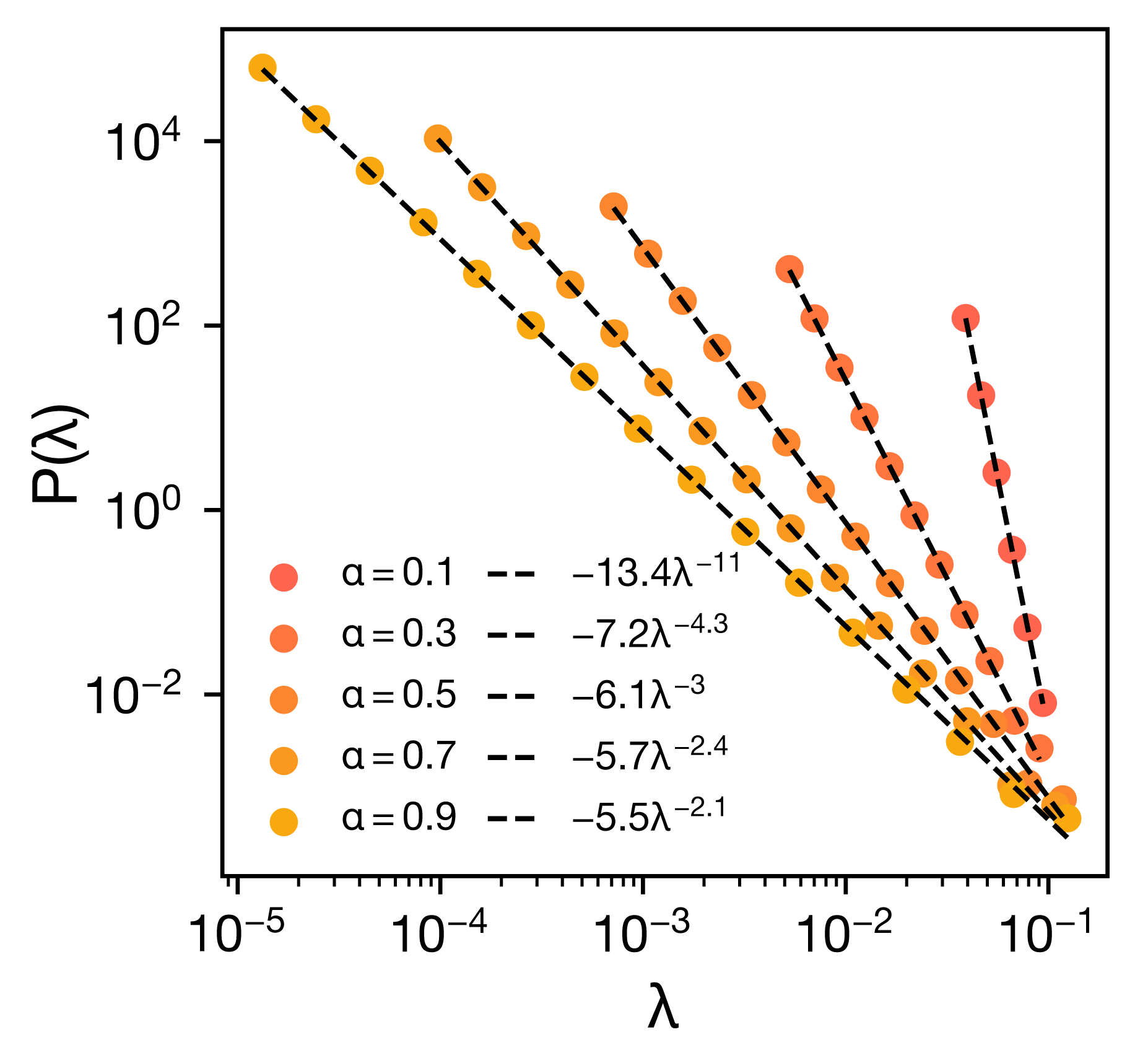}
    \caption{\textbf{Thickness distribution.} Systems with different decay exponents $\alpha$ are shown in different colors. All networks were generated with $\lambda_0 = 0.01$, and the data are averaged over 20 independent runs for 30000 time steps.}
    \label{fig:thickness_distribution}
\end{figure}
Rearranging, the infinitesimal change in $t$ is
\begin{equation}
    \mathrm{d}t/\mathrm{d}\lambda = - \lambda_0^{-1} \alpha^{-1} t^{\alpha+1}
\end{equation}
We define the probability density function
\begin{equation}
    p_{\lambda} = \frac{\mathrm{d}t}{\mathrm{d}\lambda}.
\end{equation}
Substituting $t = (\lambda/\lambda_0)^{-\frac{1}{\alpha}}$ we obtain the final thickness distribution as
\begin{equation}
    p_{\lambda} = \mathrm{d}t/\mathrm{d}\lambda = - \lambda_0^{-1} \alpha^{-1} (\lambda/\lambda_0)^{-(\alpha+1)/\alpha} = -\alpha^{-1} \lambda_0^{1/\alpha} \lambda^{-(1 + 1/\alpha)}.
    \label{eq:thickness_dis}
\end{equation}
Thus, the thickness distribution decays as a power-law with exponent $1 + 1/\alpha$ (Fig.~\ref{fig:thickness_distribution}).

\subsection{Length distribution of the Scale-Rich model}
\label{line_length_distribution}
In non-physical networks, the focus is primarily on connectivity, and geometric properties are rarely considered. In contrast, physical networks such as Scale-Rich metamaterials have structural properties—like ligament length and thickness—that directly influence both structure and function.
Ligament length is not typically measured in network models, yet it matters critically in physical systems.
Recent studies of physical network models have revealed power-law length distributions \cite{posfai2024impact,pete2024physical}, an unexpected finding in traditional network theory.
Notably, heterogeneous, heavy-tailed distributions are not unique to synthetic systems: neuronal lengths and connectivity in the brain exhibit extreme variations, captured by non-bounded distributions \cite{piazza2025physical}.
Here, we derive the ligament length distribution of the SR metamaterials.

We model the ligament length distribution as a mixture of exponential distributions.
As the network size increases, the distribution converges to the sum of two exponential random variables.
The rate parameter depends on the current number of ligament in the system.

The sum of $m$ identically distributed exponential random variables with rate $\phi$ is known to follow an Erlang($m, \phi$) distribution with shape parameter $m$ and rate parameter $\phi$. The probability density function of the Erlang distribution is given by
\begin{align}
    f(x; m, \phi) = \frac{\phi^m x^{m-1} e^{-\phi x}}{(m-1)!}, \quad \text{for } x, \phi \geq 0.
\end{align}

\noindent Similarly, the cumulative distribution function of the Erlang distribution is given by
\begin{align}
    F(x; m, \phi) = 1 - e^{-\phi x} \sum_{t=0}^{m-1} \frac{(\phi x)^t}{t!}, \quad \text{for } x, \phi \geq 0.
\end{align}

\label{sec:length_distribution}
\noindent Consider a uniformly picked point $x \in [0,1]^2$ and a direction $v \in \mathbb{R}^2$ for the new segment.
The procedure of walking in direction $v$ from the point $x$ models the event of hitting a ligament on a realization of the Scale-Rich network at time $t$ as an exponential random variable with rate $\phi_t$.
Ligament segments extend in two directions, so the length distribution $L_t$ of the segment introduced at time $t$ follows an Erlang distribution with shape parameter $2$ and rate parameter $\phi_t$
\begin{align}
    L_t \sim \text{Erlang}(2, \phi_t).
\end{align}
The corresponding length distribution is given by
\begin{align}
    p_{L_t}(x) &\sim x \phi_t^2 e^{-\phi_t x}.\label{eq:erlang}
\end{align}
We can compute various properties of the ligament-length distribution. However, it is crucial to recognize that the ligament length distribution is limited by the size of the system. This approximation is effective considering the very low probability of exceeding $\sqrt{2}$
\begin{align}
    \mathbb{P}\{ L_t \geq \sqrt{2} \} &= \int_{\sqrt{2}}^{\infty} p_{L_t}(x) \ \mathrm{d}x = e^{- \sqrt{2} \phi_t}(\sqrt{2} \phi_t + 1) \xrightarrow{t \to \infty} 0.
\end{align}
In the final configuration state, the ligament length distribution is described by a mixture of Erlang-distributed random variables with various rate parameters, all dependent on the time of introduction into the system. Let the set $\{ L_t \ : t \in [n] \}$ denote all the ligaments realized in the final configuration at iteration $n$.
We can express the cumulative distribution function of the ligament length distribution in the final state $\mathcal{L}$ as
\begin{align}
    F_{\mathcal{L}}(x) &= \mathbb{P}\{ \mathcal{L} \geq x \} = \frac{1}{n} \sum_{t=1}^n \mathbb{P}\{ L_t \geq x \} = \frac{1}{n} \sum_{t=1}^n e^{-\phi_t x} [ 1 + \phi_t x ].
\end{align}
Similarly, we can compute the probability density function as
\begin{align}
    f_{\mathcal{L}}(x) &= \frac{\partial F_{\mathcal{L}}(x)}{\partial x} = \frac{x}{n} \sum_{t=1}^{n} \phi_t^2 e^{- \phi_t x}.\label{eq:linedistribution_main}
\end{align}
In Sec.~\ref{sec:expected_length} we show that $\phi_n = b n^{\beta}$ follows a power law with exponent $\beta$ equal to either $1/2$ or $\alpha$, dependent on the initial thickness $\lambda_0$.
The value of $b$ is given by the equation $b = \mathbb{E}[L_{1}] = 2 / \ell_0$, where $\ell_0$ is the mean length of a uniformly placed segment, with both endpoints on the border of the unit cube $[0,1]^2$. The value of $\ell_0$ is given by~\cite{keller1971}
\begin{align}
    \ell_0 &= \frac{4}{\pi}\left( \log(1 + \sqrt{2}) + \frac{1 - \sqrt{2}}{3} \right) \approx 0.946,
\end{align}
resulting in $b = 2 / \ell_0 \approx 2.114$.

We can approximate the probability density function (Eq.~\eqref{eq:linedistribution_main}) by the integral
\begin{align}
    f_{\mathcal{L}}(x) &= \frac{x}{n} \sum_{t=1}^{n} \phi_t^2 \exp\{ - \phi_t x \} \\
    &\approx \frac{x}{n} \int_{1}^{n} \phi_t^2 \exp\{ - \phi_t x \} \mathrm{d} t \\
    &= \frac{x}{n} \int_{1}^{n} b^2 t^{2\beta} \exp\{ - b t^{\beta} x \} \mathrm{d} t \\
    &= \frac{x}{n} \int_{b}^{b n^{\beta}} \beta^{-1} b^{-1/\beta} u^{1 + 1/\beta} \exp\{ -u x \} \ \mathrm{d} u \\
    &= \left( \beta b^{1/\beta} \right)^{-1} \frac{x}{n} \int_{b}^{b n^{\beta}} u^{1 + 1/\beta} \exp\{ -u x \} \ \mathrm{d} u.
    \intertext{Let $\Gamma(\cdot, \cdot)$ be the incomplete Gamma function}
    f_{\mathcal{L}}(x) &= \left( \beta b^{1/\beta} \right)^{-1} \frac{x}{n} \left( x^{-1/\beta} \Gamma\left[2 + \frac{1}{\beta}, b x\right] - x^{-1/\beta} \Gamma\left[2 + \frac{1}{\beta}, b n^\beta x \right] \right) \frac{1}{x^2} \\
    &= \left( \beta b^{1/\beta} n \right)^{-1} x^{-1 - 1/\beta} \left( \Gamma\left[2 + \frac{1}{\beta}, b x\right] - \Gamma\left[2 + \frac{1}{\beta}, b n^\beta x \right] \right) .\label{eq:final}
\end{align}
The mean value of this distribution--for fixed $n$--is given by (note that this expression diverges in the limit $n \to \infty$)
\begin{align}
    \mathbb{E}[\mathcal{L}] &= \frac{1}{b (\beta - 1)} n^{-1 - \beta} \Bigg(
    2 \bigg( n^\beta - n \bigg)
    - n^\beta \Gamma\left(3, \sqrt{2} b\right) \nonumber \\
    &\quad + n \Gamma\left(3, \sqrt{2} b n^\beta\right)
    + 2^{\frac{-1 + \beta}{2 \beta}} b^{\frac{-1 + \beta}{\beta}} n^\beta \nonumber \\
    &\quad \times \bigg(\Gamma\left(2 + \frac{1}{\beta}, \sqrt{2} b\right)
    - \Gamma\left(2 + \frac{1}{\beta}, \sqrt{2} b n^\beta\right)\bigg)
    \Bigg).
\end{align}

\noindent{\bf Upper regime:} In the regime $x\approx \sqrt{2}$ and $n$ large, the function $\Gamma\left[2 + \frac{1}{\beta}, b n^\beta x \right]$ is approximately zero, giving
\begin{align}
    f_{\mathcal{L}}(x) &= \left( \beta b^{1/\beta} n \right)^{-1} x^{-1 - 1/\beta} \Gamma\left[2 + \frac{1}{\beta}, b x\right].
\end{align}
For our $b$ value and $\beta \in [1/2,1]$, the function $\Gamma\left[2 + \frac{1}{\beta}, b x\right]$ is approximately constant on the domain $[0, \sqrt{2}]$, compared to the power law behavior of $x^{-(1 + 1 / \beta)}$. Therefore, in the upper regime, we end up with a power law exponent of $\mu = 1 + 1/\beta$.

\noindent{\bf Lower regime:} Note that for small $x_0$, the incomplete Gamma function is approximately
\begin{align}
    \Gamma(s, x_0) \approx \Gamma(s) - x_0^s e^{-x_0}.
\end{align}
Meaning that for Eq.~\eqref{eq:final} we can approximate as
\begin{align}
    f_{\mathcal{L}}(x) &= \left( \beta b^{1/\beta} n \right)^{-1} x^{-1 - 1/\beta} \left( \Gamma\bigg[2 + \frac{1}{\beta}\bigg] - (bx)^{2 + \frac{1}{\beta}} e^{-bx} - \Gamma\bigg[2 + \frac{1}{\beta}\bigg] + \left(b n^{\beta} x \right)^{2 + \frac{1}{\beta}} e^{-b n^{\beta} x} \right) \\
    &= \left( \beta b^{1/\beta} n \right)^{-1} b^{2 + \frac{1}{\beta}} \left( n^{2\beta+1} e^{-b n^{\beta} x} - e^{-bx} \right) x.
\end{align}
For small $x$, the linear term dominates the exponential term, obtaining the power law scaling
\begin{align}
    f_{\mathcal{L}}(x) &\stackrel{x \approx 0}{=} \left( \beta b^{1/\beta} n \right)^{-1} b^{2 + \frac{1}{\beta}} \left( n^{2\beta+1} - 1 \right) x.
\end{align}
Therefore, in the lower regime, we end up with a power law exponent of $\mu = -1$.

\begin{figure}[htpb]
    \centering
    \includegraphics[width=.8\linewidth]{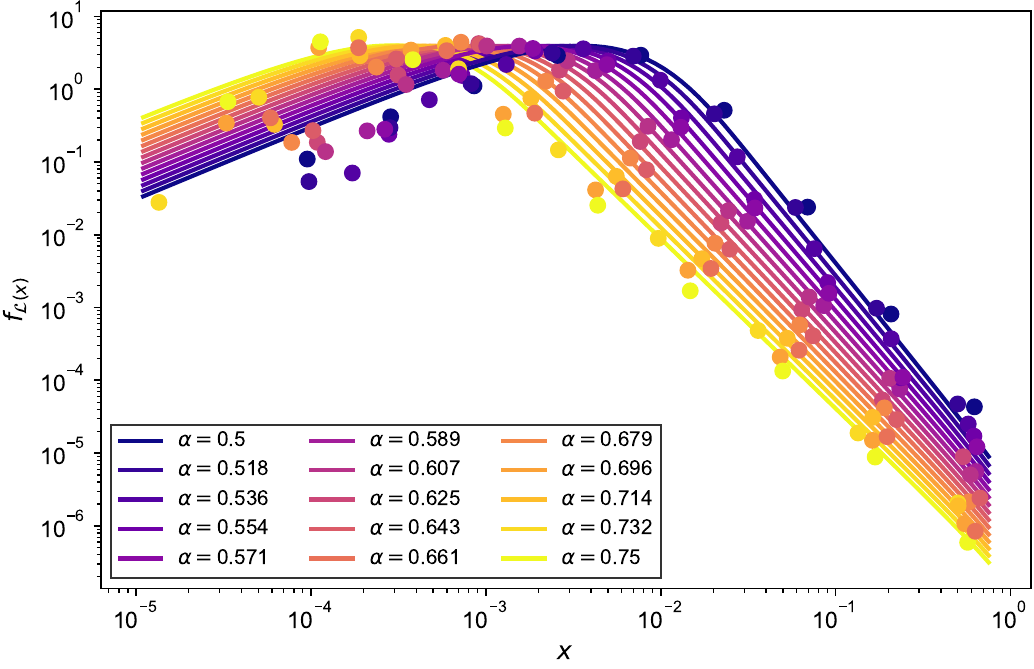}
    \caption{
    {\bf The theoretical length distribution compared with empirical measurements showing the behavior for multiple $\mathbf{\alpha}$ values on the boundary of the tunable density domain.}
    Dashed lines show the theoretical result while markers show the empirical values.}
    \label{fig:length_spectrum}
\end{figure}

Fig.~\ref{fig:length_spectrum} shows the ligament length distribution on the boundary of the phase space ($\alpha = \beta$), illustarting that the ligament length can be tuned on the boundary by $\alpha$.
This derivation reveals that the SR model produces a ligament length distribution characterized by two distinct power-law regimes with exponents $\mu = -1$ (lower regime for short ligaments) and $\mu = 1 + 1/\beta$ (upper regime for long ligaments), where $\beta=1/2$ in the tunable density regime and $\beta = \alpha$ on the boundary.
The data show distinct power-law regimes in the ligament length distribution.
This dual-regime behavior is captured by a mixture of Erlang distributions with time-dependent rate parameters, reflecting the progressive addition of ligaments during network growth.
These results confirm that the SR model generates multiple coexisting scales not only in thickness but also in ligament length.

\subsection{Degree distribution of the Scale-Rich model}
\label{degree_distribution}
Metamaterials are physical networks, where nodes and links are constrained by volume exclusion.
In Scale-Rich metamaterials, we treat the building units, the ligaments as nodes and their connections as edges.
A fundamental network characteristic is the degree (or coordination number) distribution.
A key network property is the degree (or coordination number) distribution.

\begin{figure}[htpb]
    \centering
    \includegraphics[width=1\linewidth]{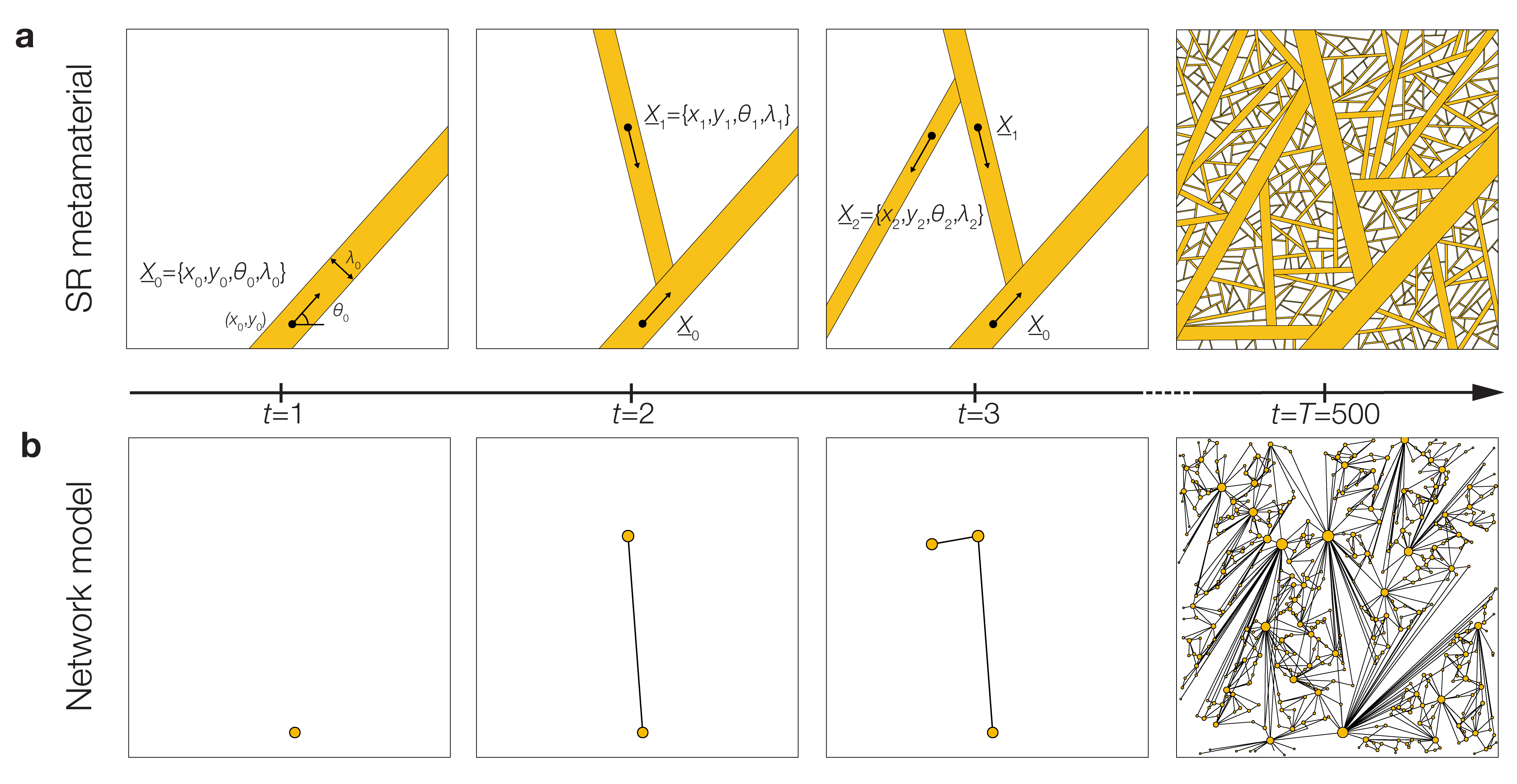}
    \caption{\textbf{The evolution of SR metamaterial with the evolution of network representation.}
    \textbf{(a)} At time step $t=1$ where an initial ligament of thickness $\lambda_0$ is generated from a randomized coordinate $(x_0,y_0)$ in a randomized orientation $\theta_0$.
    Additional ligaments at each time step $t$ (randomized coordinates and orientations) have thickness determined by $\lambda_t = \lambda_0 t^{-\alpha}$, terminating at the intersection with another ligament or the domain boundary.
    \textbf{(b)} The networks representation evolution over time.}
    \label{fig:SI_degree}
\end{figure}

A common approach in physical networks is to treat each ligament intersection as a node and the connecting segments as links \cite{paetzold2021whole, dorkenwald2024neuronal}.
In such a representation, node degrees are typically low, often 1, 2, or 3.
However, in the SR model, ligament length and thickness spans multiple orders of magnitude, motivating a representation inspired by transportation networks \cite{rosvall2005networks, pete2024physical}, where each ligament (road) is treated as a node and intersections between ligaments (roads) are treated as edges (Fig.~\ref{fig:SI_degree}).

We first derive the degree distribution for the SR model in the zero-thickness limit, followed by the case with finite ligament thickness in the tunable-density and boundary regime.
This distinction mirrors the approach used for the ligament length decay rate, where different behaviors emerged depending on the $(\alpha, \lambda_0)$ parameter configurations.

\subsubsection{Degree distribution of the Scale-Rich model without thickness ($\lambda_0 = 0$)}
\label{degree_distribution_no_thickness}
We follow the same steps as in Sec. 5.5 in~\cite{barabasi2016networkscience}.
The rate at which a node, introduced at time $i$, acquires links as a result of new nodes connecting to it at time $t$ is
\begin{align}
    \frac{\mathrm{d}k_i(t)}{\mathrm{d}t} &= 2 \frac{L_i}{\sum\limits_{j=1}^{t}L_j}
    = 2 \frac{\ell_0 i^{-\beta}}{\sum\limits_{j=1}^{t} \ell_o j^{-\beta}}
    \approx 2 i^{-\beta} \left( \int_{1}^{t} x^{-\beta} \mathrm{d}x \right)^{-1}
    \approx 2(1-\beta) i^{-\beta} t^{\beta-1}.
    \intertext{We can integrate to get }
    k_i(t) &\approx 2(1-\beta) i^{-\beta} \int_i^t x^{\beta-1} \mathrm{d}x
    = \frac{2(1-\beta)}{\beta} i^{-\beta} \left( t^{\beta} - i^{\beta} \right)
    = \frac{2(1-\beta)}{\beta} \left( \frac{t}{i} \right)^\beta.
\end{align}
\begin{figure}[htpb]
    \centering
    \includegraphics[width=\linewidth]{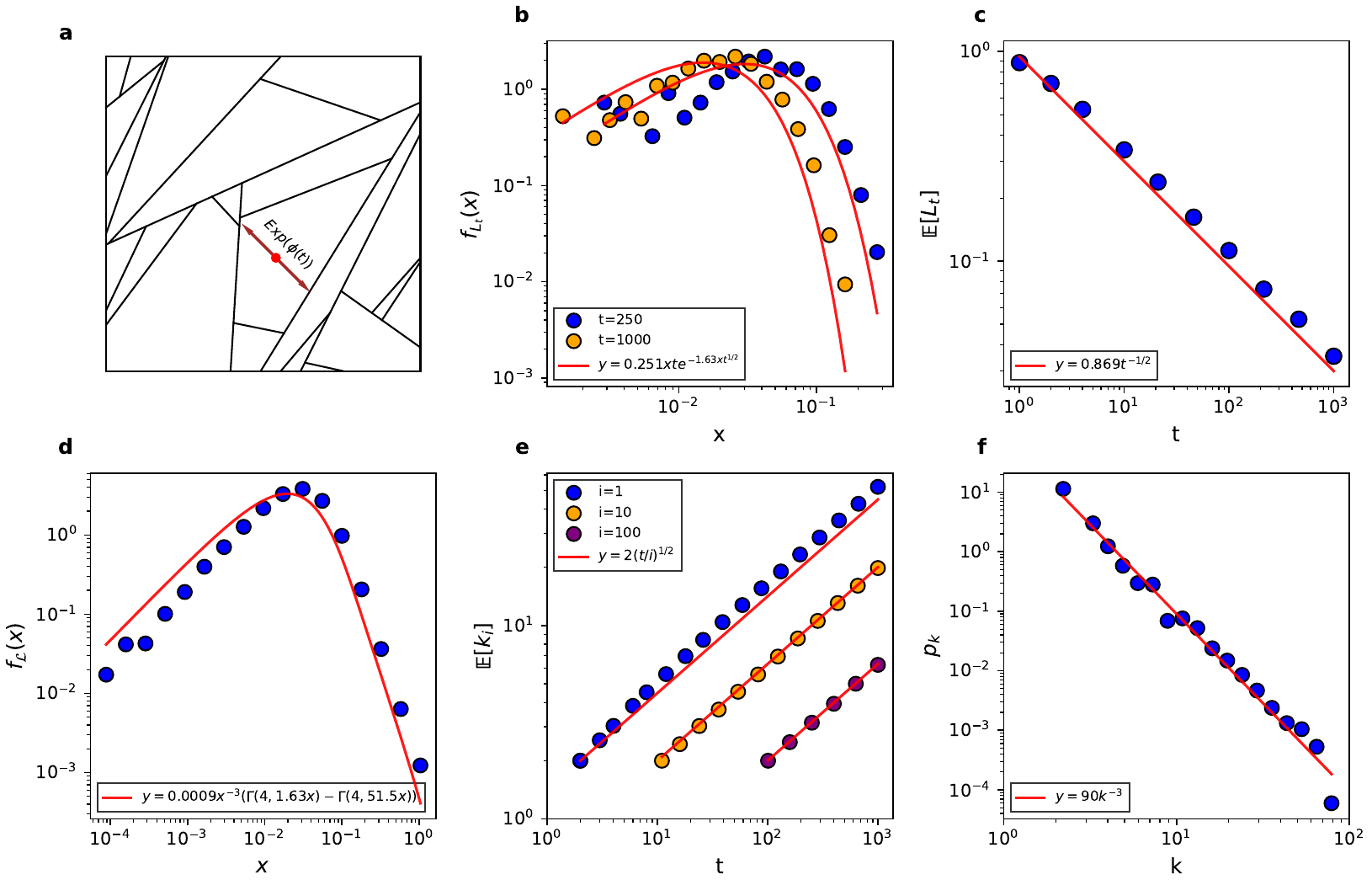}
    \caption{
    \textbf{Empirical distributions and theoretical comparisons for the Scale-Rich model without thickness ($\lambda_0 = 0$).}
    {\bf (a)} Realization of a Scale-Rich network without thickness of size 20, highlighting a nucleation point in red and illustrating the exponential model for segment lengths.
    {\bf (b)} Distributions of segment lengths introduced at times 250 and 1000, compared against the theoretical Erlang distribution in Eq.~\ref{eq:erlang}.
    {\bf (c)} Mean segment lengths for segments introduced at time $t$, shown alongside the theoretical expectation in Eq.~\ref{eq:erlang_mean}.
    {\bf (d)} Final network segment length distribution, as given by Eq.~\ref{eq:final}.
    {\bf (e)} Mean segment degree of segments introduced at times $i=1$, $i=10$, and $i=100$, shown alongside the theoretical result.
    {\bf (f)} Final network degree distribution, as given by Eq.~\ref{eq:no_thickness_degree} with $\beta = 1/2$.
    All data are aggregated from 1000 networks of size 1000.
    }
    \label{fig:sup_thickless}
\end{figure}
To find the degree distribution, we first calculate the number of nodes with degree smaller than $k$ by solving for $k_i(t) < k$. We obtain
\begin{align}
    i > t k^{-1/\beta} \left( \frac{2(1-\beta)}{\beta} \right)^{1/\beta}.
\end{align}
Therefore, the number of nodes with degree smaller than $k$ is equal to
\begin{align}
    t k^{-1/\beta} \left( \frac{2(1-\beta)}{\beta} \right)^{1/\beta}.
\end{align}
For a network of size $n$ the probability, the probability that a uniformly selected node has degree smaller than $k$ is
\begin{align}
    P_k &\sim 1 - k^{-1/\beta} \left( \frac{2(1-\beta)}{\beta} \right)^{1/\beta}.
    \intertext{Taking the derivative gives the probability density function}
    p_k &= \frac{\partial P(k)}{\partial k} \sim k^{-(1+1/\beta)}.
\label{eq:no_thickness_degree}
\end{align}
For the Scale-Rich model without thickness we have proven in Sec.~\ref{sec:expected_length} that $\beta=1/2$.
Therefore, the degree distribution follows a power law distribution with exponent $\gamma = 1 + 1/\beta = 3$

We find that in the zero-thickness limit ($\lambda_0 = 0$), the Scale-Rich model exhibits a power-law degree distribution, $P(k) \sim k^{-3}$, as well as a power-law tail in the ligament length distribution, with long ligaments following $P(\ell) \sim \ell^{-3}$ (Fig.~\ref{fig:sup_thickless}).

\subsubsection{Degree decay function of the Scale-Rich model with varying thickness ($\lambda_0 \ne 0$) in the tunable density domain}
\label{sec:degree_distribution}
When ligaments have finite thickness, the available area decreases and each ligament receives fewer connections due to reduced space.
This prompted us to derive the degree distribution and analyze whether it depends on ($\alpha, \lambda_0$), similar to ligament length.

For the Scale-Rich model with thickness ($\lambda_0 \ne 0$), the free circumference changes as the network grows, a mechanism that is not present in the Scale-Rich model without thickness ($\lambda_0 = 0$).
Note that intersections are never truly perpendicular and therefore at every timestep we will have a larger intersection than just $\lambda_t$ on both sides.
We call the appropriate scaling $\chi$ (such that we have an intersection of total $\chi \lambda_t$ length).
The rate at which the free circumference changes at time $t$, for a segment introduced at time $i$, is
\begin{align}
    \frac{\mathrm{d} W_i(t)}{\mathrm{d}t} &= -\mathbb{P}\{ \text{hit segment } i \} \cdot \chi \lambda(t) \\
    &\approx -\frac{W_i(t)}{ \sum\limits_{j=1}^{t-1} W_j(t)} \chi\lambda_0 t^{-\alpha} \\
    &\approx -\chi\lambda_0 t^{-\alpha} W_i(t) \left( 2\sum\limits_{j=1}^{t-1} L_j - \chi\sum\limits_{j=1}^{t-1} \lambda_j \right)^{-1} \\
    &= -\chi\lambda_0 t^{-\alpha} W_i(t) \left( 2\ell_0\sum\limits_{j=1}^{t-1} j^{-\beta} - \chi\lambda_0\sum\limits_{j=1}^{t-1} j^{-\alpha} \right)^{-1} \\
    &\approx -\chi\lambda_0 t^{-\alpha} W_i(t) \left( 2 \ell_0 \int\limits_{1}^{t} x^{-\beta} \, \mathrm{dx} - \chi \lambda_0 \int\limits_{1}^{t} x^{-\alpha} \, \mathrm{d}x \right)^{-1}
\end{align}
Let $\omega = \frac{\chi \lambda_0}{2 \ell_0}$. A general solution of this differential equation is given by
\begin{align}
    W_i(t) &= W_i(i) \exp\bigg\{ -\omega \int_{i}^{t} y^{-\alpha} \left( \int\limits_{1}^{y} x^{-\beta} \, \mathrm{d}x - \omega \int\limits_{1}^{y} x^{-\alpha} \, \mathrm{d}x \right)^{-1} \mathrm{d}y \bigg\} \\
    &\approx W_i(i) \exp\bigg\{ -\omega \int_{i}^{t} y^{-\alpha} \left( (1-\beta)^{-1} y^{1-\beta} - \omega (1-\alpha)^{-1} y^{1-\alpha} \right)^{-1} \mathrm{d}y \bigg\} \label{eq:circum_gen}
\end{align}
In the interior of the feasible domain we known $\alpha > \beta = 1/2$. Since $\omega < 1$, asymptotically the term $y^{1-\beta}$ denominates the denominator of the integrand. For large $t$, we have
\begin{align}
    W_i(t) &\approx W_i(i) \exp\bigg\{ -\omega (1-\beta) \int_{i}^{t} y^{\beta-\alpha-1} \mathrm{d}y \bigg\}
    \approx W_i(i) \exp\bigg\{ -\omega \frac{1-\beta}{\beta-\alpha} \left( t^{\beta-\alpha} - i^{\beta-\alpha} \right) \bigg\}.
\end{align}
We define the parameter $r = \omega \frac{1-\beta}{\alpha-\beta}$ and rewrite
\begin{align}
    W_i(t) &\approx W_i(i) \exp\bigg\{ -r i^{\beta-\alpha} \bigg\} \exp\bigg\{ r t^{\beta-\alpha} \bigg\}.
\end{align}
Let $m$ be the mean number of intersections a segment makes upon introduction.
The probability of obtaining a link at time $t$ is inversely proportional to the total free circumference in the system (for large $t$), giving
\begin{align}
    \frac{\mathrm{d} k_i(t)}{\mathrm{d}t}
    &= m \frac{W_i(t)}{\sum\limits_{j=1}^{t}W_j(t)} \\
    &\approx m W_i(i) \exp\bigg\{ r t^{\beta - \alpha} \bigg\} \exp\bigg\{ -r i^{\beta - \alpha} \bigg\} \exp\bigg\{ -r t^{\beta-\alpha} \bigg\} \left( \sum_{j=1}^{t} W_j(j) \exp\bigg\{ -r j^{\beta-\alpha} \bigg\} \right)^{-1} \\
    &\approx m 2\ell_0 i^{-\beta} \exp\bigg\{ -r i^{\beta - \alpha} \bigg\} \left( 2\ell_0 \int_{1}^{t} y^{-\beta}\exp\bigg\{ -r y^{\beta-\alpha} \bigg\} \mathrm{d}y \right)^{-1} \\
    &= m i^{-\beta} \exp\bigg\{ -r i^{\beta - \alpha} \bigg\} \left( \int_{1}^{t} y^{-\beta}\exp\bigg\{ -r y^{\beta-\alpha} \bigg\} \mathrm{d}y \right)^{-1}. \label{eq:full}
\end{align}
Note that only the last term depends on $t$, and we show that it follows a power law
\begin{align}
    \int_{1}^{t} y^{-\beta}\exp\bigg\{ -r y^{\beta-\alpha} \bigg\} \mathrm{d}y &= r^{\frac{\beta - 1}{\beta - \alpha}} (\beta - \alpha)^{-1} \left[ \Gamma\!\left(\frac{\beta - 1}{\alpha - \beta},\, r\right) - \Gamma\!\left(\frac{\beta - 1}{\alpha - \beta},\, r\,t^{-\alpha + \beta}\right) \right].
\end{align}
Note that all $t$ dependence occurs in the final term. We can write this in terms of the incomplete Gamma function $\Gamma(s,x)$, which has the equivalent form (for small $x$)
\begin{align}
    \Gamma(s,x) &= x^s \sum_{k=0}^{\infty} \frac{(-1)^k x^k}{k! (s+k)}.
    \intertext{Taking a first order approximation we get:}
    \Gamma(s,x) &= \Gamma(s) - x^s/s.
\end{align}
Rewriting in this form gives
\begin{align}
    \Gamma\!\left(\frac{\beta - 1}{\alpha - \beta},\, r\,t^{-\alpha + \beta}\right)^{-1} &\overset{\text{large } t}{\sim} \left( \frac{ (rt^{-\alpha + \beta})^{\frac{\beta - 1}{\alpha - \beta}} }{ \frac{\beta - 1}{\alpha - \beta} } \right)^{-1} \sim t^{\beta-1}.
\end{align}
Rewriting Eq.~\eqref{eq:full} gives
\begin{align}
    \frac{\mathrm{d} k_i(t)}{\mathrm{d}t} &\overset{\text{large } t}{\sim} i^{-\beta} \exp\bigg\{ -r i^{\beta - \alpha} \bigg\} t^{\beta-1}.
\end{align}
In the interior of the feasible domain, we know that $\beta$ equals $1/2$, giving the probability to obtain a link equal to $t^{-1/2}$, a power law behavior that was also observed empirically, and shown in Fig.~\ref{fig:sup_thickness}.
To obtain the expected degree at time $t$ of a link introduced at time $i$, we integrate the probability of obtaining a link with boundary condition $k_i(i)=2$, obtaining
\begin{align}
    k_i(t) &= 2 + \frac{1}{\beta}\exp\bigg\{ -r i^{1/2 - \alpha} \bigg\}\left(\left(\frac{t}{i}\right)^{1/2}-1\right).
\end{align}
Since $\alpha>\beta$ the exponential is close to one, and therefore, for large $i$, this expression behaves as $(t/i)^{1/2}$, similarly when $\lambda_0 = 0$.
This occurs because ligament thickness decays faster than the free perimeter during generation, allowing new ligaments to be added continuously in the tunable-density regime.
Consequently, the degree distribution remains unaffected by finite thickness.

\subsubsection{Degree decay function of the Scale-Rich model at the boundary of the tunable density domain}
\begin{figure}[h!]
    \centering
    \includegraphics[width=\linewidth]{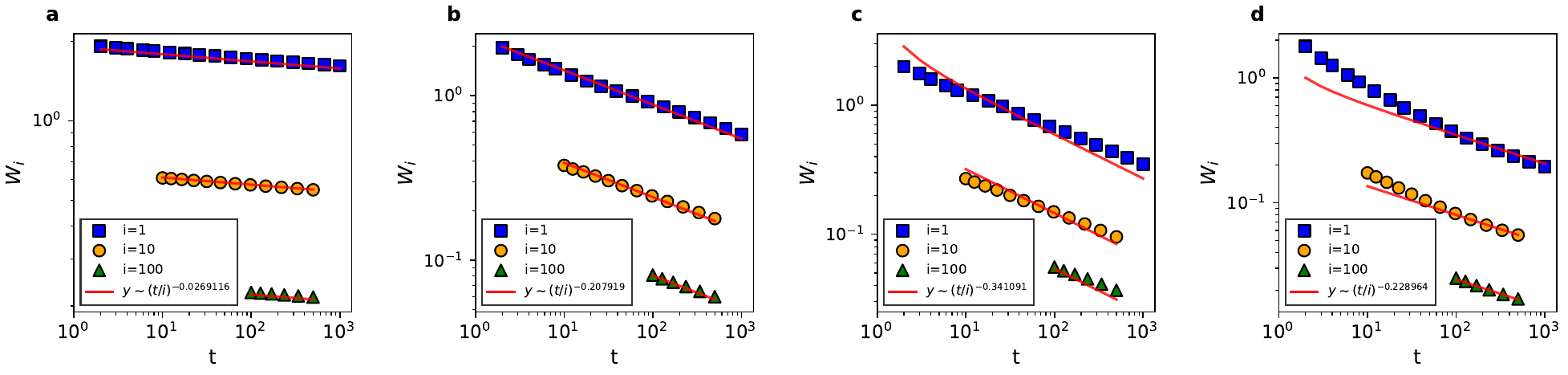}
    \caption{{\bf The free circumference of segments follows a powerlaw distribution on the boundary of the tunable density domain.}
    {\bf (a-d)} The change in free circumference of segments, at the boundary of the feasibility domain, introduced at step $i=1, 10, 100$. Data for both panels is aggregated from $1000$ networks of size $1000$ with parameters:
    {\bf (a)} $\alpha=0.5179$ and $\lambda_0=0.035072$,
    {\bf (b)} $\alpha=0.6429$ and $\lambda_0=0.244$,
    {\bf (c)} $\alpha=0.75$ and $\lambda_0=0.382793$, and
    {\bf (d)} $\alpha=0.97$ and $\lambda_0=0.586511$.
    }
    \label{fig:circumference}
\end{figure}

\begin{figure}
    \centering
    \includegraphics[width=1\linewidth]{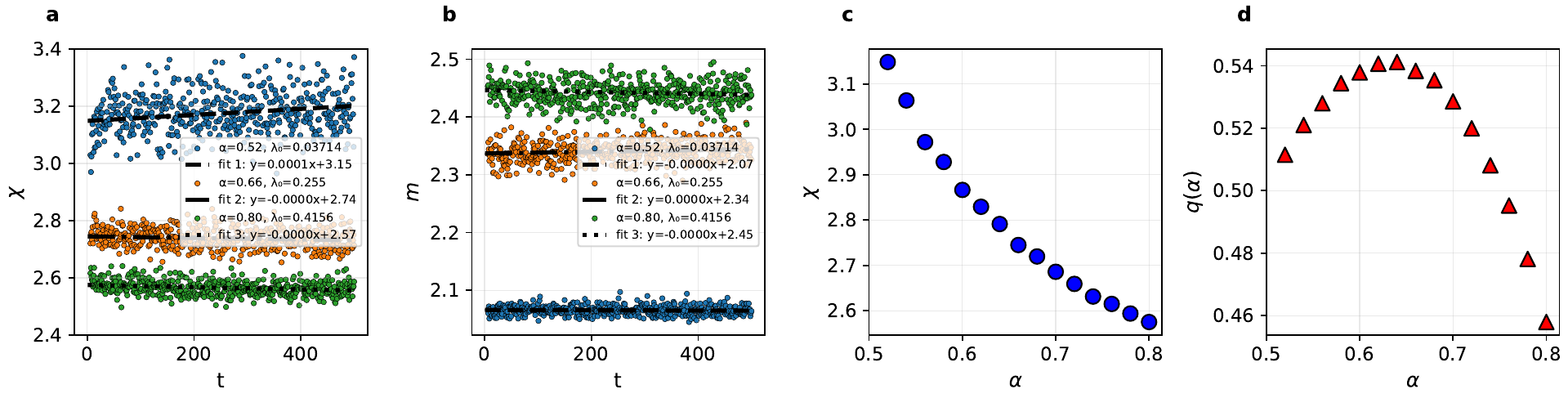}
    \caption{
    \textbf{Scaling exponent $q$ on the boundary of the tunable density domain.}
    \textbf{(a)} Mean $\chi$ values along the boundary for $\alpha = 0.52, 0.66, 0.80$ show no dependence on time $t$.
    \textbf{(b)} Mean $m$ (initial number of neighbors) along the boundary for $\alpha = 0.52, 0.66, 0.80$ show no dependence on time $t$.
    \textbf{(c)} Variation of $\chi$ with the network parameter $\alpha$, obtained from linear regression.
    \textbf{(d)} Scaling exponent $q$ computed from $\alpha$ and the corresponding $\chi$.
    For each $\alpha$, results are averaged over $1000$ networks of size $500$.
    }
    \label{fig:chi_values_plot}
\end{figure}

We next analyze the degree distribution at the boundary between the jammed and tunable-density state.
On the boundary of we have $\beta=\alpha$.
Therefore the free circumference at iteration $t$ is given by the function (Eq.~\eqref{eq:circum_gen})
\begin{align}
    W_i(t)
    &\approx W_i(i) \exp\bigg\{ -\frac{\omega(1-\alpha)}{1-\omega}\int_{i}^{t} y^{-1} \mathrm{d}y \bigg\} \\
    &= W_i(i) \left( \frac{t}{i} \right)^{\frac{\omega(1-\alpha)}{\omega-1}}.
\end{align}
Fig.~\ref{fig:circumference} shows the decay of the free circumference, for networks at the boundary between the jammed and tunable-density state, for different introduction times $i$.
Similarly to Sec.\ref{sec:degree_distribution} we determine the probability of obtaining a link using
\begin{align}
    \frac{\mathrm{d} k_i(t)}{\mathrm{d}t} &= m \frac{W_i(t)}{\sum\limits_{j=1}^{t} W_j(t)} \\
    &\approx m i^{-\alpha-\frac{\omega (1-\alpha)}{\omega-1}} \left( \int_{1}^{t} y^{-\alpha} y^{-\frac{\omega(1-\alpha)}{\omega-1}} \mathrm{d}y \right)^{-1}.
    \intertext{This expression behaves as }
    \frac{\mathrm{d} k_i(t)}{\mathrm{d} t} &\approx m i^{-\alpha-\frac{\omega(1-\alpha)}{\omega-1}} \left( t^{1-\alpha - \frac{\omega(\alpha-1)}{\omega-1}} - 1 \right)^{-1}.
    \intertext{When $\alpha$ is not too close to one, we can neglect ($-1$) term, obtaining }
    \frac{\mathrm{d} k_i(t)}{\mathrm{d}t} &\sim i^{-\alpha-\frac{\omega(1-\alpha)}{\omega-1}} t^{-1+\alpha + \frac{\omega(\alpha-1)}{\omega-1}}.
\end{align}
Substituting back $\omega$, $\lambda_0 = \ell_0^{-1}\zeta(2\alpha)^{-1}$, and simplifying gives the following exponent defined as $q$
\begin{align}
    q = -1  +\alpha + \frac{\omega(\alpha-1)}{\omega-1} &= \frac{2 (\alpha - 1)\, \zeta(2\alpha)}{\chi - 2\, \zeta(2\alpha)}.
\end{align}
The parameter $\chi$ depends on the system (Fig.~\ref{fig:chi_values_plot}).
For large $t$, the probability of obtaining a link follows
\begin{align}
    \frac{\mathrm{d}k_i(t)}{\mathrm{d}t} \sim (t/i)^{-1/2}.
\end{align}
Similar to the tunable-density regime, we observe the same behavior at the boundary, since the boundary only fully covers the area in the infinite-time limit. Taken together, these results indicate that as long as ligaments can be added without jamming, the degree distribution remains largely unaffected by physical constraints.

\subsubsection{Deriving the degree distribution for the Scale-Rich model}
To obtain the expected degree at time $t$ of a link introduced at time $i$, we integrate the probability of obtaining a link, giving
\begin{align}
    k_i(t) \sim (t/i)^{1/2}. \label{eq:th_degree_temp}
\end{align}
Solving $k_i(t) < k$, we obtain $i > t k^{-1/2}$.

Therefore, the number of nodes with degree smaller than $k$ is equal to $t k^{-2}$.

For a network of size $n$, the probability that a uniformly selected node has a degree smaller than $k$ is
\begin{align}
    P_k \sim 1 - k^{-2}.
\end{align}
Taking the derivative gives the density function
\begin{align}
    p_k = \frac{ \mathrm{\partial} P_k }{ \mathrm{\partial} k } \sim k^{-3}.\label{eq:th_degree}
\end{align}
Therefore, the degree distribution follows a power-law distribution with an exponent $\gamma = 3$.

We find that the degree distribution follows $P(k)\approx k^{-3}$,indicating the coexistence of ligaments with both few and many connections within the system.

\begin{figure}[htpb]
    \centering
    \includegraphics[width=.8\linewidth]{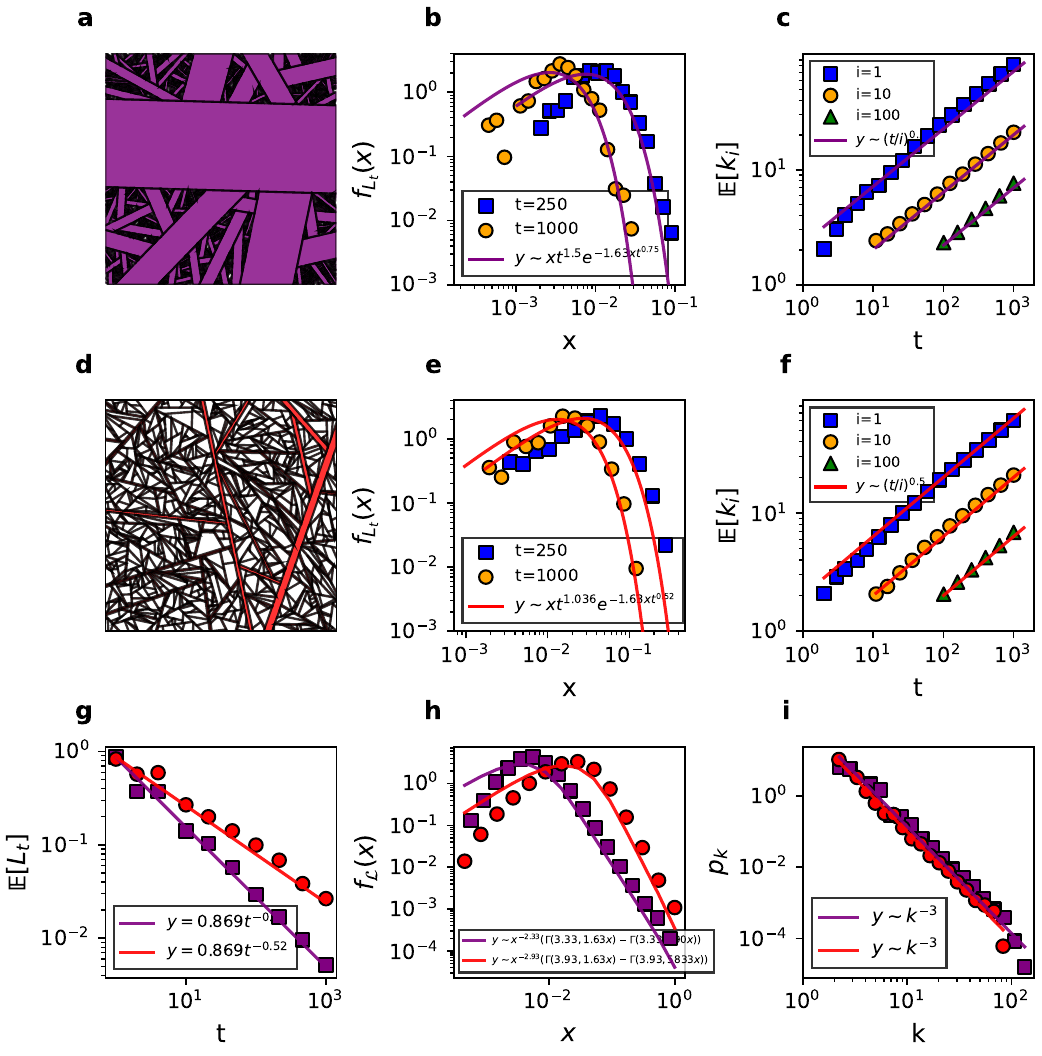}
    \caption{
    \textbf{Empirical distributions and theoretical comparisons for the Scale-Rich model with thickness.} All data are aggregated from 1000 networks of size 1000. Panels {\bf a,b,c} use data generated with thickness rate $\alpha=0.75$ and $\lambda_0=0.3828$. Panels {\bf d,e,f} use data generated with thickness rate $\alpha=0.52$ and $\lambda_0=0.035$. Both configurations are located on the boundary of the phase space. Panels {\bf g,h,i} show the combined data, indicated with purple ($\alpha$=0.75) and red ($\alpha=0.52$).
    {\bf (a)} A realization of the Scale-Rich model.
    {\bf (b,e)} Distributions of segment lengths introduced at times 250 and 1000, compared against the theoretical Erlang distribution in Eq.~\eqref{eq:erlang}.
    {\bf (c,f)} Mean segment degree of segments introduced at times $i=1$, $i=10$, and $i=100$, shown alongside the theoretical result, as given by Eq.~\eqref{eq:th_degree_temp}.
    {\bf (d)} A realization of the Scale-Rich model.
    {\bf (g)} Mean segment lengths for segments introduced at time $t$, shown alongside the theoretical expectation in Eq.~\eqref{eq:erlang_mean}.
    {\bf (h)} Final network segment length distribution, as given by Eq.~\eqref{eq:final}.
    {\bf (i)} Final network degree distribution, as given by Eq.~\eqref{eq:th_degree}.
    }
    \label{fig:sup_thickness}
\end{figure}

Taken together all the results, the ligaments lengths follow an Erlang distribution determined by $\ell_0$ and $\alpha$ (Fig.~\ref{fig:sup_thickness}b,c). The mean ligaments length introduced at time $t$ follows a power law $p_L \sim t^{-\beta}$ (Fig.~\ref{fig:sup_thickness}g). The final length distribution is a sum over gamma functions and dependent on the system parameters $\ell_0$ and $\alpha$ (Eq.~\eqref{eq:final}), yielding a power-law tail $p_{\mathcal{L}} \sim x^{1+1/\beta}$ (Fig.~\ref{fig:sup_thickness}h).
The degree distribution is determined by the mean degree of ligaments introduced at time $i$, which scales as $t^{1/2}$, independent of model parameters (Fig.~\ref{fig:sup_thickness}c). This independence leads to a universal degree distribution $p_k \sim k^{-3}$ (Fig.~\ref{fig:sup_thickness}i).

\newpage
\subsection{Generalizations of the Scale-Rich model.}
\label{subsec: model_limit}

The SR metamaterials introduced above were conceived to generate structures that lack an intrinsic length scale.
In this section, we propose a dynamical generalization of the SR model and show that, in specific limits, it maps onto well-studied models in statistical physics and materials science, like the Gilbert tessellation and fragmentation models developed to describe crack patterns and the growth in polycrystalline materials \cite{rao1995kinematic,bohn2005hierarchical, lee2025three}.
The resulting generative approach spans a wide spectrum of lattice architectures, from regular to stochastic, in both two and three dimensions.

We begin by introducing the Dynamical Scale-Rich (DSR) model that features precisely controlled nucleation points and introduction times.
Starting from a unit square ($L = 1$) in 2D, a new ligament is initiated at each time step $\Delta t$, where $\Delta t$ dictates the rate of nucleation.
A key distinction from the original SR model is that the DSR formulation enables direct control over the nucleation rate through the choice of $\Delta t$.
Ligaments originate from randomly selected points and orientations, and their thickness evolves as $\lambda(t) = \lambda_0 t^{-\alpha}$, where $\lambda_0$ is the initial thickness and $\alpha$ is the decay exponent.
The generation process consists of two steps:

(i) choose a random nucleation point $(x_t, y_t)$ within the $L^2$ space and a random angle $\theta_t$;

(ii) grow a ligament of thickness $\lambda_t$ from the nucleation point in both directions, defined by $\theta_t$, until it intersects other ligaments or the boundary.

\begin{figure}[htbp]
    \centering
    \includegraphics[width=1\linewidth]{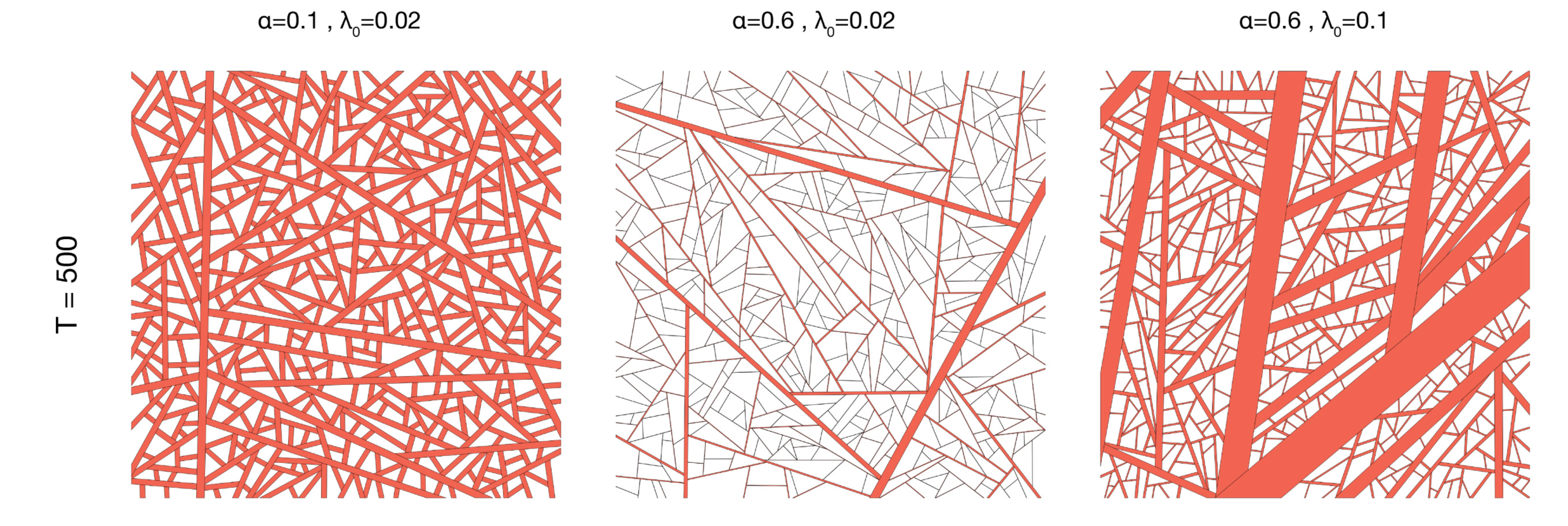}
    \caption{
    \textbf{SR metamaterials.}
    A total of $T=500$ ligament samples are generated with varying $(\alpha, \lambda_0)$ configurations. Smaller $\alpha$ values correspond to slower decay in ligament thickness.
    }
    \label{fig:sr-samples}
\end{figure}

We add a new line by repeating (i), (ii) at every $\Delta t$ interval until a predefined number of lines (time steps $T$) or until we are unable to add new ligaments, leading to a state of jamming~\cite{posfai2024impact} (Fig.~\ref{fig:sr-samples}).
By tuning $\Delta t$, we control the number of ligaments that grow simultaneously.
For large $\Delta t$, ligaments nucleate infrequently and are therefore added in a predominantly sequential manner.
In contrast, small $\Delta t$ values promote parallel growth, with many ligaments growing at the same time.
If $\Delta t$ is sufficiently large such that each ligament reaches another ligament or the boundary before the next one begins growing, the DSR model reduces to the SR model.
In the present formulation, ligaments grow with fixed velocity, although this parameter could be varied to introduce an additional level of control.
Note, that the DSR framework generates space-filling metamaterials, providing a flexible approach to construct systems of arbitrary shape, not limited to squares in two dimensions or cubes in three dimensions.

\begin{figure}[htbp]
    \centering
    \includegraphics[width=1\linewidth]{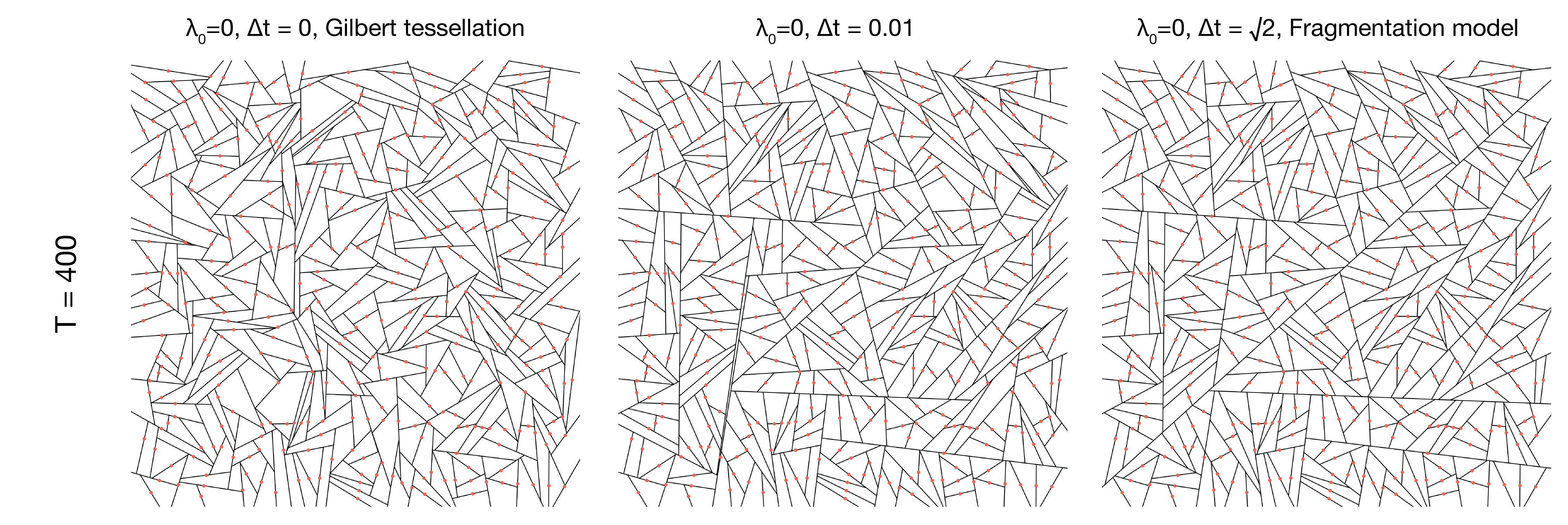}
    \caption{
    \textbf{Fragmentation model vs. Gilbert tessellation.}
    Controlling the arrival time of the ligaments, we can connect the Gilbert tessellation with the fragmentation model, which is the limit of the DSR model for $\lambda_0 = 0$. The orange dots mark the nucleation positions.
    }
    \label{fig:sr-gilbert}
\end{figure}

For $\lambda_0 = 0$ and $\Delta t \to 0$ (Fig.~\ref{fig:sr-gilbert}) the DSR model reduces to Gilbert tessellation \cite{gilbert1967random} resulting in a  Poisson distribution in $P(\ell)$ and $P(k)$  similar to the Voronoi tesselation.
In the other limit, i.e. for $\lambda_0 = 0$ and $\Delta t \to L\sqrt{2}$, the DSR model converges to the fragmentation model \cite{krapivsky1994scaling, krapivsky2010kinetic} (Fig.~\ref{fig:sr-gilbert}), that displays power-low distribution both in $P(\ell)$ and $P(k)$, except for $P(\lambda)$.
For comparison of the different models, see Table~\ref{tab:sr-limits}.

\begin{table}[]
\small
\centering
\begin{tabular}[c]{ | l | l | l | l |}
    \hline
    \textbf{Model} & \textbf{$P(\lambda)$} & \textbf{$P(\ell)$} & \textbf{$P(k)$}   \\ \hline
    \textbf{Gilbert tesselation} \cite{gilbert1967random} & $\delta(\lambda=0)$ &  Poisson & Poisson \\ \hline
    \textbf{Fragmentation} \cite{krapivsky1994scaling} & $\delta(\lambda=0)$  & $\ell^{-3}$ & $k^{-3}$ \\ \hline
    \textbf{Scale-Rich} & $\lambda^{-(1+1/\alpha)}$ & $\ell^{-(1+1/\beta)}$ where $\beta = \alpha$ or $1/2$ & $k^{-3}$ \\ \hline
    \textbf{Scale-Rich 3D plate} & $\lambda^{-(1+1/\alpha)}$ & $\ell^{-(1+1/\beta)}$ where $\beta = 2\alpha$ or $3/2$ & $k^{-2.5}$ \\ \hline
\end{tabular}
\caption{\textbf{Dynamical Scale-Rich model in the limits.} Comparison of the distributions of ligament thickness $P(\lambda)$, ligament length $P(\ell)$, and node degree $P(k)$.
The Gilbert tessellation generates networks with Poisson-distributed ligament lengths and node degrees, and ligaments of zero thickness, giving the system a characteristic scale determined by the Poisson distribution.
The fragmentation model produces networks with power-law distributions in $P(k)$ and $P(\ell)$, both with exponent 3, while the ligaments also have zero thickness.
In contrast, Scale-Rich systems exhibit tunable ligament length and thickness distributions, controlled by the parameters $(\alpha, \lambda_0)$, and display two distinct structural phases.
With their power-law degree distributions, SR systems exhibit the coexistence of multiple scales across all three distributions.}
\label{tab:sr-limits}
\end{table}

\begin{figure}[htbp]
    \centering
    \includegraphics[width=1\linewidth]{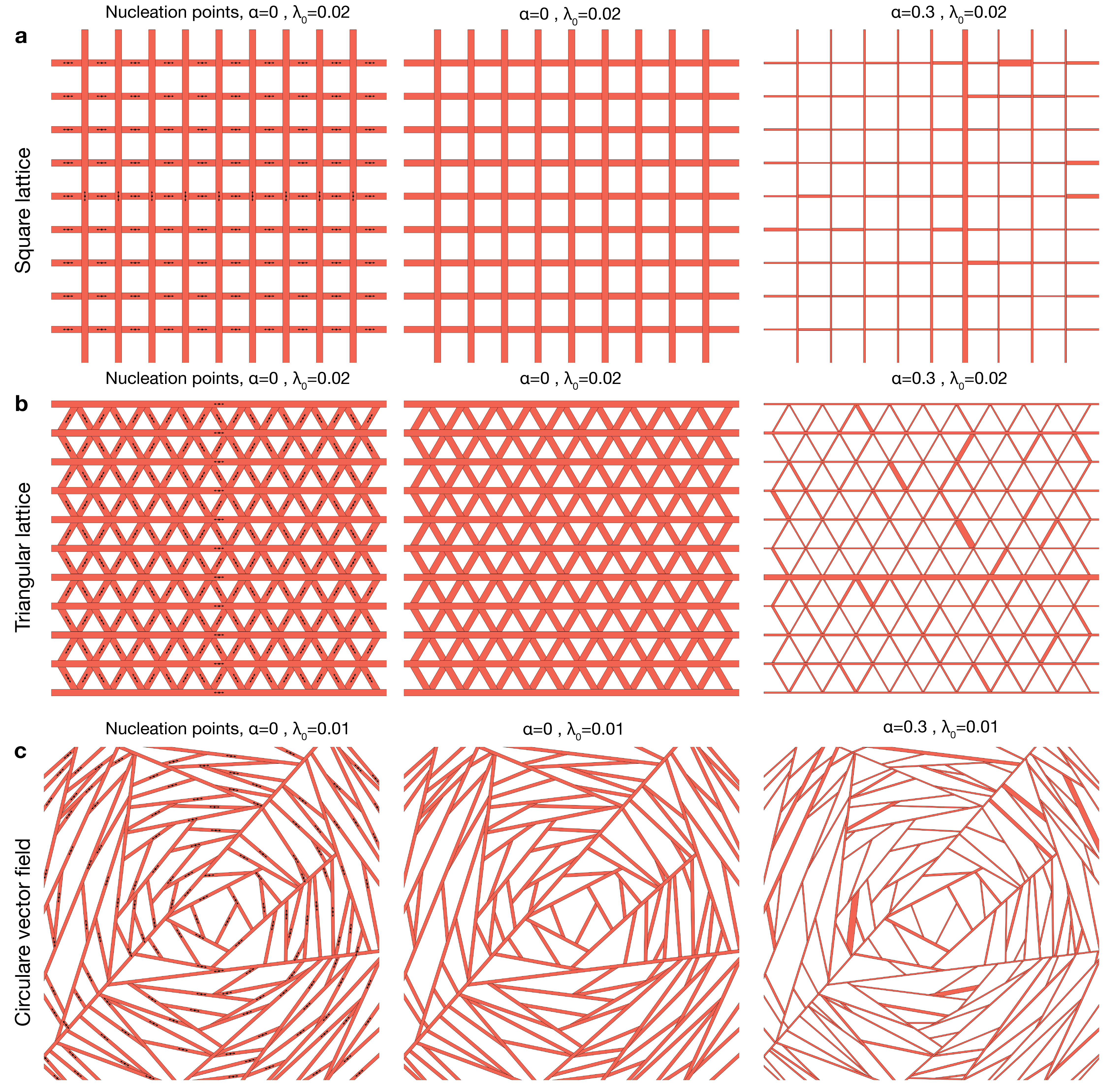}
    \caption{
    \textbf{Regular lattices generated by the SR model.}
    \textbf{(a)} Square lattice. \textbf{(b)} Triangular lattice. \textbf{(c)} Circular vector field.
    }
    \label{fig:sr-regular}
\end{figure}

Note that the DSR model can be tuned to generate more regular networks by controlling the location of nucleation points and the orientation of lines.
For example, it can generate square and triangular lattices, or approximate arbitrary vector fields (Fig.~\ref{fig:sr-regular}).
At the same time, the SR model cannot produce structures that lack long system-spanning elements, such as hexagonal honeycomb lattices.
In contrast, the Gilbert tessellation, when seeded in a controlled manner, is capable of generating such short-range systems, illustrating the difference between the two model.

\begin{figure}[htbp]
    \centering
    \includegraphics[width=1\linewidth]{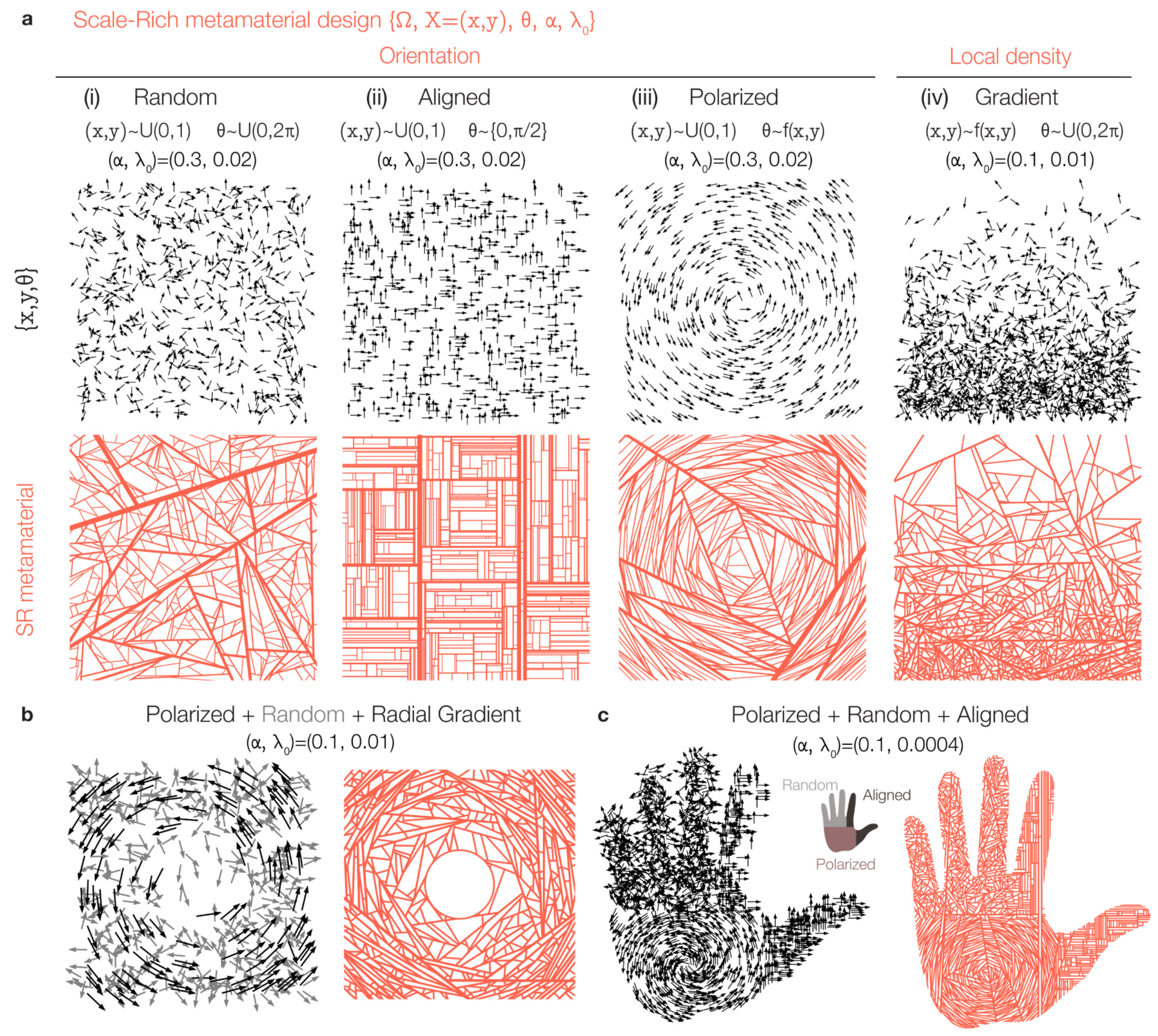}
    \caption{
    \textbf{The flexibility of SR framework.}
    \textbf{(a)}
    The SR framework enables precise control over ligament
orientation $\theta$, allowing the generation of (i) random configurations where angles are selected from a uniform distribution ($\theta = U(0,2\pi)$), (ii)
aligned networks using random selections from a prescribed set of angles ($\theta = {0,\pi/2}$), and (iii) polarized patterns by sampling angles based
on a predefined vector field ($\theta = f(x,y)$ - circular field)).
(iv) Nucleation point positions can also be controlled to vary density across the domain, enabling
gradient patterns ($(x,y) = f(x,y)$ - here gradient along the $y$ axis).
\textbf{(b)} Design strategy to protect a “payload” by varying both ligament density and orientation. The first 20\% of ligaments are oriented according to a circular vector field (black), while the remainder are assigned random orientations from a uniform distribution (grey). Nucleation points follow a radial density profile—denser at the boundary and sparser toward the center—taking inspiration from the natural microstructure of the pomelo fruit \cite{le2023influence}.
\textbf{(c)}
A non-convex “hand-shaped” domain constructed by combining three distinct orientation patterns, random, aligned, and polarized, each applied to a different region of the geometry.
    }
    \label{fig:sr-flexibility}
\end{figure}

The SR framework is less flexible than the DSR model, yet it offers a broad set of tunable parameters: the domain shape $\Omega$; nucleation point positions $\mathbf{X} = (x,y)$; ligament orientation $\theta$; thickness decay rate $\alpha$; initial thickness $\lambda_0$; and the total number of ligaments $T$.
As shown in Fig.~\ref{fig:sr-samples}, the pair $(\alpha, \lambda_0)$ governs the thickness distribution across the sample.
By prescribing $\theta$, the framework can generate a variety of orientation patterns: random networks by sampling angles from a uniform distribution (Fig.~\ref{fig:sr-flexibility}a(i)); aligned configurations by selecting from a predefined set of angles (Fig.~\ref{fig:sr-flexibility}a(ii), e.g., $\theta={0,\pi/2}$); or polarized structures by drawing angles from a vector field (Fig.~\ref{fig:sr-flexibility}a(iii), circular field).
The nucleation point density can also be controlled to create spatially varying architectures, such as gradient patterns (Fig.~\ref{fig:sr-flexibility}a(iv), gradient along $y$).
These parameters may be tuned independently or simultaneously.
Fig.~\ref{fig:sr-flexibility}b illustrates a design that protects an embedded “payload” by adjusting ligament density and orientation.
The framework further extends to non-convex geometries: Fig.~\ref{fig:sr-flexibility}c demonstrates an SR structure generated within a hand-shaped domain, combining random, aligned, and polarized regions.

\begin{figure}[htbp]
    \centering
    \includegraphics[width=1\linewidth]{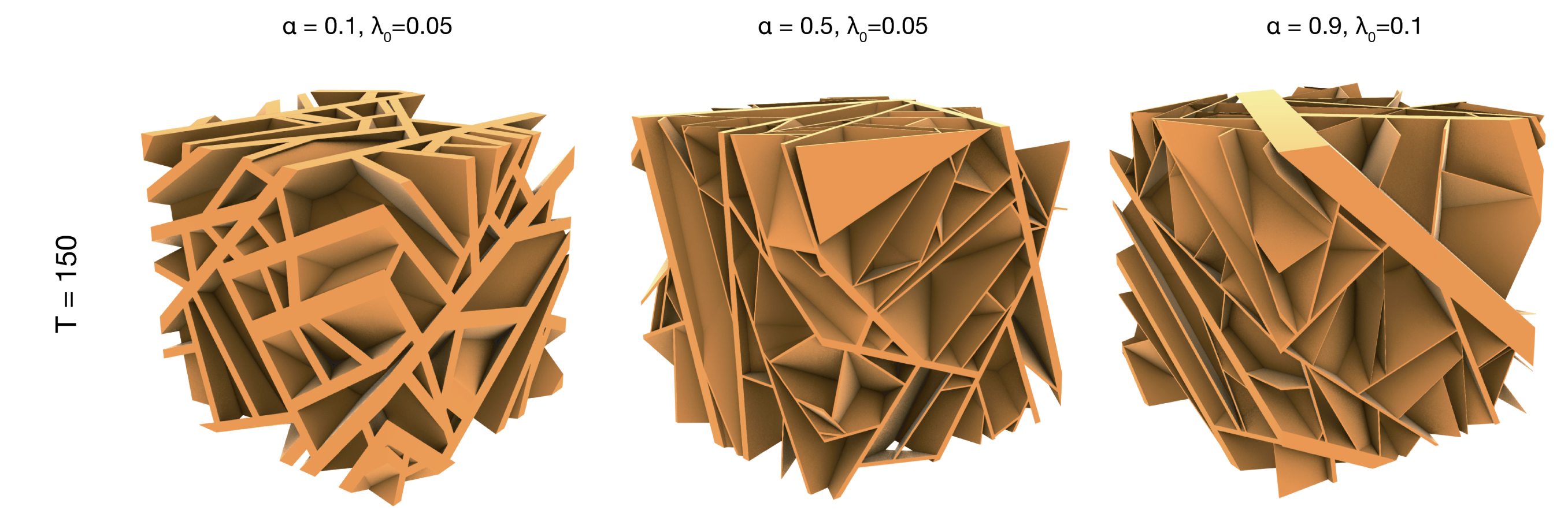}
    \caption{
    \textbf{SR model with plates in 3D.}
    A total of $T=500$ plate samples generated with varying $(\alpha, \lambda_0)$ configurations.
    }
    \label{fig:sr-3Dplates}
\end{figure}

In 3D, the construction follows the same iterative scheme, with ligaments replaced by plates.
At each time step, a plate of thickness $\lambda(t) = \lambda_0 t^{-\alpha}$, where $\lambda_0$ is the initial thickness and $\alpha$ the decay exponent, is introduced with a random angle combination $(\theta, \phi)$ and grown until it intersects either another plate or the system boundary \ref{fig:sr-3Dplates}.
This results in a closed cell structure.
The 3D SR model with $\lambda_0 = 0$ reduces to the STIT  model (STable under Iteration of Tessellations)
~\cite{thale2010new}.

\begin{figure}[htbp]
    \centering
    \includegraphics[width=1\linewidth]{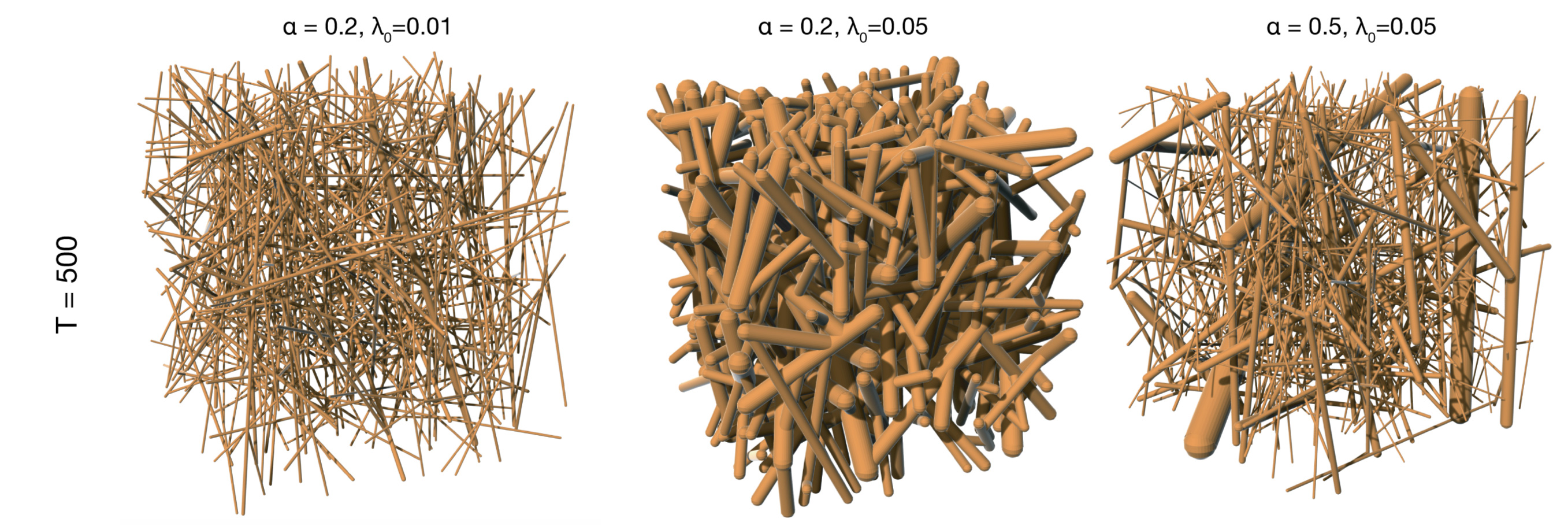}
    \caption{
    \textbf{SR model with 3D cylinders.}
    Cylindrical structures are placed with varying $(\alpha, \lambda_0)$ parameters. A total of $T=500$ samples are generated.
    }
    \label{fig:sr-3Dbars}
\end{figure}

The SR framework can also be extended to non–space-splitting variants, where the elements do not recursively partition the domain into smaller subregions but instead remove volume from it.
In three dimensions, for example, one can introduce cylindrical units within a cubic domain. At each time step, a cylinder of thickness $\lambda(t) = \lambda_0 t^{-\alpha}$, where $\lambda_0$ denotes the initial thickness and $\alpha$ the decay exponent, is placed at a random orientation $(\theta, \phi)$ and grown until it intersects either another cylinder or the system boundary (Fig.~\ref{fig:sr-3Dbars}).
In this case, the cylinders occupy parts of the available space but do not subdivide it.

\clearpage
\subsection{3D plate Scale-Rich model}\label{sec:3D_plates_model}
While we focused on the 2D version of Scale-Rich model, the framework can be extended to 3D for generating closed cell metamaterials \cite{Crook2020}.
Here, we study the behavior the 3D plate model, deriving the plate area and the contact network degree distribution, where the nodes are the plates and edges are the connection points (Fig.~\ref{fig:sr-3Dplates}).

To generate a 3D network composed of intersecting plates with finite thickness, we initialize the process within a domain $D \in \mathbb{R}^3$.
At each generative step, a random point inside the cube is uniformly selected as the reference point for a new plate.
If this point lies within any of the previously inserted plate, it is rejected and another random point is drawn.
Otherwise, a random unit vector is sampled from the unit sphere and assigned as the normal vector of the new plate (Fig.~\ref{fig:3D_PD}a).
Each inserted plate has a finite thickness, which decays over time following the power-law $\lambda_t$ of the $t$-th inserted plate is defined as
$\lambda_t = \lambda_0 \cdot t^{-\alpha}$, where $\lambda_0$ is the initial thickness, and $\alpha$ is the thickness decay exponent.
The intersection of these randomly oriented and progressively thinner plates gives rise to a scale-rich three-dimensional structure.

To uncover the characteristics of the 3D plate Scale-Rich model, we first characterize how plate size and network density evolve as functions of the decay exponent $\alpha$ and the initial plate thickness $\lambda_0$.
We begin with the zero-thickness limit ($\lambda_0 = 0$), which isolates the purely geometric fragmentation dynamics, independent of physical thickness constraints.
We then extend the analysis to finite-thickness plates, revealing how thickness, plate size, and connectivity co-evolve to define the architecture of Scale-Rich metamaterials in 3D.

\subsubsection{Expected plate area of the model without thickness ($\lambda_0 = 0$)}
We first consider the case where the structure consists solely of plates with zero thickness ($\lambda_0 = 0$) during generation.

Let $M$ be the volume of the domain.
At each generative step, a plate is inserted, dividing one of the polyhedra into two parts.
Thus, at time $t$, the number of polyhedra is $t+1$. Consequently, the expected volume $\langle V_t\rangle$ of a single polyhedron at time $t$ is
\begin{equation}
    \langle V_t \rangle = \frac{M}{t+1}.
\end{equation}
For large $t$, the relation can be approximated as $\langle V_t\rangle \sim t^{-1}$.

According to Crofton’s formula, the expected area generated by random planar slices through a convex body is proportional to the total surface area of that body~\cite{santalo2004integral}.
In our model, both plate orientations and positions are sampled uniformly at random, and thus, on average, the remaining void region can be approximated by a convex body with similar extent in all spatial directions—effectively a cube.
For such a three-dimensional void, the surface area scales as $6V_t^{2/3}$.
Therefore, the expected area $\langle A_t \rangle$ of a plate introduced at time $t$ follows $\langle A_{t} \rangle  \sim V_t^{\frac{2}{3}}$.
Since $V_t \sim t^{-1}$, the expected area equals $\langle A_t \rangle \sim t^{-2/3}$.

\subsubsection{Expected plate area of the model with thickness ($\lambda_0 \ne 0$)}
The plate added in the $t$th generative step has thickness $\lambda_t = \lambda_0 t^{-\alpha}$.
The density $\rho_t$ of the first $t$ plates equals
\begin{equation}
    \rho_t = \sum_{j=1}^t \lambda_j A_j.
\end{equation}
The mean volume of an empty polygon equals
\begin{equation}
    \langle M_t \rangle
    = \frac{M - \langle \rho_t \rangle}{t+1}
    = \frac{M - \sum\limits_{j=1}^t \lambda_j \langle A_j\rangle}{t+1}.
\end{equation}
Using Crofton's formula again, we obtain
\begin{align}
    \langle A_t \rangle = \langle M_t \rangle^{2/3} &= \left( \frac{M - \langle \rho_t \rangle}{t+1} \right)^{2/3}. \label{eq:3D_mean_A}
\end{align}

\noindent In the same manner as in Eq.~\eqref{eq:2D_left_volume} and Eq.~\eqref{eq:2D_left_volume_approx} for the 2D case, we have
\begin{align}
    M - \langle \rho_t \rangle = \lambda_0 A_0 \sum_{j=t+1}^{\infty} j^{-\alpha-\beta} \approx \frac{\lambda_0 A_0}{\beta + \alpha - 1} (t+1)^{1- \beta - \alpha}.
\end{align}
Using Eq.~\eqref{eq:3D_mean_A} we obtain
\begin{align}
    \langle A_t \rangle \sim t^{-\frac{2}{3}(\alpha + \beta)}.
\end{align}
From the scaling relation derived for the plate area with $\lambda_0=0$, the plate's area is following the decay $\langle A_t \rangle = A_0 t^{-\beta}$, $t^{-\frac{2}{3}(\alpha + \beta)} = t^{-\beta}$, which provides the requirement $\beta = 2\alpha \quad \text{in the case that the entire domain is covered.}$

Further, the initial thickness to cover the entire domain should satisfy
\begin{align}
    M = \lambda_0 A_0 \sum_{j=1}^{\infty} t^{-\alpha - \beta} = \lambda_0 A_0 \sum_{j=1}^{\infty} t^{-3 \alpha} = \lambda_0 A_0 \zeta(3 \alpha).
\end{align}
To cover the entire domain we should take the initial thickness as
\begin{align}
    \lambda_0 = \frac{M}{A_0 \zeta(3\alpha)}.
\end{align}
\noindent This defines the phase space (Fig.~\ref{fig:3D_PD}b)
\begin{equation}
    \rho_\infty =
    \begin{cases}
        M & \lambda_0 \geq (A_0 \zeta(3\alpha)))^{-1} \quad (\text{jammed}), \\[1mm]
        \lambda_0 A_0 \,\zeta(\alpha + \beta) < M & \lambda_0 < (A_0 \zeta(3\alpha)))^{-1} \quad (\text{tunable density}).
    \end{cases}
\end{equation}
In the regime where the system asymptotically covers the entire domain, we have shown that the exponents satisfy $\beta = 2\alpha$.
In contrast, when the coverage remains incomplete in the large-$t$ limit, i.e., $\rho_\infty < M$, we obtain the following relation by combining Eq.~\eqref{eq:3D_mean_A} and Eq.~\eqref{eq:3D_powerlaw}
\begin{align}
    \langle A_t \rangle = \left( \frac{M - \rho_\infty}{t+1} + \frac{\rho_\infty - \langle \rho_t \rangle}{t+1} \right)^{2/3} = \left( \frac{M - \rho_\infty}{t+1} + \frac{A_0 \lambda_0}{\alpha + \beta - 1} t^{-\alpha - \beta} \right)^{2/3}.
\end{align}
For large $t$, the term inside the square root is dominated by the $t^{-1}$ term.
We need $\beta$ to satisfy the relation $t^{-2/3} = t^{\beta}$, leading to $\beta = 2/3 \quad \text{in the case that $\rho_{\infty} < M$.}$

\noindent  Taken together, introducing finite plate thickness causes the system to accumulate volume during growth.
This gives rise to two distinct outcomes: a jammed state, where the remaining voids shrink until no additional plates can be inserted, and a tunable-density state, where enough free volume persists to allow continued growth.
Consequently, the average plate area $\langle A_t \rangle$ exhibits two distinct scaling behaviors depending on the $(\alpha, \lambda_0)$ parameters
\begin{equation}
    \langle A_t \rangle \approx
    \begin{cases}
        t^{-3\alpha} & \text{ when }  \beta = 2\alpha \text{ \quad\, (boundary)},\\[1mm]
        t^{-(\alpha+2/3)} &\text{ when } \beta = 2/3 \quad \text{ (tunable-density regime)}.
    \end{cases}
\end{equation}

\subsubsection{Density}
Let $\rho_{\infty}$ be the asymptotic density of the system, that follows
\begin{align}
    \langle \rho_t \rangle = \rho_{\infty} - \sum_{j=t+1}^{\infty} \langle A_t \rangle \lambda_t = \rho_{\infty} - \frac{A_0 \lambda_0}{\beta + \alpha - 1} (t+1)^{1 - \beta - \alpha}. \label{eq:3D_powerlaw}
\end{align}

\subsubsection{Plate thickness and area distribution.}
The thickness distribution of the plates can be derived in the same way as in Eq.~\eqref{eq:thickness_dis}.
Following the derivation of the thickness distribution, the probability density function of the area distribution is
\begin{align}
    p_A(x) \sim
    \begin{cases}
        x^{-\left(1 + \frac{3}{2}\right)} = x^{-2.5}, & \quad \text{for the tunable-density region}, \\
        x^{-\left(1 + \frac{1}{2\alpha}\right)}, & \quad \text{at the boundary}.
    \end{cases}
\end{align}

\begin{figure}[hptb]
\centering \includegraphics[width=\linewidth]{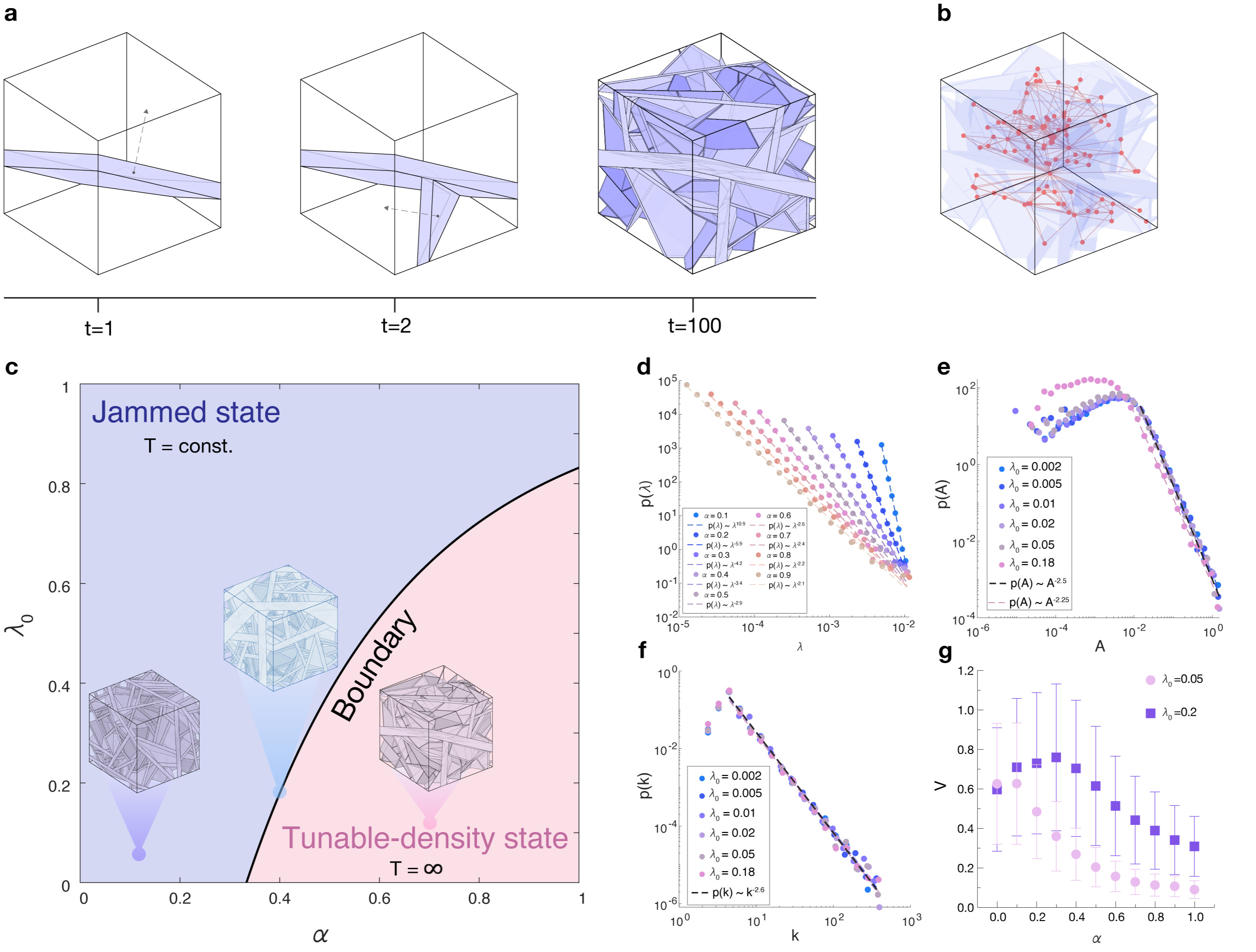}
\caption{\small \textbf{3D plate Scale-Rich metamaterial model.} \textbf{a}. Sequential snapshots ($t=1, 2, 100$) illustrate the stochastic insertion of directed plates into a 3D cube, forming a space-filling network structure over time. \textbf{b}. The resulting network formed by plate-plate intersections, with nodes (red spheres) representing the centroids of polyhedra and edges (black lines) indicating intersections. \textbf{c}. Phase diagram in the $(\alpha, \lambda_0)$ parameter space reveals two distinct regimes: a \emph{jammed state} with a finite number of plates, and a \emph{variable-volume state} with adaptive space utilization and an infinite number of plates. The boundary represents a transition point where the structural constraints balance insertion and saturation, allowing the addition of arbitrarily many planes without reaching geometric frustration.
\textbf{d}. Distribution of thicknesses $\lambda$, following a power-law $p(\lambda) \sim \lambda^{-1-1/\alpha}$.
\textbf{e}. Distribution of plates' areas $A$, showing a universal scaling $p(A) \sim A^{-2.5}$ across variable-volume state space, and $p(A) \sim A^{-1-1/2\alpha}$ across boundary.
\textbf{f}. Degree distribution $p(k)$ of the intersection network across different $\lambda_0$ values exhibits a robust power-law $p(k) \sim k^{-2.6}$, indicating scale-free connectivity.
\textbf{g}. The average volume fractions corresponding to initial thicknesses of $0.05$ and $0.2$, with error bars representing the standard deviation over five simulations. All simulations were conducted with a predefined maximum network size of $2000$ nodes.}
\label{fig:3D_PD}
\end{figure}

\noindent This derivation reveals that the 3D plate SR model produces a power law plate area (Fig.~\ref{fig:3D_PD}e) and thickness (Fig.~\ref{fig:3D_PD}d) distribution, confirming that the SR model generates multiple coexisting scales not only in 2D but also in the 3D plate model.

\subsection{Universal degree scaling from geometric mean-field analysis}
Here, we present a general framework for deriving the degree distribution of the Scale-Rich model across different spatial dimensions, demonstrating that the power-law exponent depends on the dimension.

\noindent \textbf{Derivation of degree distribution for 3D networks with 2D planes.} At time $t$, the system contains $t$ planes, each with thickness $\lambda_0 i^{-\alpha}$ and area $A_0 i^{-\beta}$. The total occupied volume is
\begin{equation}
V(t) = A_0 \lambda_0 \sum_{i=1}^t i^{-\alpha-\beta},
\end{equation}
so the fraction of unoccupied space is $1 - V(t)$.
The average remaining volume accessible to each plate is therefore
\begin{equation}
\langle V \rangle_{\text{left}} = \frac{1 - V(t)}{t+1}.
\end{equation}

\noindent When a new plate is introduced, the probability that it attaches to the $i$-th plane is proportional to the fraction of local voids adjacent to that plate
\begin{equation}
\mathbb{P}\{\text{hit plane $i$}\} = \kappa_3 \frac{k_i(t)\,\langle V \rangle_{\text{left}}}{(t+1)\,\langle V \rangle_{\text{left}}}
\approx \kappa_3 \frac{k_i(t)}{t}, \quad t \to \infty,
\end{equation}
where $k_i(t)$ is the degree of the $i$-th plate, and $\kappa_3$ is a geometric coefficient determined by the typical shape of residual voids, representing the probability that a newly introduced plane will contact plate $i$ once a void adjacent to it has been selected.
In three dimensions, residual voids are typically hexahedral, implying $\kappa_3 = 2/3$. This leads to the recursive growth law
\begin{equation}
    k_i(t+1) - k_i(t) = \kappa_3 \frac{k_i(t)}{t}.
\end{equation}
In the continuum limit, this can be written as the differential equation
\begin{equation}
    \frac{\mathrm{d}k_i}{\mathrm{d}t} = \kappa_3 \frac{k_i(t)}{t}.
\end{equation}
Separating variables gives
\begin{equation}
    \frac{\mathrm{d}k_i}{k_i(t)} = \kappa_3 \frac{\mathrm{d}t}{t},
\end{equation}
which integrates to
\begin{equation}
    k_i(t) = C_1\, t^{\kappa_3} + C_2.
\end{equation}
Imposing the initial condition $k_i(i) \approx 2$ yields the asymptotic scaling form
\begin{equation}
    k_i(t) \sim \left(\frac{t}{i}\right)^{\kappa_3} = \left(\frac{t}{i}\right)^{2/3}, \qquad t \gg i.
\end{equation}

\noindent This result in $P(k)\approx k^{-2.5}$ (Fig.~\ref{fig:3D_PD}f).
Notably, the degree exponent is smaller than $3$, so the second moment diverges.
In other words, the 3D system exhibits an extreme heterogeneity, making it more anisotropic than the 2D SR material, and potentially enabling extreme robustness and efficient stress distribution throughout the structure.

\noindent\textbf{Degree distribution of 2D networks.} An analogous argument applies in two dimensions, where residual voids are quadrilaterals and $\kappa_2 = 1/2$.
Consequently, the degree growth law becomes
\begin{equation}
k_i(t) \sim (t/i)^{1/2}.
\end{equation}

\noindent\textbf{Generalization to $D$ dimensions.} By symmetry, the typical residual void in a $D$-dimensional hypercubic system is bounded by $2(D-1)$ facets out of a total of $2D$. Thus the geometric coefficient is
\begin{equation}
    \kappa_D = \frac{D-1}{D},
\end{equation}
leading to the universal degree scaling
\begin{equation}
    k_i(t) \sim \left(\frac{t}{i}\right)^{\frac{D-1}{D}}.
\end{equation}

\subsection{Computational complexity of the SR model}\label{sec:implementation}
We implemented the Scale-Rich model as follows. For a given point  $\mathbf{x} \in (0,1)^2$ in 2D and direction $\theta$, let $s_{\mathbf{x},\theta}$ denote the ligament obtained by extending a ray from $\mathbf{x}$ in directions $\theta$ and $\pi-\theta$ until it intersects an existing ligament.
Let $P_t$ represent the set of polygons present at time $t$ that remain uncovered, with the initial condition $P_0 = \{[0,1]^2\}$.
Let $S_t$ denote the collection of all segments introduced up to time $t$, and let $Q_t$ denote the set of boundary segments, where each element corresponds to the pair of boundary lines associated with a segment.

\begin{figure}[h!]
    \centering
    \includegraphics[width=1\linewidth]{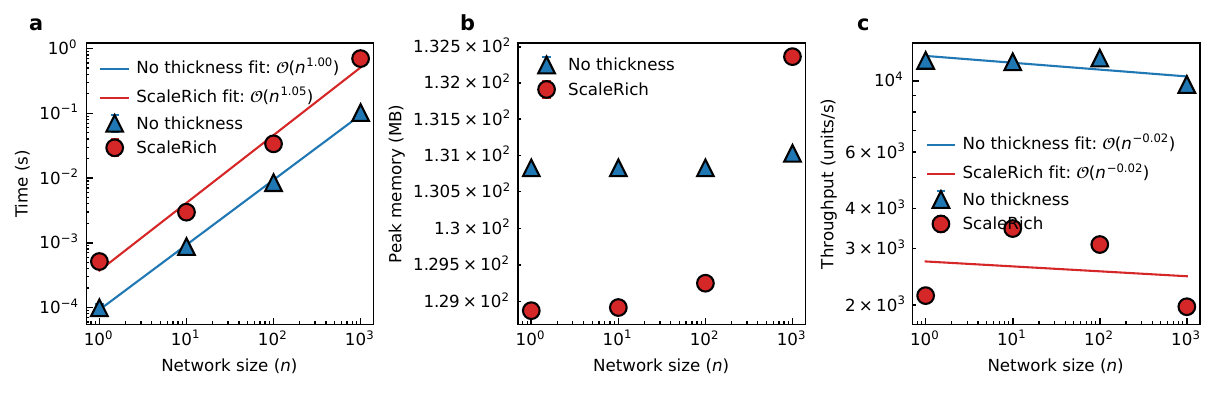}
    \caption{\textbf{Computational complexity scaling with network size.}
    {\bf (a)} Average runtime (mean~$\pm$~s.e.m.) for networks generated without thickness (blue triangles) and for the SR model (red circles) with parameters $\alpha=0.75$, $\lambda_0 = 1/\zeta(2\alpha)$, plotted against network size $n$. Solid lines denote best-fit power-law models, revealing near-linear scaling with exponents of $\mathcal{O}(n^{1.09})$ and $\mathcal{O}(n^{1.11})$.
    {\bf (b)} Peak resident set size (RSS) memory consumption as a function of $n$, showing approximately constant maximum usage.
    {\bf (c)} Effective throughput, defined as the number of network segments generated per second, which decreases gradually with $n$ consistent with the observed runtime scaling.
    Error bars in all panels represent variability across 1000 independent realizations.}
    \label{fig:complexity}
\end{figure}

At each time step $t$:
\begin{enumerate}
    \item Select a domain $D \in P_t$ with probability proportional to its area.
    \item Sample a point $\mathbf{x}$ uniformly at random from $D$.
    \item Sample a direction $\theta$ uniformly at random.
    \item Let $\mathbf{v}$ be the vector started at $\mathbf{x}$ with angle $\theta$. Let $\mathbf{n}$ denote the unit normal vector to $\mathbf{v}$.
    \item Define the offset vectors $s_{-} = s_{\mathbf{x} - \lambda_t/2 \mathbf{n}}$ and $s_{+} = s_{\mathbf{x} + \lambda_t/2 \mathbf{n}}$, where $\lambda_t$ is the thickness of the ligament at times step $t$. Note that $s_-$ and $s_+$ are parallel to $\mathbf{v}$.
    \item Record the initial neighbors of both ligaments by computing their intersections with the boundary of $D$.
    \item Store $s_{-}$ and $s_{+}$ in $Q_t$.
    \item Update $P_t$ and $S_t$ accordingly.
\end{enumerate}
By storing all polygons and previously added segments, the computational cost can be reduced substantially. Rather than computing all intersection points with previously introduced segments, which requires time complexity $O(n^2)$, we restrict computations to intersections with the faces of the selected polygon. This results in a significant efficiency gain. Figure~\ref{fig:complexity} illustrates the computational time (in seconds) as a function of network size.


\clearpage
\section{The mechanical behavior of Scale-Rich metamaterials}

\subsection{Material properties of the constituent resin.}\label{sec:SI-material_properties}
For the nonlinear large-deformation experiments, we printed samples using an acrylate-based polymer resin commercially available as Anycubic Tough Resin 2.0. To determine the material properties of the constituent material, we conducted uniaxial compression experiments on monolithic prismatic specimens of height 40 mm and square cross-section 13.33 mm $\times$ 13.33 mm at a strain rate of $5 \times 10^{-3}$ s$^{-1}$, equivalent to the strain rate used in experiments on the architectures. The elastic modulus across three specimens is found to be $1173 \pm 36$ MPa, and the 0.2\%-offset yield strength is found to be $29.9 \pm 1.4$ MPa.

\begin{figure}[hbpt]
    \centering
    \includegraphics[width=0.6\linewidth]{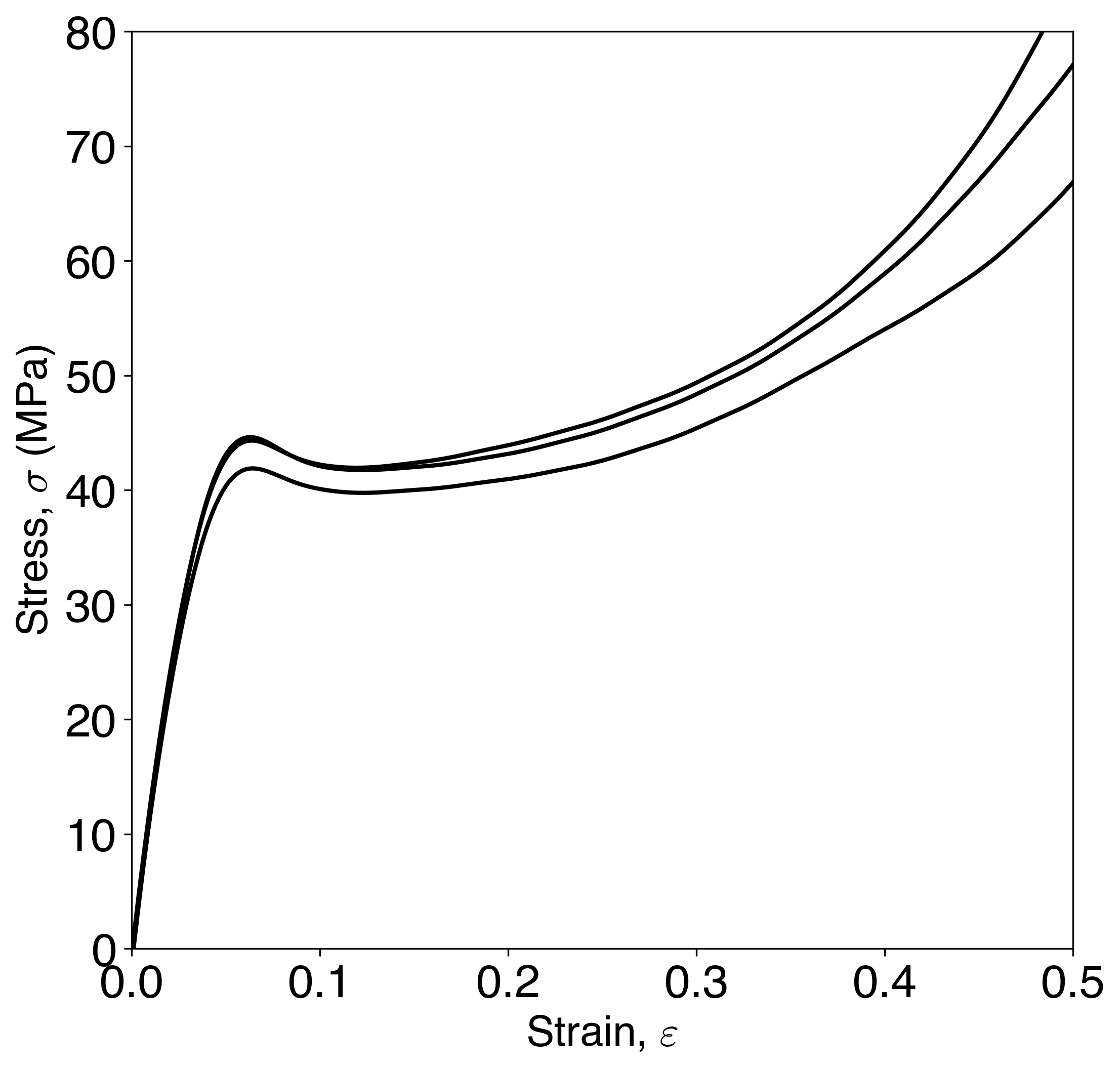}
    \caption{\textbf{Constituent resin properties.} Stress-strain data from uniaxial compression on monolithic resin pillars used in vat photopolymerization (VPP) 3D printing for fabricating the samples.}
    \label{fig:SI-greenResin-compression}
\end{figure}

The material is elastic-plastic, undergoing slight softening after an initial load peak at the onset of inelastic deformation, followed by hardening as the macroscopic strain is increased (Fig.~\ref{fig:SI-greenResin-compression}). This is consistent with the observed behavior of crosslinked polymers printed using vat photopolymerization 3D-printing \cite{saelen2023mechanical}.

As a result of the printing process, the relative density of each as-fabricated sample is higher than the nominal value; in the main text, we use the as-fabricated density values (obtained by measuring the mass and volume of each specimen), except when the word ``nominal'' (or the symbol $\bar{\rho}_\mathrm{nom}$) is explicitly used.

\clearpage
\subsection{Fabrication limit}
\label{sec:SI-fabrication_limit}
\begin{figure}[ht]
    \centering
    \includegraphics[width=0.7\linewidth]{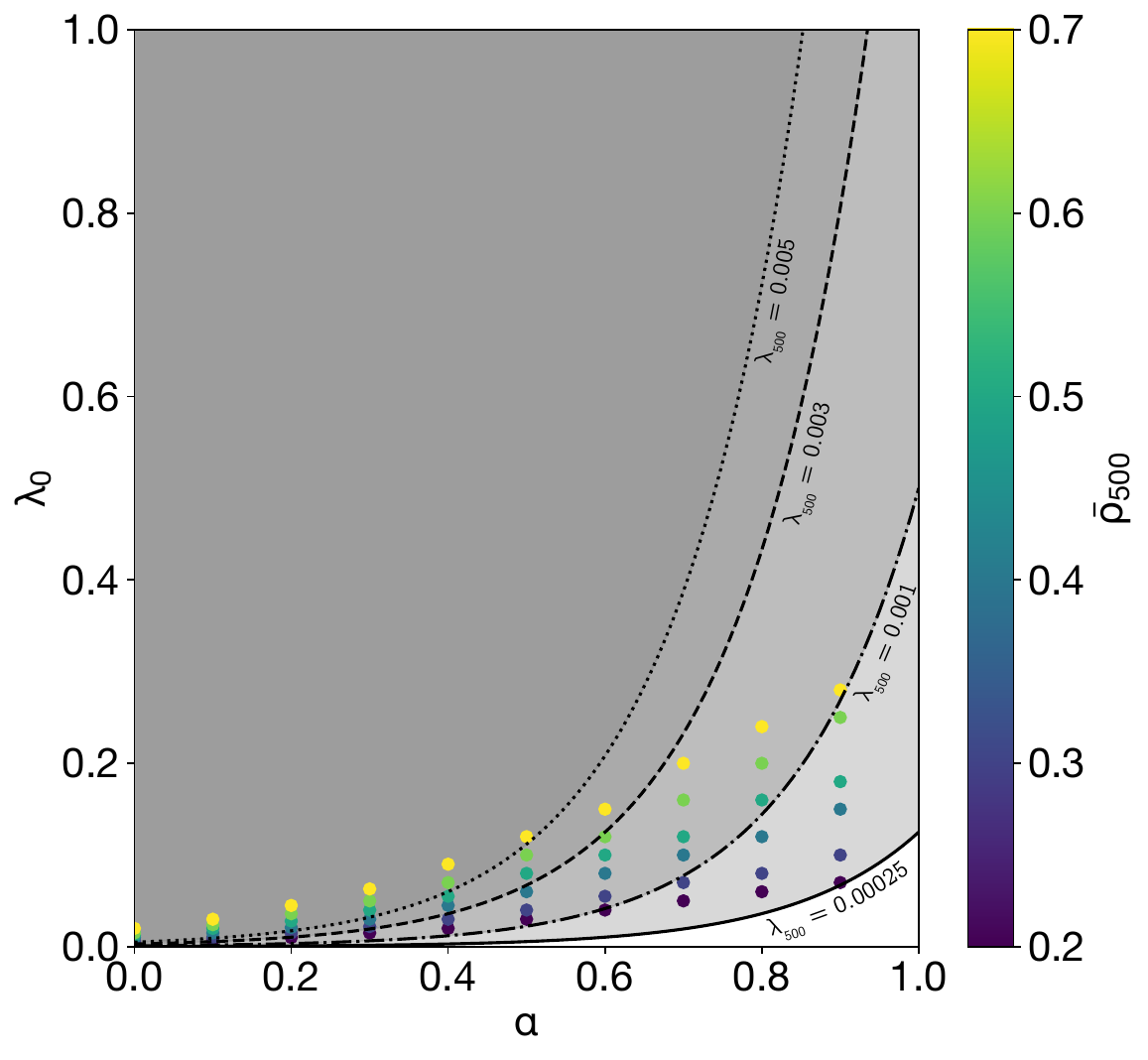}
    \caption{\textbf{Scale-Rich material fabrication.} Accessible printing regimes in the  $(\alpha, \lambda_0)$ parameter space at $T=500$ steps. The grey shaded regions above each line correspond to parameter combinations achievable with printing technologies of minimum resolution
    $\lambda_{500}$, where $\lambda_{500}$ denotes the thinnest printable feature as a fraction of the unit domain size.
    }
    \label{fig:fabrication}
\end{figure}

If the fabrication constraint for an SR material is set by the smallest resolvable feature size, the experimentally accessible domain is smaller than the theoretically feasible one. Considering the $T = 500$ ligament model used in the experiments, we determined the achievable $(\alpha, \lambda_0)$ parameter space for a given fabrication resolution.  Fig.~\ref{fig:fabrication} overlays curves of constant minimum feature size $\lambda_{500}$ over the $(\alpha, \lambda_0)$ phase diagram, together with points associated with constant density (at $T = 500$). For convenience, here we take $L = 1$. As the minimum feature size is decreased, more of the $(\alpha, \lambda_0)$-landscape, particularly lower densities at higher $\alpha$, becomes accessible.

\clearpage
\subsection{Convergence study}\label{sec:SI-convergence-study}
\begin{figure}[hbt!]
    \centering
    \includegraphics[width=0.6\linewidth]{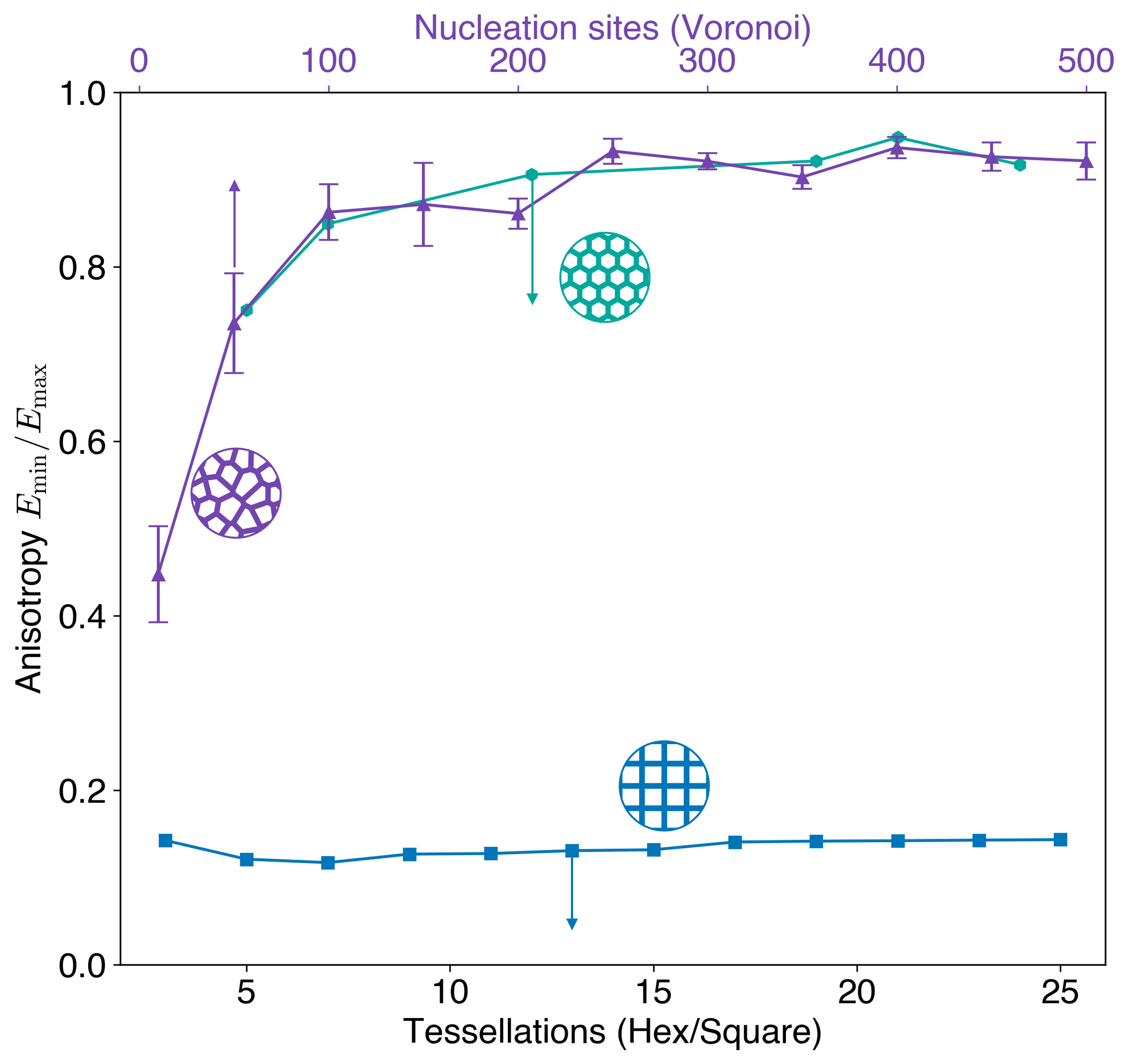}
    \caption{\textbf{Unit cell count convergence study.} Elastic anisotropy versus tessellation count (hex, square lattices) or number of nucleation sites (Voronoi lattice).}
    \label{fig:SI-regular-convergence}
\end{figure}

In the reference systems, a sufficiently large number of unit cells was selected to ensure a representative volume and minimize any boundary effects. To determine the requisite minimum number of unit cells, we conducted a convergence study over the unit cell count (Fig.~\ref{fig:SI-regular-convergence}), in each case keeping the overall relative density constant. In particular, we find that for tessellations of $N \geq 20$ unit cells, the linear elastic mechanical properties of the square and hexagonal lattice converge. For the Voronoi lattice, the critical number of nucleation sites is $N = 250$.
\vspace{-0.1em}
\vspace{-0.1em}

\begin{figure}[hbt!]
    \centering
    \includegraphics[width=.6\linewidth]{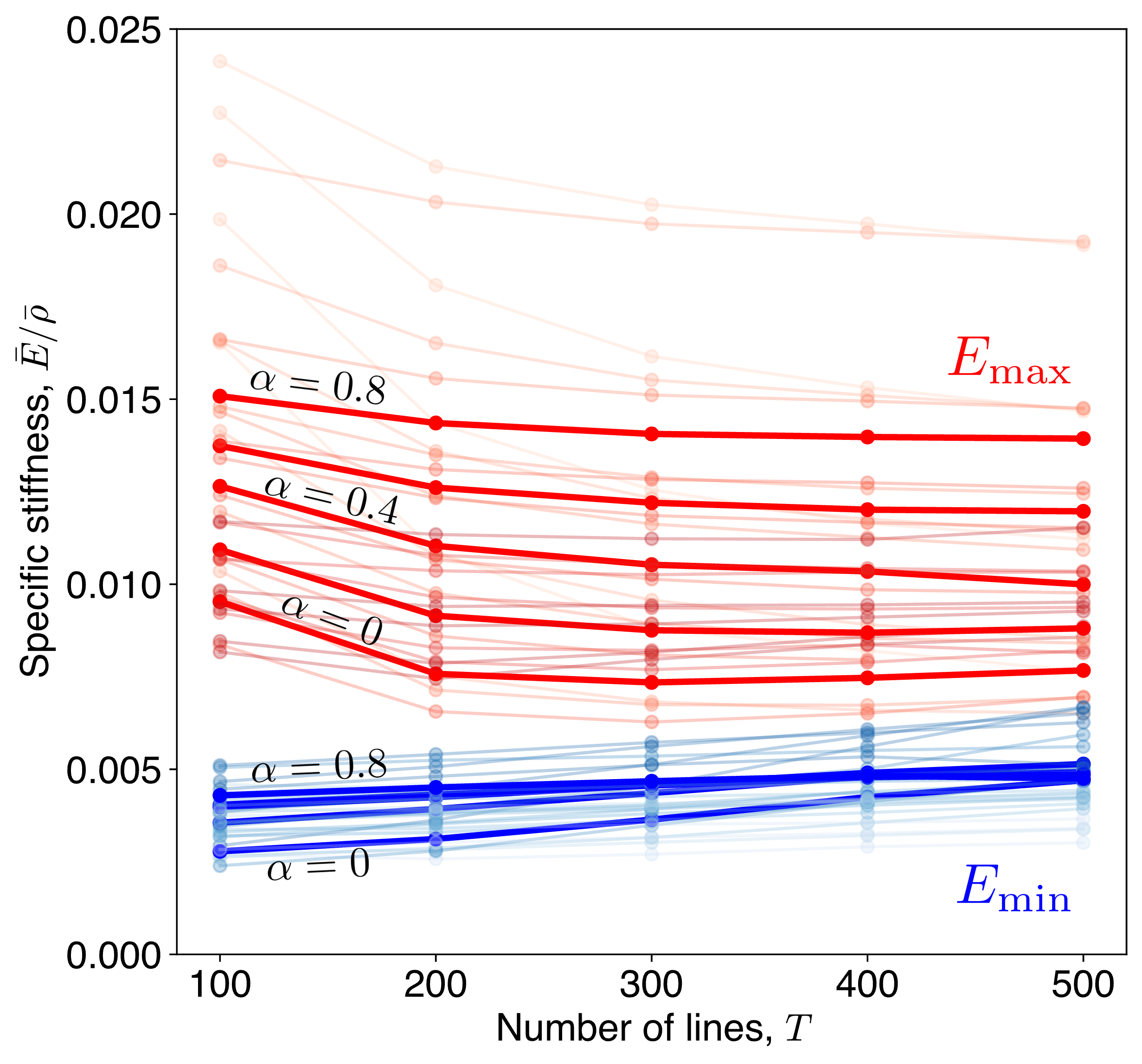}
    \caption{\textbf{Convergence study over $T$, the number of ligaments, in the SR networks.} Lightly-shaded lines connect individual samples of common $\lambda_0$; the dark lines are averages of all samples sharing a common value of $\alpha$.}
    \label{fig:SI-SR-convergence}
\end{figure}

Similarly, we focused on SR networks with $T = 500$ lines, at which point the mechanical properties converge (Fig.~\ref{fig:SI-SR-convergence}). To evaluate this, we measured the maximum and minimum specific stiffness of a given tessellation as the number of lines was varied from $T = 100$ to $T = 500$. In the figure, the lightly shaded lines represent individual samples (each having constant $\lambda_0$), and the darkly shaded lines are an average over all samples of common $\alpha$.

\clearpage
\subsection{Mesh convergence}
\begin{figure}[hbt!]
    \centering
    \includegraphics[width=.8\linewidth]{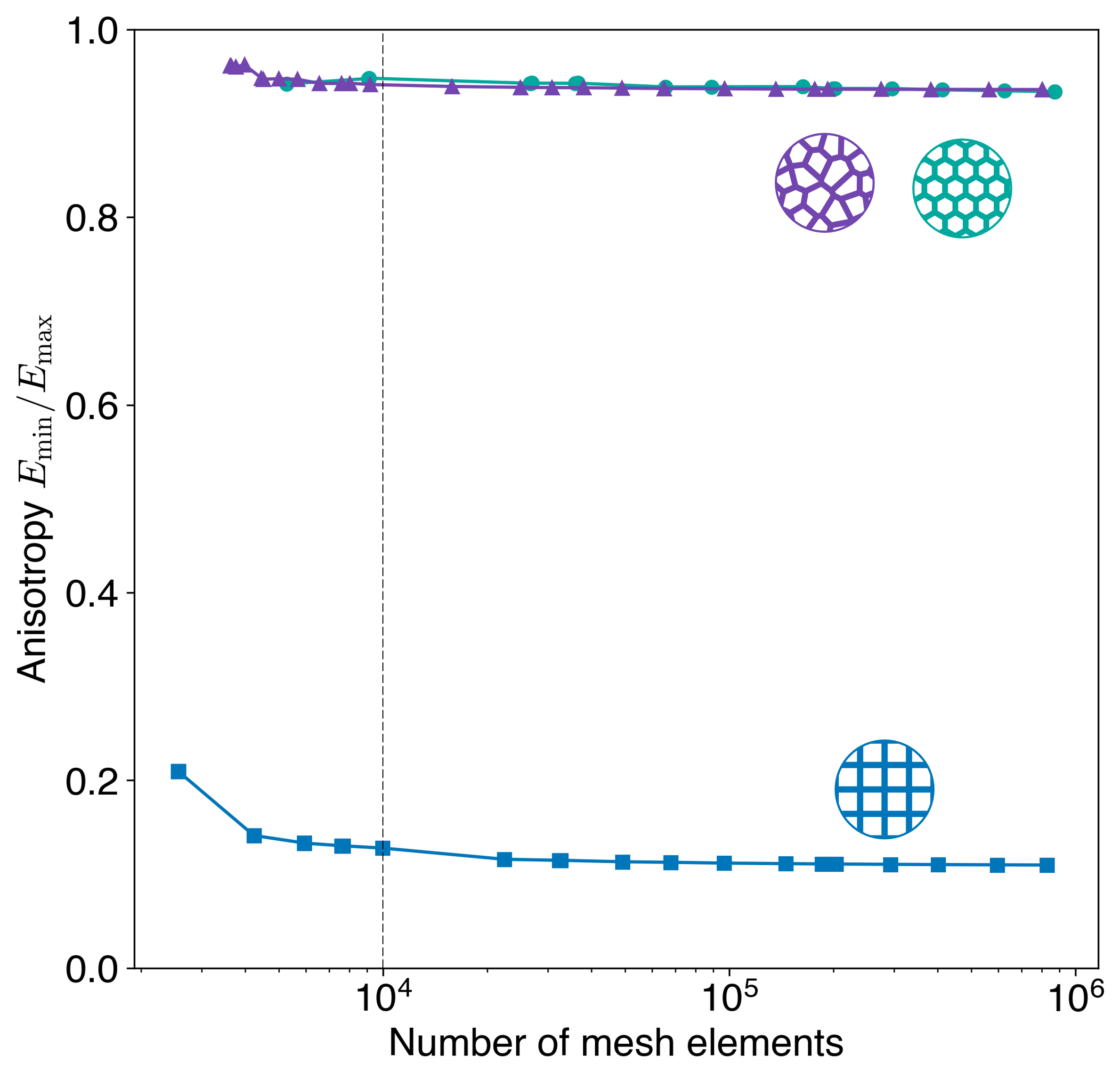}
    \caption{\textbf{Mesh convergence study.} Elastic anisotropy versus number of mesh elements.}
    \label{fig:SI-mesh-convergence}
\end{figure}

To ensure the simulation results accurately yield a convergent set of properties, we conducted a mesh convergence study, i.e., we varied the number of mesh elements used to discretize a particular geometry and repeated the simulation with all other parameters held constant (Fig.~\ref{fig:SI-mesh-convergence}).
We found that the elastic anisotropy---our figure of merit---was roughly constant for $M \geq 10^4$ elements; further discretization produces a negligible change in the computed elastic anisotropy at the expense of computational complexity.

\clearpage
\subsection{Density as a function of $\alpha$, $\lambda_0$ for the specialization $T = 500$}\label{SI:density}

For the finite case of $T = 500$ as studied experimentally, we constructed a phase diagram showing the relative density (fill fraction) in function of the model parameters $\alpha$ and $\lambda_0$, computed by summing the areas of all ligaments and normalizing by the bounding-box area,

$$\bar{\rho} = \frac{1}{L^2}\sum\limits_{i=1}^{T = 500} \ell_i \lambda_i.$$

\begin{figure}[hbt!]
    \centering
    \includegraphics[width=.8\linewidth]{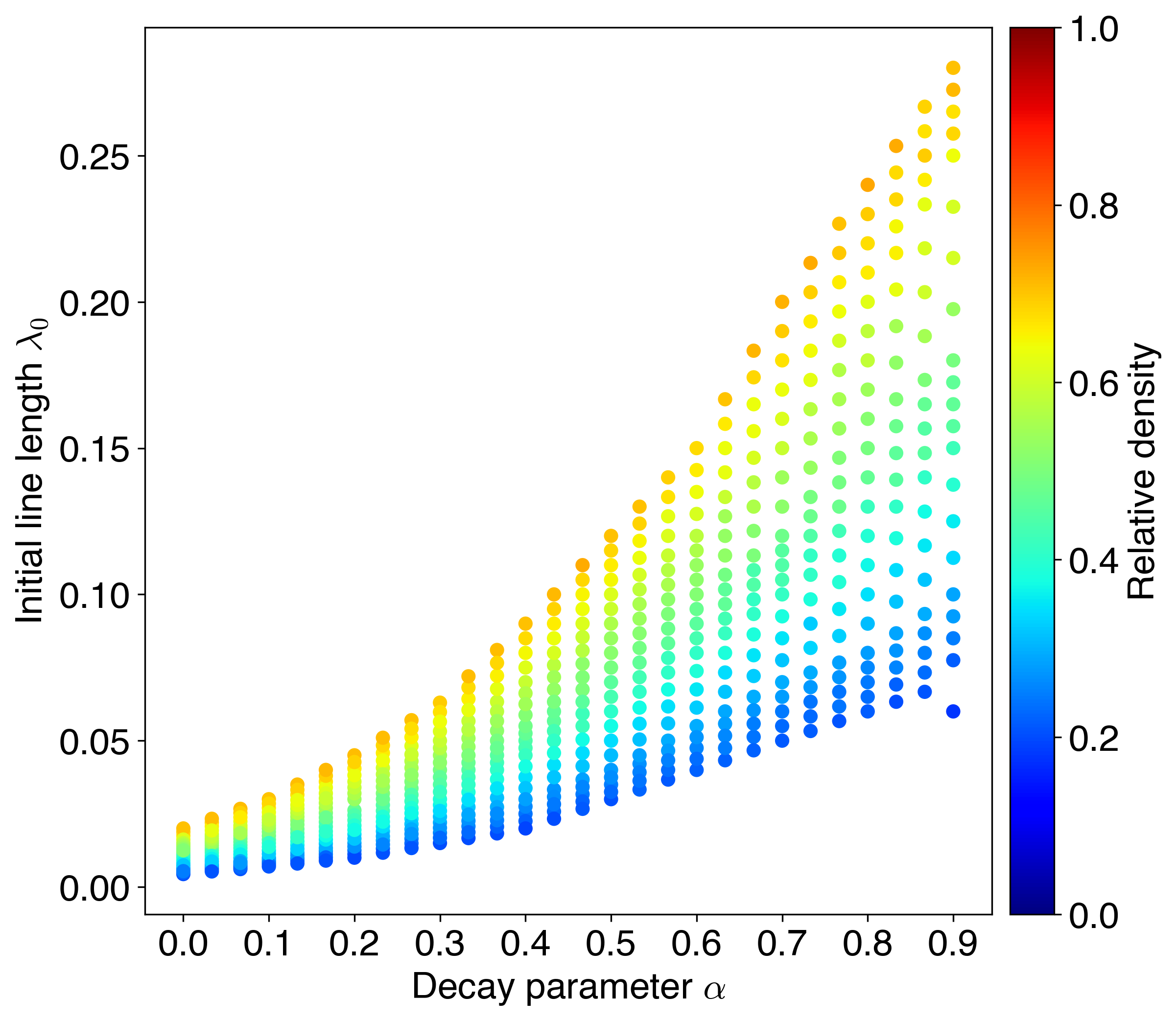}
    \caption{\textbf{Density in the $\alpha, \lambda_0$ phase space.} Relative density on the ($\alpha, \lambda_0$) phase diagram for fixed $T = 500$ lines.}
    \label{fig:SI-network_rd}
\end{figure}

\clearpage
\subsection{Geometric anisotropy}\label{sec:SI-geometric anisotropy}

\begin{figure}[hbt!]
    \centering
    \includegraphics[width=1\linewidth]{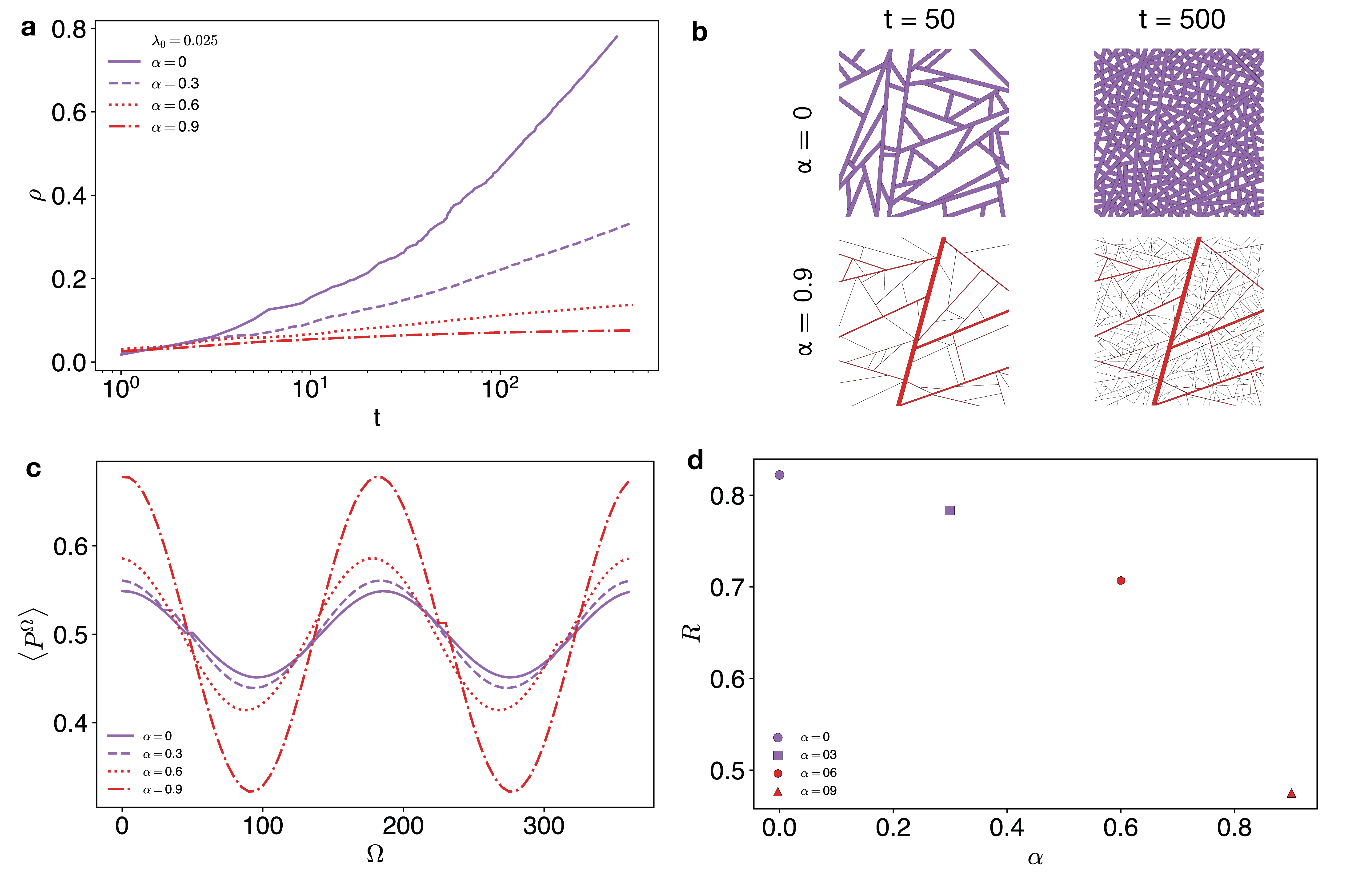}
    \caption{\small \textbf{Density and anisotropy of the SR model.} (a) Density evolution of the SR model. For $\alpha =0$ the system becomes jammed at $T = 500$. (b) Samples at different time steps, $t$. (c) Representative distribution of sample orientation as a function of director angle for a variety of $\alpha$. (d) Geometric anisotropy metric $R = \langle P^{\Omega} \rangle_\mathrm{min}/\langle P^{\Omega} \rangle_\mathrm{max}$ as a function of $\alpha$.}
    \label{fig:density_evo}
\end{figure}

The SR model is characterized by a spontaneous orientation symmetry breaking, making the system inherently anisotropic.
To quantify this, we compute the modified second Legendre polynomial for each ligament $i$, $$P_i^{\Omega} = \ell_i \lambda_i \cos^2{(\Omega-\theta_i)},$$ where $\ell_i \lambda_i$ is the area of ligament $i$ and $\Omega-\theta_i$ is the angle between the ligament’s orientation $\theta_i$ and the orientation of a reference (director) vector $\Omega$ (Fig.~\ref{fig:density_evo}c).
The overall sample orientation is given by $$\langle P^{\Omega} \rangle = \frac{\sum\limits_{i=1}^{T} \ell_i \lambda_i \cos^2{(\Omega-\theta_i)}}{\sum\limits_{i=1}^{T} \ell_i \lambda_i}$$ for $\Omega \in [0\degree,360\degree]$, resulting in the anisotropy parameter $R = \langle P^{\Omega} \rangle_\mathrm{min}/\langle P^{\Omega} \rangle_\mathrm{max}$, where $\langle P^{\Omega} \rangle_\mathrm{min}$ and $\langle P^{\Omega} \rangle_\mathrm{max}$  represent the minimum and maximum sample orientation, respectively (Fig.~\ref{fig:density_evo}d).
If the lines preferentially align with a specific direction, the system is considered anisotropic ($R \ll 1$); conversely, if the lines are uniformly distributed without favoring any direction, the system is isotropic ($R \approx 1$).

An equivalent measure of geometric anisotropy can be computed the second-order orientation tensor $\mathbf{M}$, constructed for a given geometry as
$$\mathbf{M} \equiv \sum_{t = 1}^T A_t \, \mathbf{d}_t \otimes \mathbf{d}_t,$$ where $A_t$ is the area of the $t$-th ligament and $\mathbf{d}_t$ is its unit direction vector (i.e., a unit vector in the direction $\Omega_i$). The tensor $\mathbf{M}$ is symmetric and thus admits a spectral decomposition $$\mathbf{M} = \sum_{i=1}^2 \,\mu_i \,\mathbf{m}_i \otimes \mathbf{m}_i$$ with eigenvalues $\mu_1 \leq \mu_2$. The ratio $\eta_G \equiv \mu_1 / \mu_2$, $0 \leq \eta_G \leq 1$, is then a measure of geometric anisotropy in the sample.

A strong correlation ($R^2 \approx 0.84$) with the elastic anisotropy $\eta_E \equiv E_\mathrm{min} / E_\mathrm{max}$ is found by multiplying the ratio $\mu_1 / \mu_2$ by the relative density of the sample, $\bar{\rho}$, raised to the power 0.643, which is the other major nondimensional parameter responsible for controlling the system's behavior. (We find that the result is independent of the decay rate $\alpha$.) This suggests that the heterogeneity of ligament directions and areas is a physical source of elastic anisotropy.

\begin{figure}[hbt!]
    \centering
    \includegraphics[width=.6\linewidth]{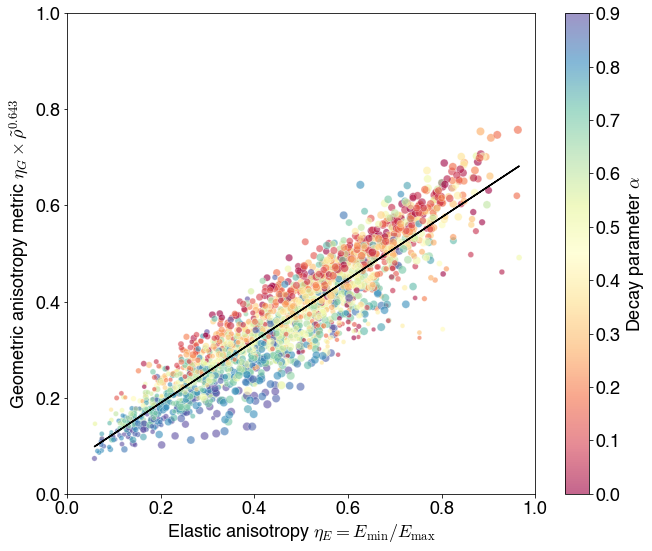}
    \caption{Geometric anisotropy, calculated from beam orientations and areas, versus elastic anisotropy, calculated from FE simulations. The black line represents the best fit ($R^2 = 0.84$).}
    \label{fig:anisotropy-elastic-geometric}
\end{figure}

\begin{figure}[hbt!]
    \centering
    \includegraphics[width=.6\linewidth]{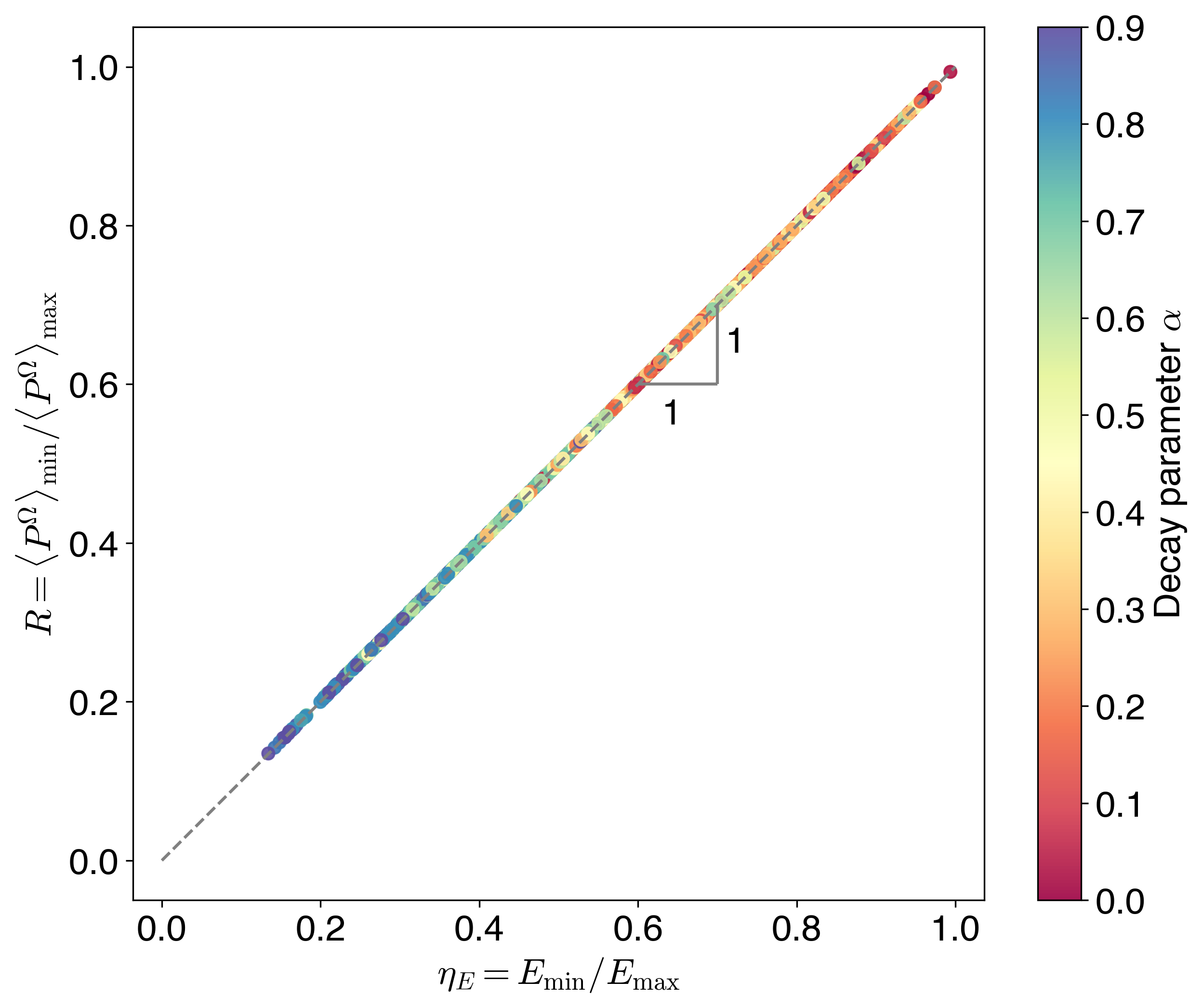}
    \caption{Two equivalent measures of geometric anisotropy: $R$, using the modified second Legendre polynomial, and $\eta_E$, using the material's orientation tensor. The line $R = \eta_E$ is plotted for reference.}
    \label{fig:anisotropy-elastic-geometric}
\end{figure}

\clearpage
\subsection{Experimentally measuring the directional stiffness}\label{sec:SI-experiment-stiffness}

To experimentally measure the directional stiffness of SR architectures, we first generated a master structure within a large square domain of side length $\sqrt2 L$. From this structure, we extracted samples using an oriented square stencil of side length $L$. We extracted samples in $30\degree$ increments from $0\degree$ to $150\degree$; the structure has $180\degree$ symmetry which we exploited in plotting the results.

We printed the extracted samples using an acrylate photopolymer resin (Formlabs Clear v4), which was chosen over the Anycubic resin used in the nonlinear experiments because of its increased stiffness; consequently, the linear-elastic regime was more clearly evident. We normalized the stiffness of the SR samples against the modulus of the monolithic polymer material, measured to be $E_s = 2130 \pm 4$ MPa, to obtain experimental values of $E_d/E_s$.

\begin{figure}[hbt!]
    \centering
    \includegraphics[width=0.7\linewidth]{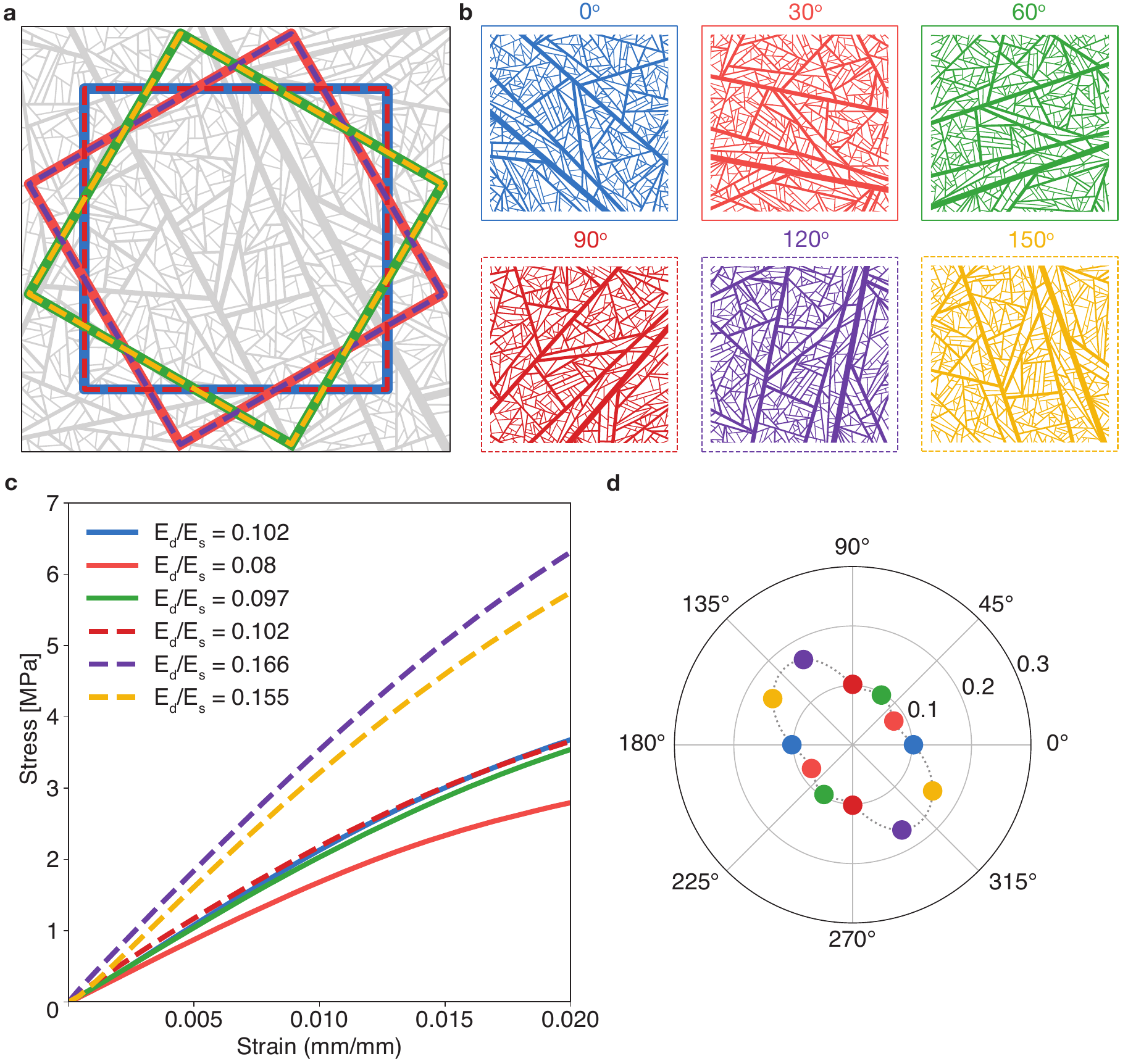}
    \caption{ \textbf{Methodology to experimentally measure the directional stiffness.} \textbf{(a)} Generated a large domain of size $(\sqrt{2}L)^2$, from which an oriented subdomain of size $L^2$ was extracted, rotated in increments of 30$\degree$ from $0\degree$ to $150\degree$.
    \textbf{(b)} Extracted samples at different orientations.
    \textbf{(c)} Stress-strain curves from quasi-static uniaxial compression.
    \textbf{(d)} Normalized directional stiffness, $E_d/E_s$, with mirrored data points (assuming two-fold symmetry) and interpolation for visualization.
}
    \label{fig:experiment-stiffness}
\end{figure}

\clearpage
\subsection{Error bars for the nonlinear stress-strain plots of reference geometries}\label{sec:SI-nonlinear-errorbars}
The figure below reproduces Fig.~\ref{fig:nonlinear}a for the reference geometries with error regions (each consisting of $\pm 1$ standard deviation) around the mean, which is represented with a thick solid line.

\begin{figure}[hbt!]
    \centering
    \includegraphics[width=.6\linewidth]{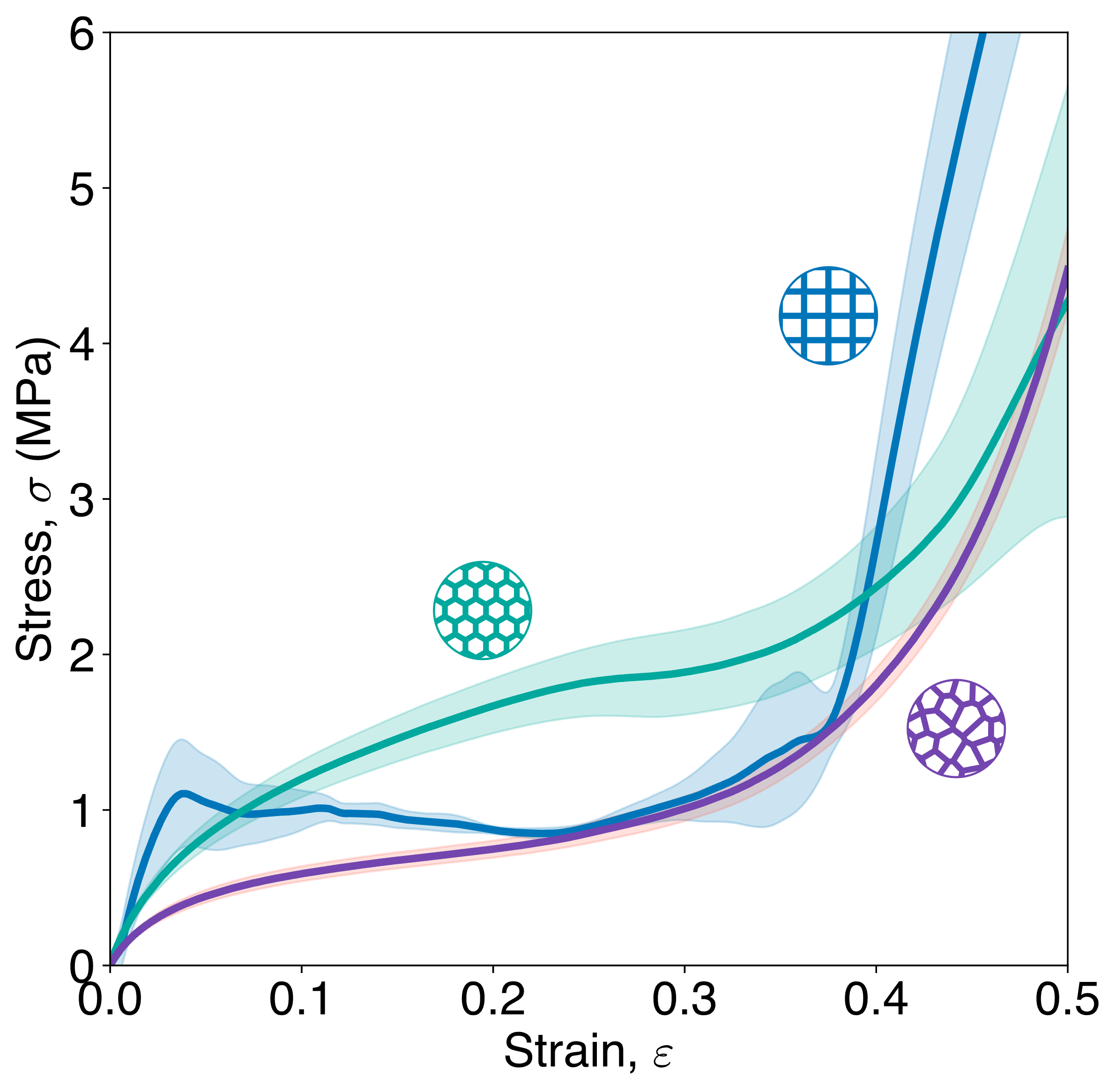}
    \caption{\textbf{Repeated nonlinear experiments on the reference geometries.} Nonlinear stress-strain curves of the reference geometries with error regions from experiments on $N = 3$ replicates.}
    \label{fig:errorbars}
\end{figure}

\clearpage
\subsection{Localization metric $\mathcal{L}$}\label{sec:curly_L}
Given the set of Jacobians $$\{J_1, J_2, \dots, J_N\}, \quad J_i \equiv A_{i} / A_{i,0},$$ where $A_{i,0}$ is the initial size of each cell (i.e., pore) in the microstructure and $A_i$ is its current size, the localization metric $\mathcal{L}$ is defined in the main text as the scalar quantity $$\mathcal{L} \equiv \frac{\var{(\mu_1, \mu_2, \dots, \mu_N)}}{\var{(J_1, J_2, \dots,\ J_N)}}.$$ The quantity $\mu_i$ is the local average Jacobian associated with each cell $i$,
$$\mu_i \equiv \frac{1}{M} \sum_{j \in \mathcal{B}_r(i)} J_j.$$
This mean is computed over the neighborhood $\mathcal{B}_r(i)$ (containing $M$ cells), which for each cell $i$ is defined to consist of all cells whose initial centroids are located within a radius of $r = 0.1L$ from the centroid of $i$.  Here $L$ is the size of the entire sample. The use of \textit{reference-configuration} (initial) centroid positions ensures that the definition of all neighborhoods remains fixed despite the progression of deformation.

\begin{figure}[h]
    \centering
    \includegraphics[width=.9\linewidth]{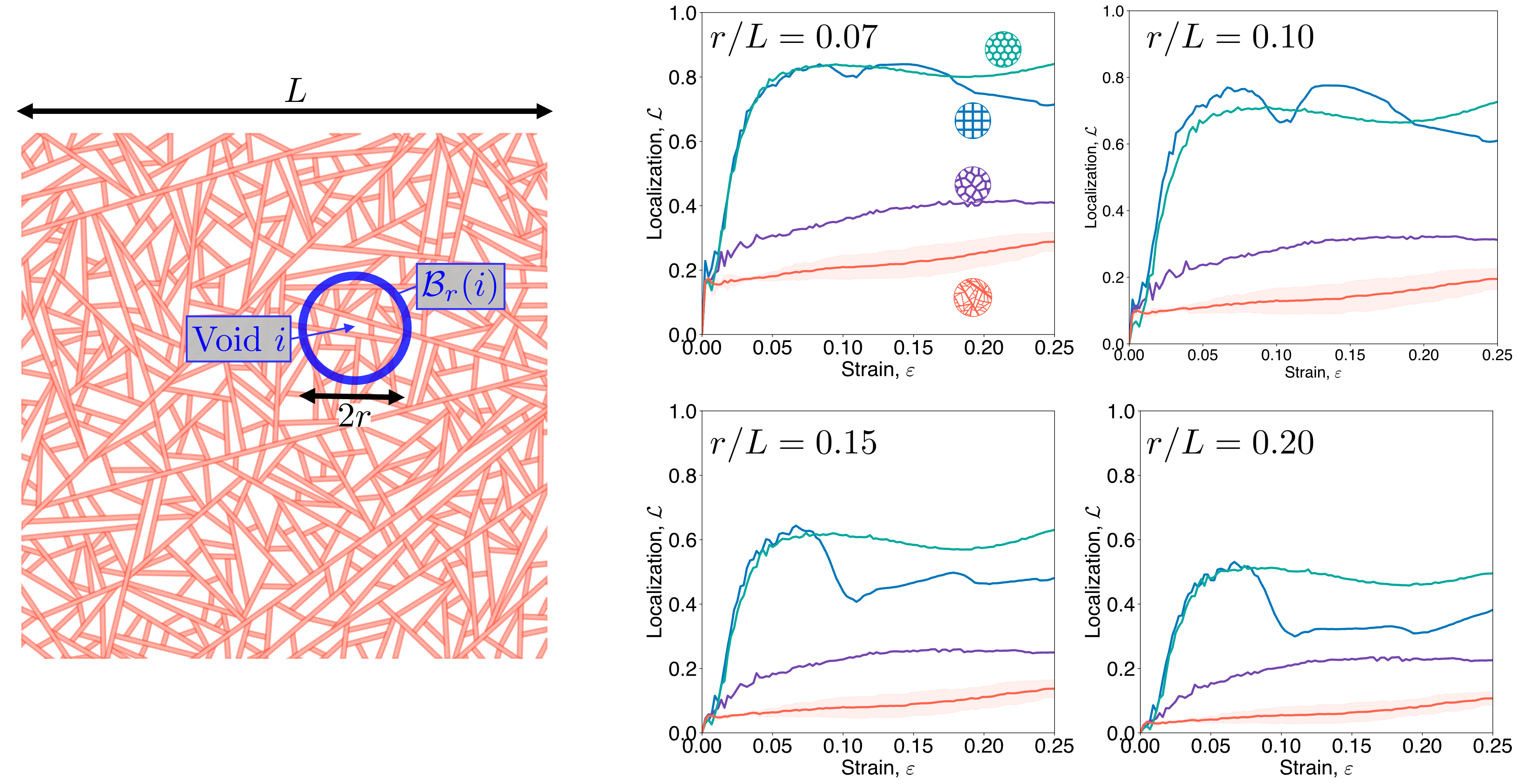}
    \caption{\textbf{Effect of neighborhood size on $\mathcal{L}$.} (left) Illustration of the dimensionless radius $r/L$ used to compute the neighborhood $\mathcal{B}_M(i)$ of a given void $i$. (right) Effect of $r/L$ on the localization across geometries.}
    \label{fig:SI-rOverL}
\end{figure}

The pitch (spacing) of unit cells in the square lattice sets a minimum physical value of $r/L = 0.07$. In general, we see that the overall trends are independent of the value of $r/L$ (Fig.~\ref{fig:SI-rOverL}); importantly, the SR samples display the least amount of localization for all values of $r/L$. For analysis, we choose a physically reasonable domain $r/L = 0.1$.

\begin{figure}[h]
    \centering
    \includegraphics[width=.9\linewidth]{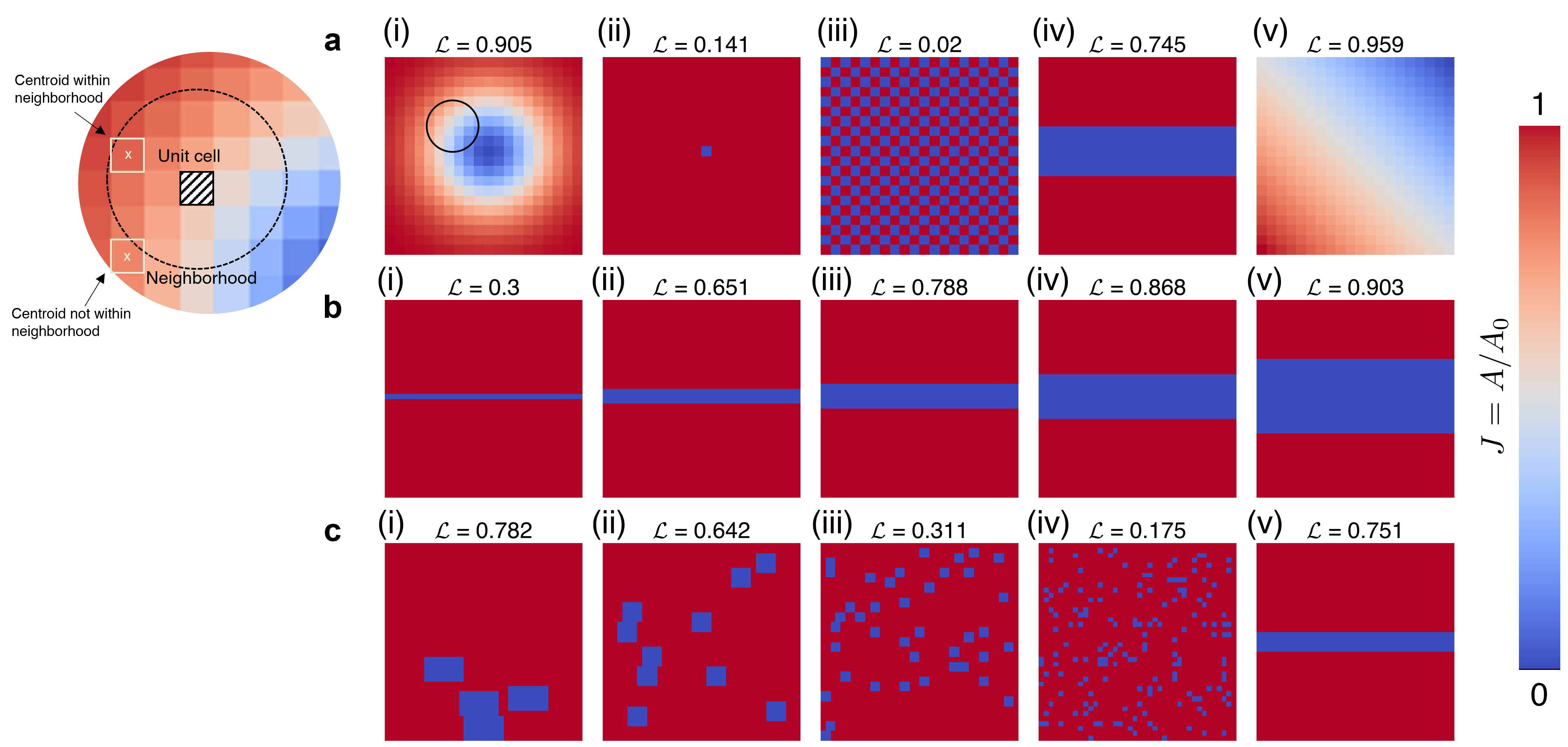}
    \caption{\textbf{Illustrative examples of localization.} Illustrations of the localization metric $\mathcal{L}$ applied to hypothetical scenarios on a $20 \times 20$ grid.}
    \label{fig:SI-L}
\end{figure}

To further demonstrate how this metric quantifies localization, we present several additional examples of $\mathcal{L}$ measurements for hypothetical illustrative scenarios, Fig.~\ref{fig:SI-L}. In each case the domain consists of a $20 \times 20$ array of ``unit cells'' with each unit cell assigned a value $0 \leq J \leq 1$. Panel (a) contrasts strongly localized cases (e.g., (i) and (v), where damage is widespread) with the near-ideal, ``checkerboarded'' case (iii), which yields a correspondingly low $\mathcal{L}$. In particular, case (iii) is designed to illustrate that $\mathcal{L}$ remains low despite the presence of damage in the material --- the homogeneous \textit{distribution} of damage in this scenario results in delocalization, which is captured by our metric. Panel (b) shows that increasing the width of a compaction-band-like feature raises $\mathcal{L}$, reflecting progressive localization as seen in experiments. Panel (c) shows five cases in which all cells take \textit{only} either $J=0$ or $J=1$ (i.e., with no intermediate values) under the constraint of identical total damage (i.e., the same total number of blue cells). Conventional localization metrics based \textit{solely} on $\var(J_1,\dots,J_N)$ or $\mathrm{std}(J_1,\dots,J_N)$ would assign the same $\mathcal{L}$ to all cases, despite their clear differences in spatial distribution. By incorporating local average quantities, i.e., the $\mu_i$, our metric $\mathcal{L}$ distinguishes these patterns.

\clearpage
\subsection{Nonlinear behavior for $\alpha > 0$}\label{sec:high-alpha-behavior}
\begin{figure}[hbt!]
    \centering
    \includegraphics[width=1\linewidth]{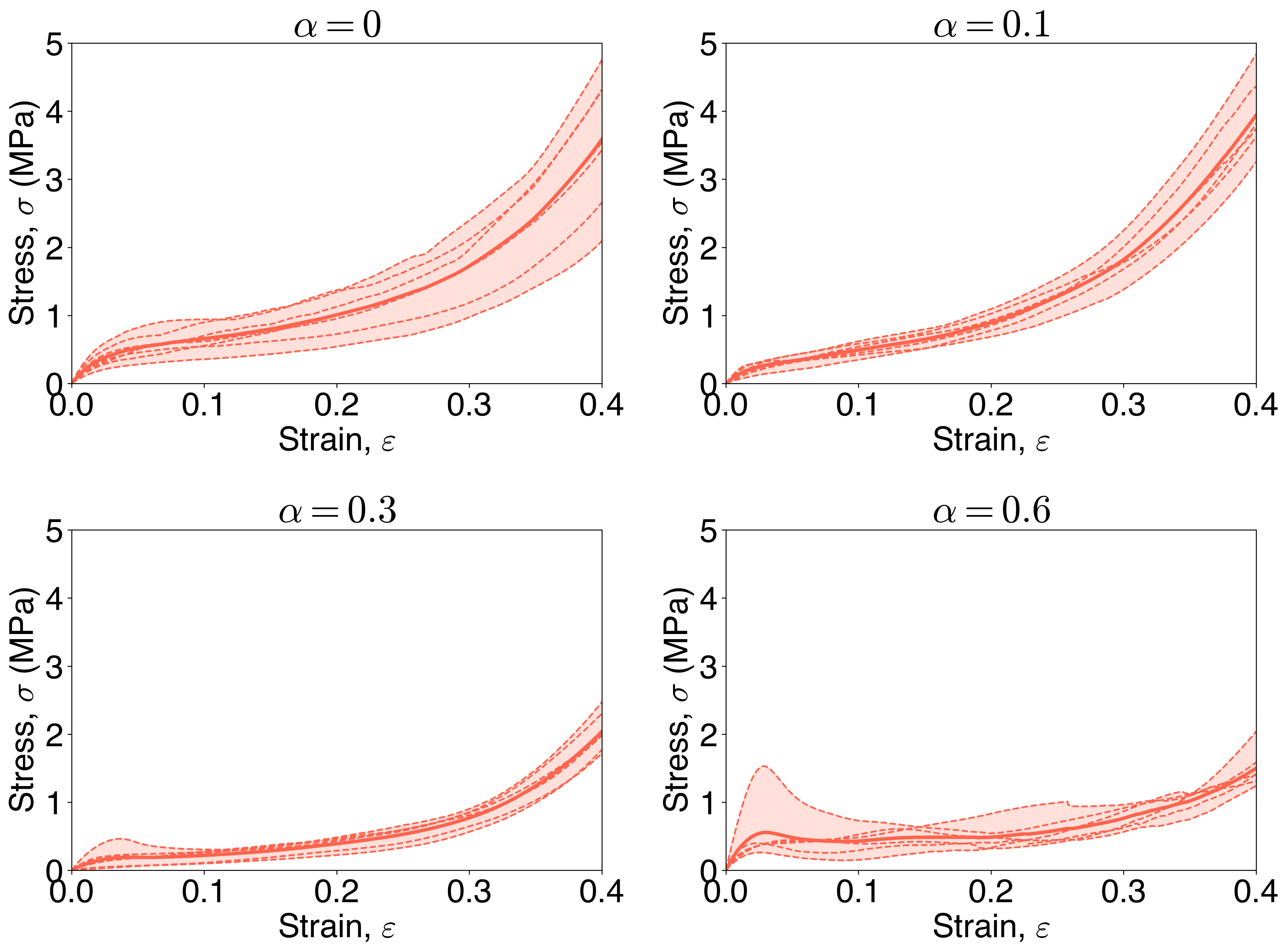}
    \caption{\textbf{Stress-strain curves of SR samples with higher values of $\alpha$.}}
    \label{fig:SI-high-alpha-stress-strain}
\end{figure}

To understand the nonlinear behavior of the SR geometries at higher values of $\alpha$, we conducted the same uniaxial compression experiments (at a constant strain rate of $\dot{\varepsilon} = 5 \times 10^{-3}$ s$^{-1}$). Fig.~\ref{fig:SI-high-alpha-stress-strain} shows the results for six specimens for $\alpha = 0.1$, $\alpha = 0.3$, $\alpha = 0.6$, all at 42\% nominal relative density, together with the $\alpha = 0$ specimens for the main text, repeated here for comparison. Each dashed curve represents one individual specimen, the shaded region represents the attainable regime of behavior across all specimens, and the solid line is the mean behavior.

In general, as $\alpha$ increases, the influence of the first few ligaments dominates. An important consequence of this fact is visible in the plots for $\alpha \geq 0.3$: when the first beam is oriented perpendicular to the direction of compression, the sample behavior is dominated by buckling. In contrast, for the same values of initial parameters ($\alpha$, $\lambda_0$), samples with largest beams oriented parallel to the compression direction experience monotonic (stable) stress-strain behavior. For extreme values, namely as $\alpha \to 1$, the influence of the largest line is dominant such that the finite specimen is not a representative volume element. For this reason, we restrict our discussion of behavior of the SR materials in the main text to $\alpha = 0$, when these aspects are not present.

\vspace{-0.3em}

\clearpage
\subsection{Bloch analysis and wave velocity calculations}\label{sec:bloch_waveSpeed}
\begin{figure}[hbt!]
    \centering
    \includegraphics[width=\linewidth]{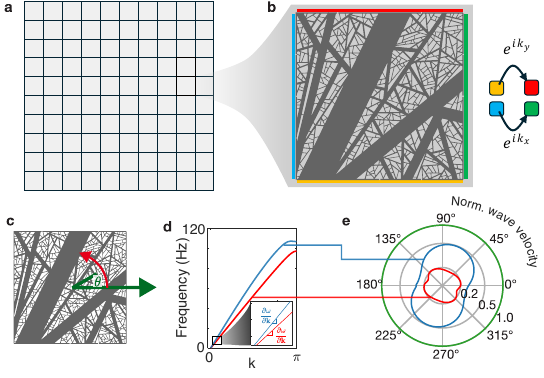}
    \caption{\textbf{Bloch analysis of SR metamaterials.} {\bf (a)} Tiling approach to SR materials, treating each SR sample. {\bf (b)} Closeup of an SR unit cell, showing thin, disconnected borders are added to allow for periodic relations between extrema for Bloch analysis. The left (bottom) and right (top) edges are related to each other by a phase factor $k_x$ ($k_y$) equivalent to the wavenumber in that direction. {\bf (c)} The dispersion relation was calculated along all directions, and {\bf (d)} the slope of the quasi-shear (red) and quasi-longitudinal (blue) curves, in the long-wavelength limit, was taken as the wave velocity for those modes. {\bf (e)} We take these results along all directions and conveniently represent them in a polar plot of directional velocity, revealing the anisotropic nature of SR metamaterials.}
    \label{fig:SI-BAWV}
\end{figure}
\begin{figure}[hbt!]
    \centering
    \includegraphics[width=\linewidth]{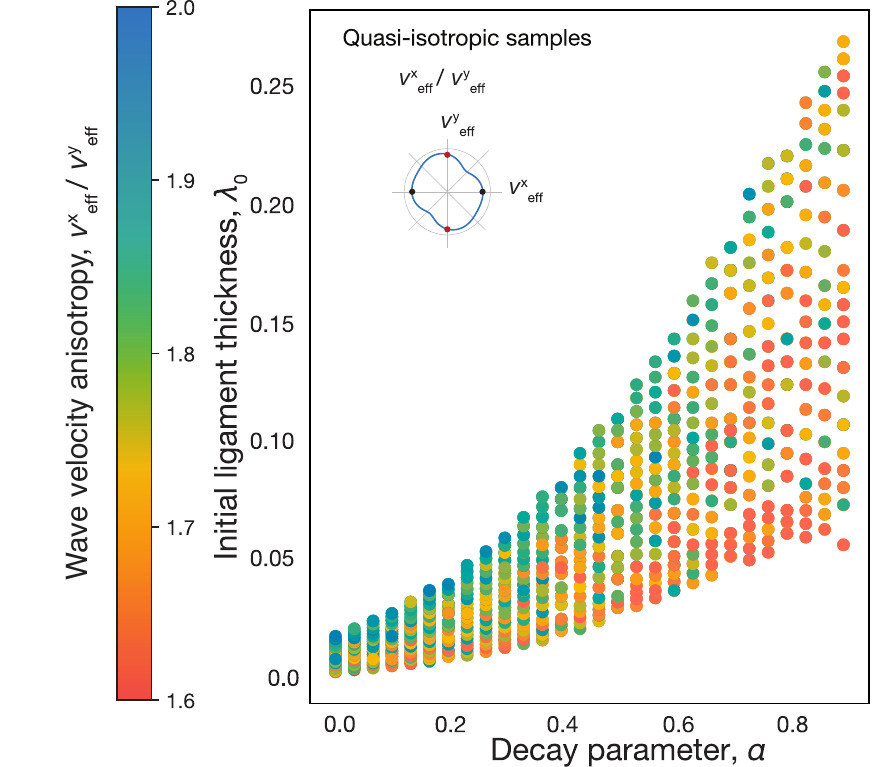}
    \caption{\textbf{Wave anisotropy in the $\alpha, \lambda_0$ phase space.} $\alpha$-$\lambda$ phase diagram showing wave velocity-based anisotropy of SR metamaterials, where the markers are colored by the ratio of minimum to maximum directional velocity ($\frac{v_{min}}{v_{max}}$). Note that, when compared to the stiffness anisotropy phase diagram for the quasi-static case, this figure appear less anisotropic (i.e., marker colors skewed to the color red); this is partly because of the added borders to help enforce periodicity, which while thin and disconnected, have an inevitable effect on the elastodynamic properties of the structure.}
    \label{fig:SI-WVphaseDiag}
\end{figure}

\begin{figure}[hbt!]
    \centering
    \includegraphics[width=\linewidth]{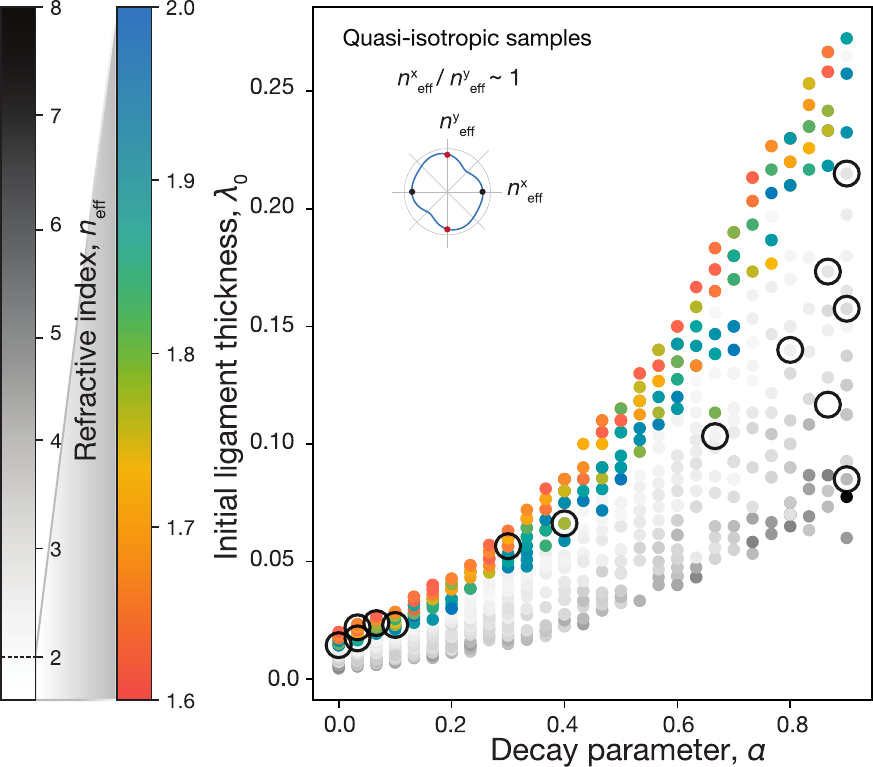}
    \caption{\textbf{Refractive index in the $\alpha, \lambda_0$ phase space.} ($\alpha$-$\lambda_0)$ phase diagram showing the span of refractive indices achieved by SR metamaterials, where the marker colors match that of Fig.~\ref{fig:waveProp}.}
    \label{fig:SI-refractiveInd}
\end{figure}

SR metamaterials can also be interpreted as cells which can be tiled to generate a larger scale metamaterial (Fig.~\ref{fig:SI-BAWV}a). Because the SR framework does not enforce periodicity along edges at extrema, we add a border around each sample which allows us to interface a single tile with adjacent units (Fig.~\ref{fig:SI-BAWV}b), introducing a feature useful in agnostic tiling---for now we consider the case where the tessellation is homogeneous. The borders are thin ($0.5\%$ of a sample's characteristic length) to avoid altering the mechanical behavior of the stochastically-generated interior, and are disconnected into four portions to avoid the creation of lines of self-stress which may bear load or act as waveguides, akin to the case of the square lattice.

To understand the vibration response of randomly generated SR metamaterials, we apply Bloch-periodic boundary conditions (Fig.~\ref{fig:SI-BAWV}b) and extract allowed vibrational frequencies and propagation modes of a wave traveling through a lattice composed of an infinite tessellation. This tessellation is composed of SR unit cells repeated along Bravais lattice vectors $\mathbf{\ell}_1=[1,0]$ and $\mathbf{\ell}_2=[0,1]$. In the finite element framework, this amounts to discretizing the geometry into $N$ nodes with $d \times N$ degrees of freedom $\mathbf{u}$---where $d=2$ for two-dimensional geometries---and assembling the corresponding stiffness matrix \textbf{K} and mass matrix \textbf{M}. Under tiling, repeated cells can be indexed by $\mathbf{m}=[m_1,m_2]$ so that a node $p_0$ in cell $\mathbf{m}=[0,0]$ can be mapped to its corresponding location at an arbitrary cell by $p(\mathbf{m})=p_0+m_1 \mathbf{\ell}_1 + m_2 \mathbf{\ell}_2$. Under Bloch-periodic boundary conditions the displacement of nodes at extreme ends of the unit cell along the tessellation directions $\mathbf{\ell}_i$ are related to each other by $\mathbf{u}_{(+)}=\mathbf{u}_{(-)} e^{\mathbf{k}}$, where $\mathbf{k}=[k_1,k_2]$ is the wavevector defining the wave front normal of an incident plane wave. This results in the introduction of a wavevector dependence into the equation of motion $\mathbf{K}_r(\mathbf{k}) - \omega \mathbf{M}_r(\mathbf{k}))\mathbf{u}_r=0$, where the $r$ subscripts denotes Bloch-periodic reduction. Solving this eigenvalue problem yields eigenvalues $\omega$ and eigenvectors $\mathbf{u}$ which can be physically interpreted as the allowed vibrational frequencies of the system and corresponding deformation modeshape, for a chosen $\mathbf{k}$.

Given the breadth of anisotropy of SR samples, we explore the direction-dependent wave velocities by solving the eigenvalue problem for $\mathbf{k}=0.1(\cos(\theta),\sin(\theta))$ at $\theta=[0,2\pi)$, which amounts to solving the long-wavelength response of the system to waves in every direction (Fig.~\ref{fig:SI-BAWV}c).  We repeat this for a large number of samples stochastically generated across $\alpha=[0,0.9]$. At the continuum limit the phase velocity $v_p=\omega/||\mathbf{k}||$ and group velocity $v_g=\partial \omega/\partial \mathbf{k}$ are nearly identical as the dispersion relation approaches linearity (Fig.~\ref{fig:SI-BAWV}d-e). Generally, the first and second modes are recognized as shear and longitudinal waves, where the direction of displacement polarization is transverse and parallel to the wavefront, respectively. However, in the case of SR samples we see a high degree of anisotropy commensurate to that found in the statics case, which results in a mixing of these two modes; we can refer to them instead as quasi-shear and quasi-longitudinal, respectively. However, more interestingly, when we focus on the maximum and minimum quasi-longitudinal wave velocities of the samples, we see a large breadth of velocities can be achieved with this framework by simply tuning $\alpha$ and $\lambda$. The range of this anisotropy can be appreciated by studying the phase diagram in Fig.~\ref{fig:SI-WVphaseDiag}a which represents wave velocity anisotropy as the ratio of the minimum and maximum quasi-longitudinal wave $v_\mathrm{min}/v_\mathrm{max}$; the ratio spans the range $0.2$ (highly anisotropic) to $0.98$ (almost perfectly isotropic). The resulting achievable refractive index library is shown in Fig.~\ref{fig:SI-refractiveInd}.
\clearpage

\clearpage
\subsection{Square and hexagonal lattice wave velocities}\label{sec:sqHex_wv}
\begin{figure}[hbt!]
    \centering
    \includegraphics[width=\linewidth]{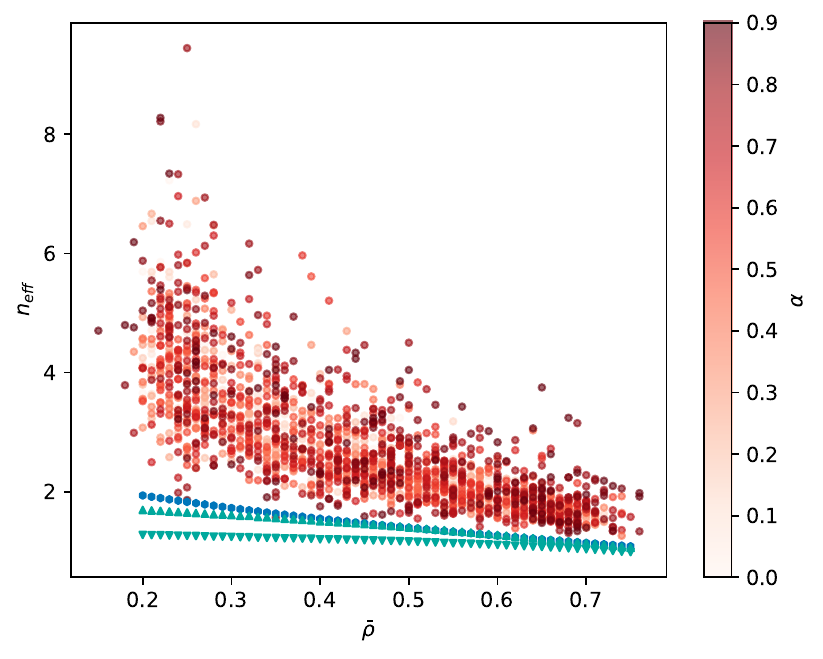}
    \caption{\textbf{Refractive indices for control lattices.} Refractive index $n$ as a function of relative density for the isotropic hexagonal lattice (blue circles) and directional square lattice (triangles). Refractive index for the SR metamaterials (red circles) is also plotted for comparison, with marker colors graded with $\alpha$.}
    \label{fig:SI-sqHexN}
\end{figure}

We calculated the directional wave velocities of the hexagonal and square lattices -- using the Bloch analysis technique described in the previous section -- for relative densities ranging $\bar{\rho} = 20-75\%$. In Fig.~\ref{fig:SI-sqHexN} we plot the refractive index $n_\mathrm{eff}$ for the isotropic hexagonal lattice and the minimum and maximum refractive index of the directional square lattice. Refractive index values for SR unit cells (plotted for comparison) achieve a much wider $1.5 \leq n_\mathrm{eff} \leq 8.3$ than the range for the hexagonal and square lattices combined ($1.0 \leq n_\mathrm{eff} \leq 1.9$), showing that SR metamaterials posses superior wave control tunability to standard lattices.

\clearpage
\subsection{Elastic response of 3D plate-based SR metamaterials}
\label{sec:3D_plates}
\begin{figure}[h]
    \centering
    \includegraphics[width=0.8\linewidth]{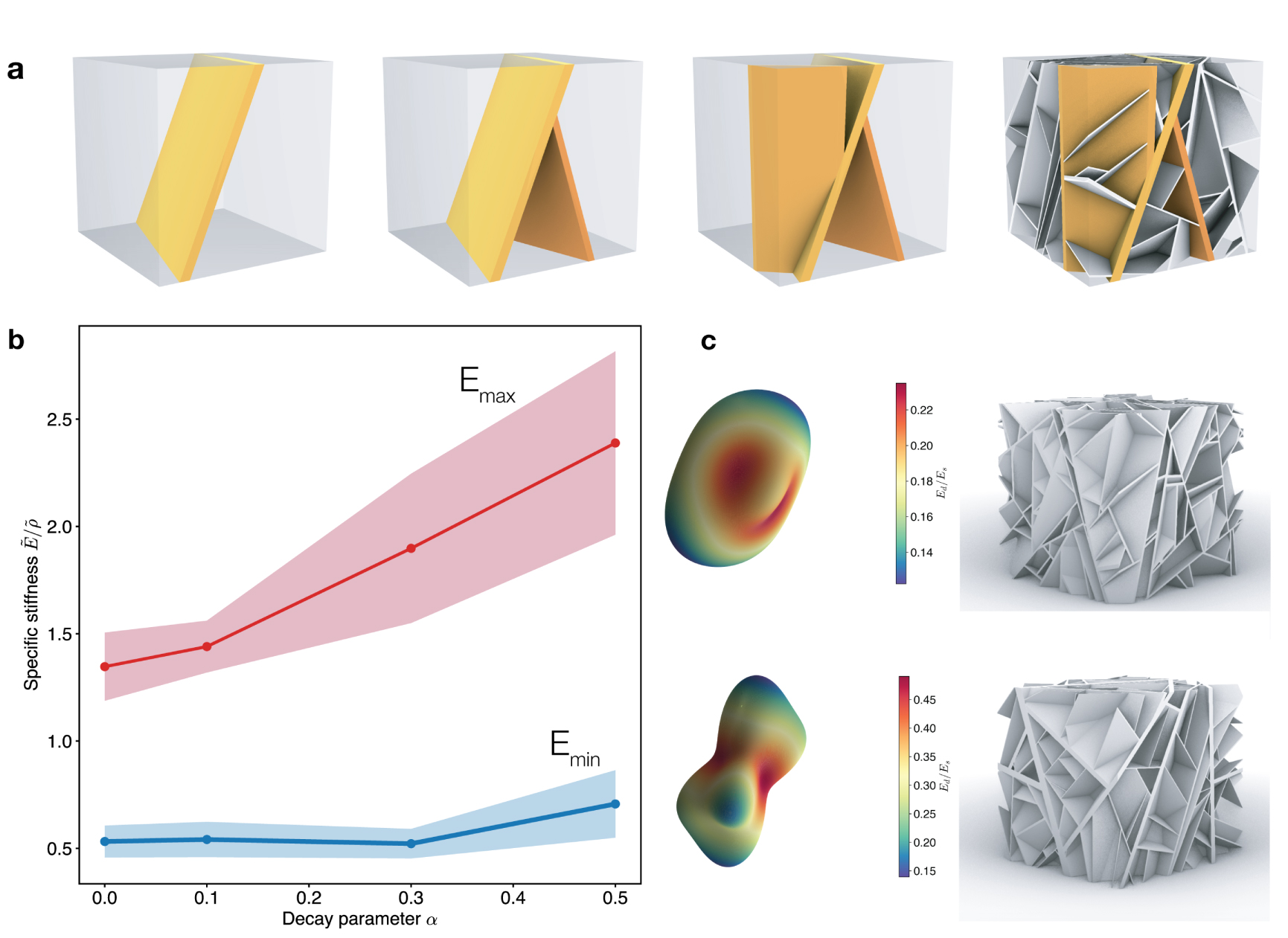}
    \caption{ \textbf{3D plate-based SR metamaterials.} \textbf{(a)} Generation steps $T = 1$, $2$, $3$, and $100$. \textbf{(b)} Minimum and maximum stiffness as a function of $\alpha$, showing a similar trend to our 2D systems. \textbf{(c)} Examples of directional stiffness surfaces for systems of different $\alpha$.}
    \label{fig:3D_plates}
\end{figure}

The extension of the 2D ligament-based Scale-Rich model to 3D consists of generating finite-thickness plates in a three-dimensional domain, e.g., a unit cube (Fig.~\ref{fig:3D_plates}a),  as described in SI~\ref{subsec: model_limit}.

To evaluate the directional stiffness of the plate-based geometries, we extended our linear-elastic homogenization scheme to 3D finite-element simulations. The geometries were meshed with ten-node, quadratic tetrahedral continuum elements (C3D10) and 3D affine boundary conditions were applied to compute the full homogenized elastic tensor $\mathbb{C}$ for each specimen. From a set of three axial and three shear simulations, we extracted the directional stiffness \textit{surfaces} for each geometry (Fig.~\ref{fig:3D_plates}c), normalizing against the elastic properties of the constituent material. From these surfaces we quantified the maximum and minimum specific stiffnesses (Fig.~\ref{fig:3D_plates}b), finding that the trends observed for the 2D ligament-based samples extend to three dimensions.

\end{document}